\documentclass[usenatbib]{mn2e}
\usepackage[pdftex]{graphicx}
\usepackage[fleqn]{amsmath}
\usepackage{lscape}

\defcitealias{2012MNRAS.426.1475U}{U12}
\defcitealias{2013MNRAS.428..389P}{P13}

\begin{document}

\title[Globular Clusters of NGC 4278]{The SLUGGS Survey: Wide Field Imaging of the Globular Cluster System of NGC 4278}
\author[Usher et al.]{Christopher~Usher,$^{1}$\thanks{E-mail: cusher@astro.swin.edu.au} Duncan~A.~Forbes,$^{1}$ Lee~R.~Spitler,$^{2,3}$ Jean~P.~Brodie,$^4$ \newauthor Aaron~J.~Romanowsky,$^{4,5}$, Jay~Strader,$^{6}$ Kristin~A.~Woodley$^{4}$ \\
$^1$Centre for Astrophysics and Supercomputing, Swinburne University of Technology, Hawthorn, VIC 3122, Australia\\
$^2$Department of Physics and Astronomy, Macquarie University, North Ryde NSW 2109, Australia\\
$^3$Australian Astronomical Observatory, PO Box 915, North Ryde, NSW 1670, Australia\\
$^4$University of California Observatories, 1156 High Street, Santa Cruz, CA 95064, USA\\
$^5$Department of Physics and Astronomy, San Jos\'e State University, One Washington Square, San Jose, CA 95192, USA \\
$^6$Department of Physics and Astronomy, Michigan State University, East Lansing, Michigan 48824}
\maketitle

\begin{abstract}

We use multi-pointing HST ACS and wide field Subaru Suprime-Cam imaging to study the globular cluster system of the L* elliptical galaxy NGC 4278.
We have also obtained a handful of new globular cluster spectra with Keck/DEIMOS.
We determine the globular cluster surface density profile and use it to calculate the total number of globular clusters, finding the system to be slightly more populous than average for galaxies of its luminosity.
We find clear evidence for bimodality in the globular cluster colour distribution and for a colour--magnitude relation in the blue subpopulation (a `blue tilt').
We also find negative radial colour gradients in both colour subpopulations of equal strength which are similar in strength to those reported in other galaxies. 
The sizes of NGC 4278's globular clusters decrease with redder colours and increase with galactocentric radius.
The ratio of the sizes of blue to red globular clusters is independent of galactocentric radius demonstrating that internal effects are responsible for the size difference with colour.

\end{abstract}


\section{Introduction}

Found in galaxies of all shapes and sizes, globular clusters (GCs) are fossils of galaxy formation as they are among the oldest objects in the universe (ages $> 10$ Gyr, e.g. \citealt[][]{2001ApJ...563L.143F, 2005A&A...439..997P, 2005AJ....130.1315S}).
Extensive optical imaging studies \citep[e.g.][]{2001AJ....121.2950K, 2001AJ....121.2974L} have shown that almost all massive galaxies have bimodal GC colour distributions.
Although the origin of the colour bimodality has been debated \citep[e.g.][]{2006Sci...311.1129Y}, large spectroscopic studies \citep[e.g.][]{2007AJ....133.2015S, 2008MNRAS.386.1443B, 2011MNRAS.417.1823A, 2012MNRAS.426.1475U, 2012ApJ...759L..33B} have now shown that most galaxies have bimodality metallicity distributions.
Typically the metal rich, red subpopulations show similar spatial, kinematic and stellar population properties to the galaxy stellar bulges while the metal poor, blue subpopulations have been tentatively connected to the galaxy's halo \citep[and references therein]{2012MNRAS.425...66F}.
Due to the low surface brightness of galaxy halos and since GCs are much brighter than individual stars, investigations of blue GCs are one of the few ways (other methods include planetary nebula and X-ray hot gas) to study halos beyond the local group.

Here we study the giant elliptical NGC 4278 ($M_{K} = -23.8$, \citealt{2011MNRAS.413..813C}), part of the Coma I group \citep{1977ApJ...213..345G, 1996AJ....112.1409F}.
NGC 4278 is part of the galaxy sample of the SAGES Legacy Unifying Globulars and Galaxies Survey\footnote{http://sluggs.swin.edu.au} (SLUGGS, Brodie et al. in prep.), an ongoing chemodynamical study of nearby early-type galaxies and their GC systems.
NGC 4278 hosts a weak radio AGN with observed radio lobes \citep{2005ApJ...622..178G} that alternates between LINER-like and Seyfert-like nuclear activity \citep{2010A&A...517A..33Y}.
The galaxy also hosts a massive, warped H I disc \citep{2006MNRAS.371..157M}, an old, $\alpha$-element enhanced and solar metallicity \citep{2010MNRAS.408...97K} stellar population with no sign of ongoing star formation \citep{2010MNRAS.402.2140S}.
The intermediate luminosity elliptical galaxy NGC 4283 ($M_{K} = -21.8$ \citep{2011MNRAS.413..813C}) is 3.5 arcmins away from NGC 4278, a separation of 16 kpc at our adopted distance to NGC 4278 of 15.6 Mpc ($m - M = 30.97$), the same distance adopted by \citet{2011MNRAS.413..813C} for their ATLAS$^{\text{3D}}$ survey.
The distances, which are based on the surface brightness fluctuation (SBF) distances of \citet{2001ApJ...546..681T}, are consistent with both galaxies being at the same distance.
At the adopted distance, NGC 4278's effective radius of $R_{e} = 32$ arcsec \citep{2011MNRAS.413..813C} is equivalent to 2.4 kpc. 

The GC system of NGC 4278 was first studied by \citet{1981AJ.....86.1627H} using Canada-France-Hawaii Telescope photographic plates.
\citet{1996ApJ...467..126F} and \citet{2001AJ....121.2950K} used Hubble Space Telescope (HST) Wide Field Planetary Camera 2 (WFPC2)  $VI$-band imaging to study the GC system but did not find clear evidence of bimodality.
\citet{2009ApJS..181..605B} and \citet{2010ApJ...725.1824F} used the WFPC2 imaging to study the connection between low-mass X-ray binaries and GCs in NGC 4278.
\citet{2011A&A...525A..19C, 2011A&A...525A..20C, 2012A&A...539A..54C} used William Herschel Telescope Long-slit Intermediate Resolution Infrared Spectrograph $K$-band imaging and HST Advanced Camera for Surveys (ACS) $gz$-band imaging of NGC 4278 as part of a larger of study of the near-infrared (NIR) colours of GCs.
As part of a SLUGGS study of GC kinematics, \citet[hereafter P13]{2013MNRAS.428..389P} used Keck Deep Imaging Multi-Object Spectrograph (DEIMOS) spectroscopy to measure the radial velocity of 256 GCs in NGC 4278.
In addition to finding clear colour bimodality, they found that the red GCs do not rotate while the blue GCs do.
As part of a SLUGGS study of GC metallicity distributions, \citet[hereafter U12]{2012MNRAS.426.1475U} used the same spectroscopy to measure metallicities for 150 GCs and found a bimodal metallicity distribution.

This paper adds to the handful of multi-pointing HST imaging studies of GC systems such as those of NGC 4594 \citep{2006AJ....132.1593S, 2010MNRAS.401.1965H}, NGC 4365 \citep{2012MNRAS.420...37B} and NGC 4649 \citep{2012ApJ...760...87S}. 
The radial range provided by five ACS pointings of NGC 4278 allow us to study how GC colour and size change with radius.
We use the same instrument and filters as well as similar exposure times as the ACS Virgo Cluster Survey \citep{2004ApJS..153..223C}, ACS Fornax Cluster Survey \citep{2007ApJS..169..213J} and other multi-pointing ACS studies such as \citet{2012MNRAS.420...37B} and \citet{2012ApJ...760...87S}, allowing for direct comparisons.
Compared to previous multi-pointing HST GC studies of early-type galaxies, NGC 4278 is a significantly less luminous galaxy. 
NGC 4278 has a $K$-band luminosity of  $6.7 \times 10^{10} L_{\sun}$ while NGC 4594, NGC 4365, NGC 4649 have $K$-band luminosities of $1.9 \times 10^{11} L_{\sun}$, $2.5 \times 10^{11} L_{\sun}$ and $3.1 \times 10^{11} L_{\sun}$ respectively \citep{2011MNRAS.413..813C}.
Using the galaxy densities in the \citet{1988cng..book.....T} catalogue, NGC 4278's environment (1.25 Mpc$^{-3}$) is intermediate to that of NGC 4594 (0.32 Mpc$^{-3}$) and of NGC 4365 and NGC 4649 (2.93 and 3.49 Mpc$^{-3}$ respectively.)

In this paper we use HST ACS and Subaru Suprime-Cam wide field imaging to study the properties of a large number of GCs in NGC 4278.
We first describe the imaging reduction before explaining the selection of GC candidates (Section~\ref{data}).
We study the spatial distribution of NGC 4278's GCs and determine the size of the GC system (Section~\ref{surfacedensity}).
In this section we also investigate the contribution of the nearby galaxy NGC 4283 to NGC 4278's GC system.
We then study the colour distribution including second order effects such as the relationships between magnitude and colour and between galactocentric distance and colour (Section~\ref{colour}).
We also use the superior angular resolution of ACS to measure GCs sizes, allowing us to study the relations between GC half light radii and colour, galactocentric distance and magnitude (Section~\ref{size}).
In Section~\ref{summary} we summarise our results.

\section{Data acquisition and analysis}
\label{data}
\begin{figure*}
	\begin{center}
   
		\includegraphics[width=504pt]{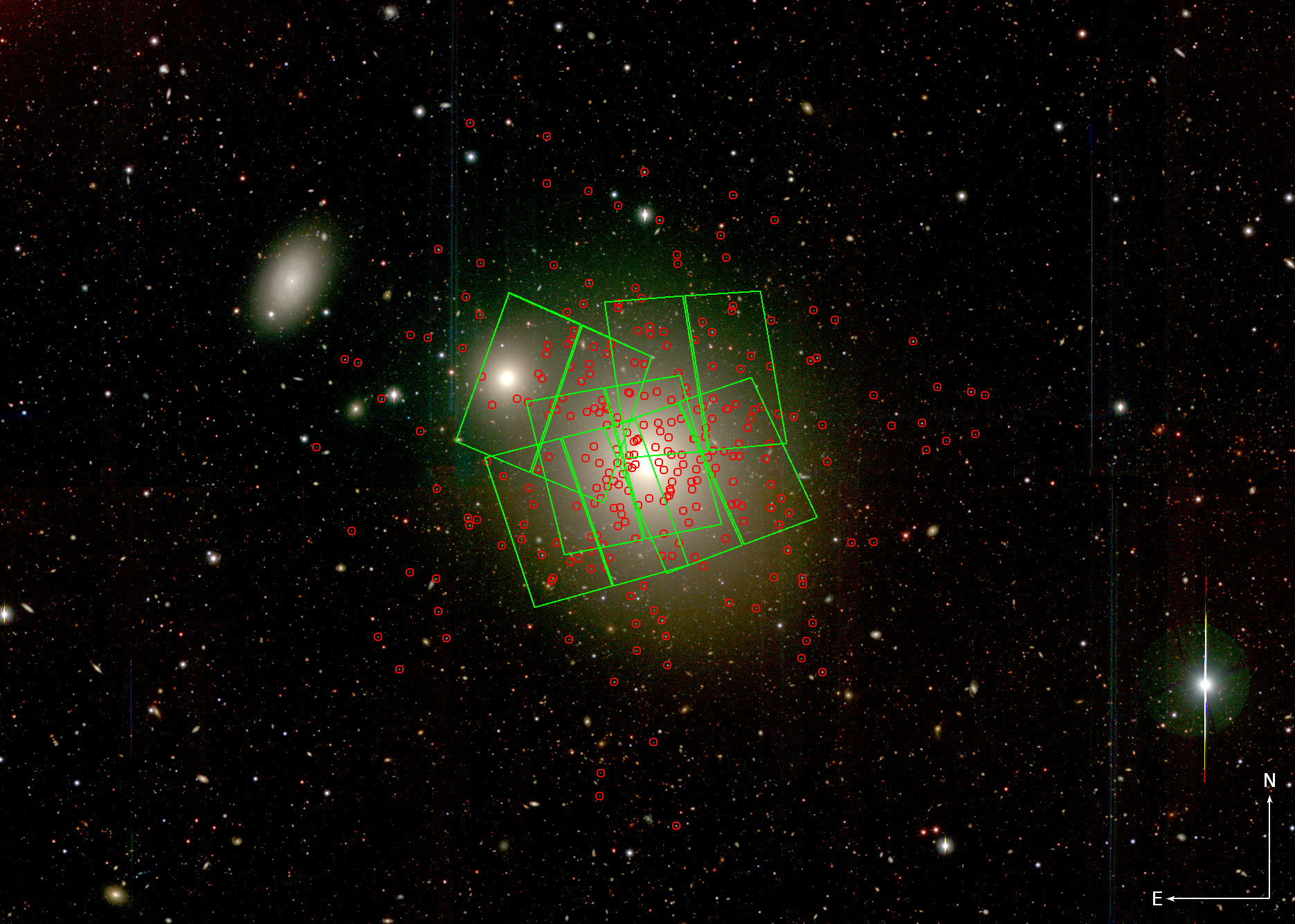}
		\caption{\label{fig:globs} Central 28 $\times$ 20 arcmin (125 $\times$ 90 kpc) of the Suprime-Cam three filter imaging of NGC 4278.
The green lines show the outlines of five ACS pointings.
The red circles show the locations of spectroscopically confirmed GCs.
The fainter galaxy covered by ACS imaging is NGC 4283; the disc galaxy to the north east is NGC 4286.}
	\end{center}
\end{figure*}

\subsection{HST ACS imaging}
HST ACS wide field camera images in two bands, F475W and F850LP (hereafter $g$ and $z$), for five pointings were downloaded from the Hubble Legacy Archive.
Details of the images are given in Table~\ref{tab:acs} and the footprints of the pointings are shown in Figure~\ref{fig:globs}.
The ACS imaging extends to 4.5 arcmin from the centre of NGC 4278.
The filters and exposure times are similar to those of the ACS Virgo Cluster Survey \citep{2004ApJS..153..223C} and the ACS Fornax Cluster Survey \citep{2007ApJS..169..213J} allowing a direct comparison between our results and those of these two large GC studies.
Four of these pointings were previously analysed in \citetalias{2013MNRAS.428..389P}.
A custom pipeline was used to measure aperture photometry \citep{2006AJ....132.1593S, 2006AJ....132.2333S}.
The photometry was calibrated to the AB photometric system using the zero points of \citet{2005PASP..117.1049S}.
The magnitudes were measured using 5 pixel (0.25 arcsec) apertures.
For objects in multiple pointings, we averaged the magnitudes measured in each pointing.
In both filters the difference between the magnitudes measured using a 5 pixel aperture and a 10 pixel aperture is dependent on the half light radius size of the object.
Using the sizes calculated below, a quadratic relation was derived between the half light radius and the aperture difference for each band.
This relation was used to correct each of the measured magnitudes from 5 pixel to 10 pixel apertures.
The magnitudes were corrected from 10 pixels to infinity using the corrections provided by \citet{2005PASP..117.1049S}.
The uncertainty in the aperture correction was added in quadrature to the photometric error to give the quoted uncertainty for each total magnitude.
Finally, the magnitudes were corrected for extinction using the dust maps of \citet{1998ApJ...500..525S}.

\begin{table}
	\caption{\label{tab:acs}HST ACS Observations}
	\begin{tabular}{c c l r c} \hline
		\multicolumn{2}{c}{Pointing Centre} & \multicolumn{2}{c}{Exposures} & Program\\ 
		RA. [deg] & Dec. [deg] & Filter & Time [s] & ID \\ \hline
		185.038292 & 29.279528 & F475W & 676 & 10835\\
		&& F850LP & 1240\\
		185.002458 & 29.275750 & F475W & 702 & 10835\\
		&& F850LP & 1200\\
		185.053667 & 29.262000 & F475W & 702 & 10835\\
		&& F850LP & 1200\\
		185.008542 & 29.312139 & F475W & 702 & 10835\\
		&& F850LP & 1200\\
		185.066708 & 29.303611 & F475W & 758 & 11679\\
		&& F850LP & 1287\\ \hline
	\end{tabular}
\end{table}

\label{sizemeasurements}
To help distinguish GCs from stars the superior angular resolution of the ACS data was used to measure the sizes of slightly resolved objects.
The \textsc{daophot} package in \textsc{iraf} was used to generate empirical point spread functions for each band from bright stars in the images.
Each pointing and each filter was measured individually using \textsc{ishape} \citep{1999A&AS..139..393L} using circular \citet{1962AJ.....67..471K} profiles with a concentration parameter fixed at $r_{t}/r_{c} = 30$.
The uncertainty on the measured sizes was provided by \textsc{ishape}.
The weighted average of the $g$ and $z$ sizes from all pointings was taken as the true half light radius.

\subsection{Subaru Suprime-Cam imaging}
To study the GC system out to large galactocentric radii, Subaru Suprime-Cam \citep{2002PASJ...54..833M} imaging of NGC 4278 was obtained from the SMOKA data archive \citep{2002ASPC..281..298B} in the $BVI$ filters.
The Suprime-Cam imaging was observed on 2002 February 13 and covers a field of view of 36 $\times$ 29 arcmin centred on NGC 4278.
Details of the exposures and the seeing are given in Table~\ref{tab:supcam} and the imaging is shown in Figure~\ref{fig:globs}.

\begin{table}
	\caption{\label{tab:supcam}Suprime-Cam Observations}
	\begin{tabular}{c c c c} \hline
		Filter & Exposures & Total & Mean Seeing \\ \hline
		$B$ & $3 \times 200$ s & 600 s & 1.4 arcsec \\
		$V$ & $3 \times 150$ s & 450 s & 0.9 arcsec \\
		$I$ & $3 \times 120$ s & 360 s & 0.9 arcsec \\ \hline
	\end{tabular}
\end{table}

A modified version of the Suprime-Cam \textsc{SDFRED} package \citep{2002SPIE.4847..322Y,2004ApJ...611..660O} was used to reduce the raw images.
Standard aperture photometry techniques using \textsc{IRAF} \textsc{DAOPHOT} package \citep{1992ASPC...25..297S} were used to extract the photometric data for point like objects in the images.
The astrometry was calibrated to the USNO-B2 astrometric system.

Photometry from the Sloan Digital Sky Survey (SDSS) was used to find the photometric zero points for the Suprime-Cam imaging.
Point-like objects with good Suprime-Cam photometry and GC like colours more than 1 arcmin from the centre of NGC 4278 were matched with stars in Data Release 7 of the SDSS catalogue \citep{2009ApJS..182..543A}.
From 280 matching objects with magnitudes $17.6 < i_{SDSS} < 22$, Johnson-Cousins $BVI$ magnitudes in the Vega system were calculated using the transformations provided in \citet{2007AJ....134..973I} which are based on the \citet{2000PASP..112..925S} extensions of the \citet{1992AJ....104..340L} standards to fainter magnitudes.
A zero point including a $(V - I)$ colour term was fitted using weighted least squares to the difference between the instrument magnitudes and the magnitudes calculated from the SDSS photometry.
While the colour term is small for the $V$ and $I$ bands, it is significant for the $B$ band.
The uncertainties in the zero point fits were used to estimate the uncertainties in the corrected magnitudes and were added in quadrature to the photometric uncertainties.
Finally, the magnitudes were corrected for extinction using the dust maps of \citet{1998ApJ...500..525S}.

To directly compare the ACS photometry to the Suprime-Cam photometry we used 168 spectroscopically confirmed GCs with both ACS and Suprime-Cam photometry to determine the following relations:
\begin{equation}\label{eq:vigz}
(g - z) = (1.72 \pm 0.03) \times (V - I) + (-0.71 \pm 0.03)
\end{equation}
\begin{equation}
z = I + (-0.25 \pm 0.04) \times (V - I) + (0.62 \pm 0.05) .
\end{equation}
The root mean squared difference of both these relations is 0.07 mag and is consistent with the photometric errors.
 
\subsection{Globular cluster selection}
While the bulk of point-like objects in both the ACS imaging and the Suprime-Cam imaging are GCs, contamination by foreground stars and background galaxies must be removed to produce a clean sample of GC candidates.
We used the spectroscopically confirmed GCs to guide our selection of GCs.

\begin{figure*}
	\begin{center}
		\includegraphics[width=504pt]{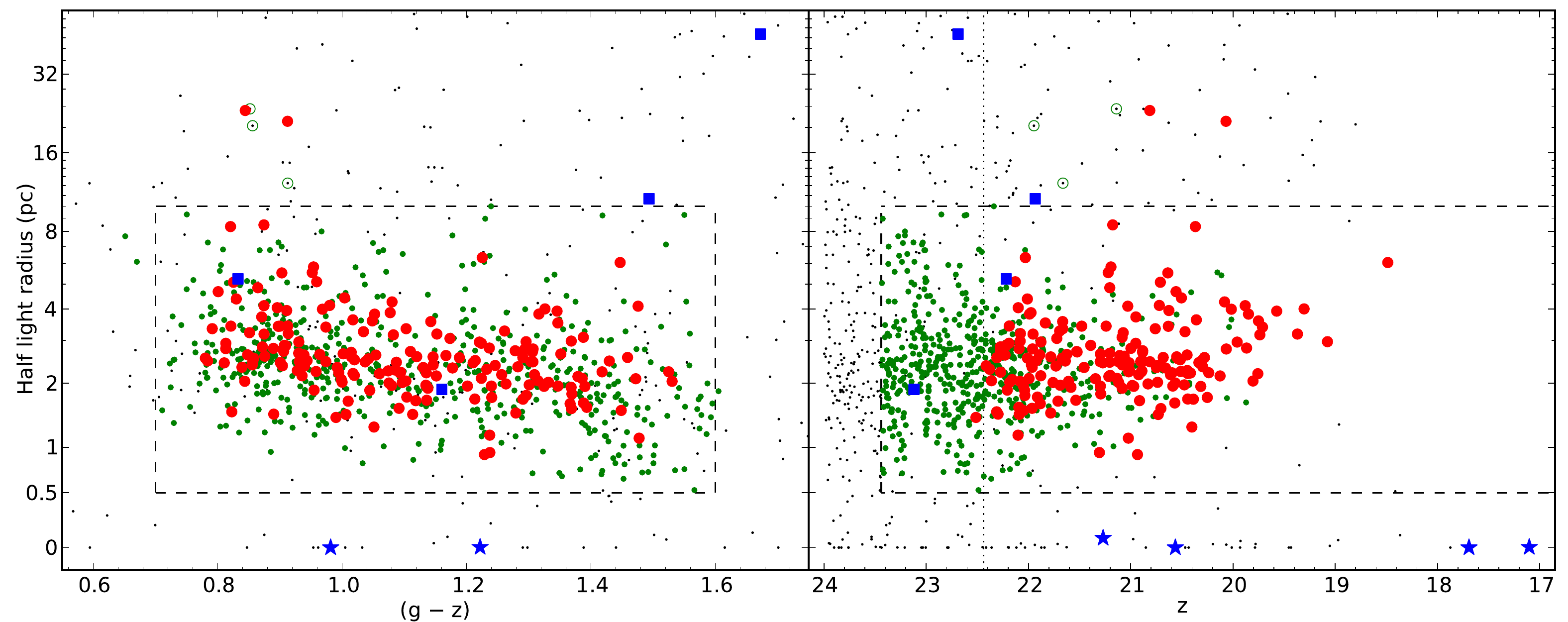}
		\caption{\label{fig:acsselection}
ACS GC selection.
\emph{Left} Half light radius versus colour selection for the ACS imaging.
The small black dots are objects in the ACS imaging brighter than $z = 24$, the small green circles are all objects meeting the selection criteria and the large red circles are spectroscopically confirmed GCs and UCDs.
Blue stars are spectroscopically confirmed stars and blue squares are spectroscopically confirmed galaxies.
The circled points are UCD candidates.
The dashed line box shows the size and colour criteria used to select GC candidates.
\emph{Right} Half light radius versus magnitude selection for the ACS imaging.
The symbols are identical to the left panel and the dashed line box shows the size and magnitude criteria used to select GC candidates.
The dotted vertical line is the turnover magnitude of $z = 22.44$.
Both panels use an inverse hyperbolic sine scale for the half light radii.
The brightest two stars were observed as guide stars; the remaining two stars were observed to fill the slit masks.
} 
	\end{center}
\end{figure*}

GC candidates were selected from objects in the ACS imaging on the basis of their colour, size and magnitude.
The colour--size selection is displayed the left side of Figure~\ref{fig:acsselection} and the magnitude--size selection on the right side of Figure~\ref{fig:acsselection}.
Before this selection was made, objects with irregular morphologies were removed by hand.
Objects whose photometric errors put them within the colour range $0.7 < (g - z) < 1.6$ were considered to be candidates.
Using the colour--metallicity transformation provided by \citetalias{2012MNRAS.426.1475U} the limits correspond to [Z/H] $= -2.12$ and [Z/H] $= +0.60$ respectively while using the colour--metallicity relation of \citet{2006ApJ...639...95P} the limits correspond to [Fe/H] $= -2.61$ and [Fe/H] $= +0.18$.
Objects with half light radii between 0.5 pc and 10 pc were selected as GC candidates.
The lower limit of 0.5 pc includes smaller red GCs while the upper limit was chosen to reject background galaxies.
\citet{2011AJ....142..199B} adopted 10 pc as the dividing line between GCs and UCDs.

\begin{figure}
	\begin{center}
   
		\includegraphics[width=240pt]{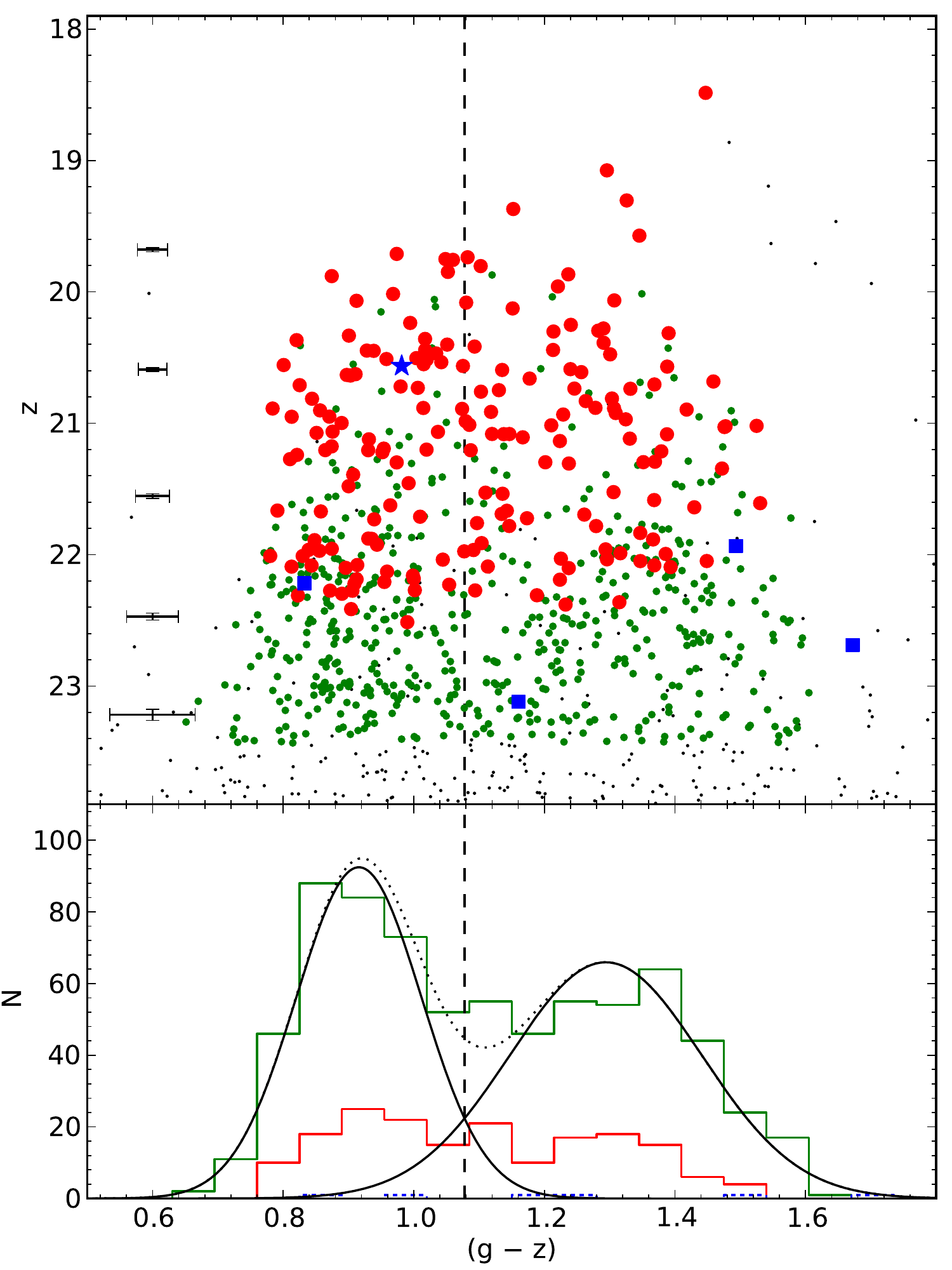}
		\caption{\label{fig:acscolourmag}ACS GC candidates.
\emph{Top panel} ACS colour--magnitude diagram.
Black points are objects in the ACS catalogue; green small circles are GC candidates selected by size, colour and magnitude; red circles are spectroscopically confirmed GCs; blue stars indicate spectroscopically confirmed stars; blue squares indicate spectroscopically confirmed galaxies.
Mean error bars are shown on the left.
\emph{Bottom panel} ACS colour histograms.
The green histogram is all GC candidates; the red histogram is spectroscopically confirmed GCs; the dashed blue histogram is spectroscopically confirmed stars and galaxies.
Over-plotted in black are the Gaussian fits from \textsc{GMM}.
The black dashed line shows the colour cut $(g - z) = 1.08$ used to divide the sample into red and blue subpopulations.}

	\end{center}
\end{figure}

Objects were selected as GC candidates if they are brighter than $z = 23.44$.
This limit corresponds to one magnitude fainter than the $z$ band turnover magnitude of the GC luminosity function of $z = 22.44$ which corresponds to the absolute magnitude of $M_{z} = -8.53$ \citep{2010ApJ...717..603V} at the distance of NGC 4278.
One magnitude corresponds to the dispersion of the GC luminosity function for a galaxy of NGC 4278 luminosity \citep{2010ApJ...717..603V}.
A colour magnitude diagram of the ACS GC candidates is shown in Figure~\ref{fig:acscolourmag}.
The ACS GC candidates are listed in Table~\ref{tab:acsgcc}, the full version of which appears online.

Objects brighter than the turnover magnitude that are larger than our half light radius cut of 10 pc but pass the colour cuts were visually inspected.
Extended objects with round, regular morphologies are listed in Table~\ref{tab:ucdc} as UCD candidates.
We do not include these UCD candidates in further analysis.

\begin{figure}
	\begin{center}
   
		\includegraphics[width=240pt]{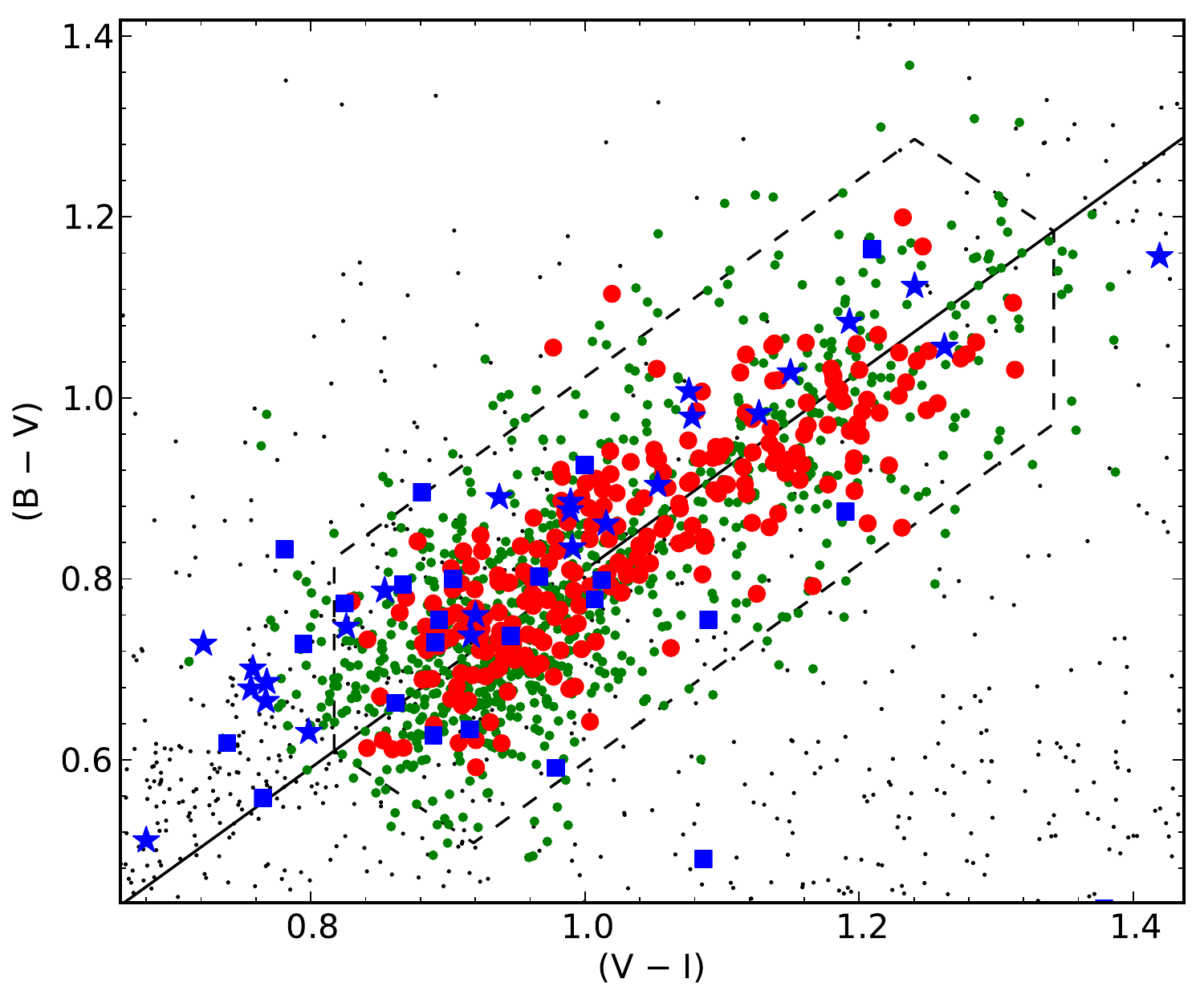}
		\caption{\label{fig:subcolourcolour} Colour--colour selection for the Suprime-Cam imaging.
The small black dots are point like objects detected in all three Suprime-Cam bands, the small green circles are all objects meeting the selection criteria and the large red circles are spectroscopically confirmed GCs.
Blue stars are spectroscopically confirmed stars and blue squares are spectroscopically confirmed galaxies.
The solid line shows the $(B - V)$ to $(V - I)$ relation fit to spectroscopically confirmed GCs while the dashed lines are the colour cuts applied.
The two spectroscopically confirmed GCs which fail the Suprime-Cam selection criteria have poor Suprime-Cam photometry due to crowding or image artifacts.
Both objects are covered by ACS imaging and satisfy the ACS selection criteria.}
	\end{center}
\end{figure}

\begin{figure}
	\begin{center}
   
		\includegraphics[width=240pt]{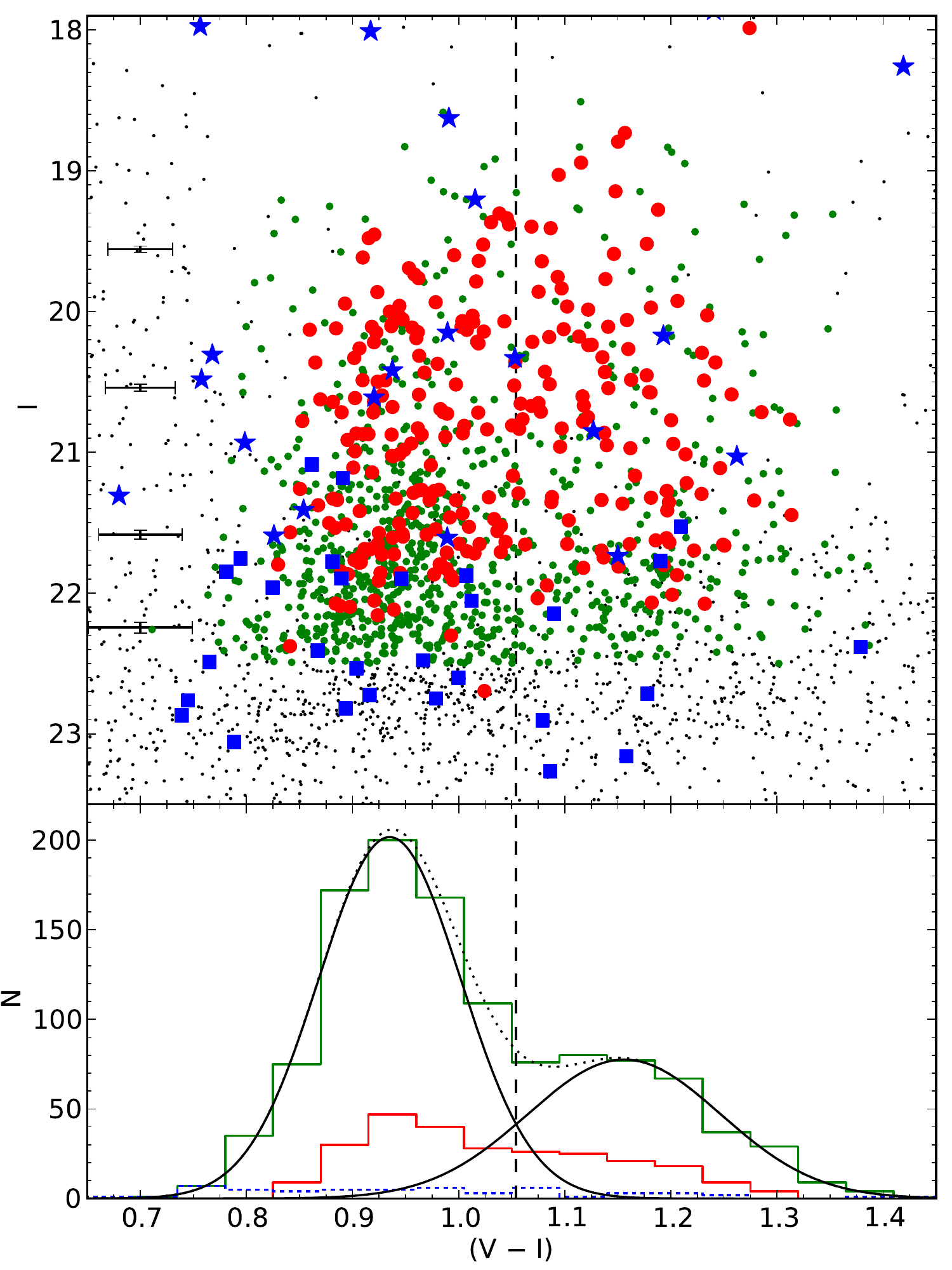}
		\caption{\label{fig:subcolourmag}Suprime-Cam GC candidates.\emph{Top panel} Suprime-Cam colour--magnitude diagram.
Black points are all objects in the Suprime-Cam catalogue; green small circles are GC candidates selected by colour and magnitude; red circles are spectroscopically confirmed GCs; blue stars indicate spectroscopically confirmed stars; blue squares show spectroscopically confirmed galaxies.
Mean error bars are shown on the left.
\emph{Bottom panel} Suprime-Cam colour histograms.
The green histogram is GC candidates; the red histogram is spectroscopically confirmed GCs; the dashed blue histogram is spectroscopically confirmed stars and galaxies.
Over-plotted in black are the Gaussian fits from \textsc{GMM}.
The black dashed line shows the colour cut $(V - I) = 1.05$ used to divide the sample into red and blue subpopulations.}

	\end{center}
\end{figure}

GC candidates were selected from objects in the Suprime-Cam imaging on the basis of their colour and magnitude.
First, extended objects such as background galaxies where rejected by using the difference between measured magnitudes of different sized apertures.
The selection in $(B - V)$ versus $(V - I)$ space used to select candidates is shown in Figure~\ref{fig:subcolourcolour}.
A line was fit through the spectroscopically confirmed GCs: 
\begin{equation}
(B - V) =  1.09 \times (V - I) - 0.28 .
\end{equation}
Objects whose uncertainties put them within three times the scatter (0.05 mag) of the line, within $1.43 < (B - I) < 2.53$ and within $0.82 < (V - I) < 1.34$ were selected.
The $(B - I)$ and $(V - I)$ limits correspond to $0.70 < (g - z) < 1.60$.
Objects passing the colour cuts were considered GC candidates if their magnitudes were in the range $18.5 < I < 22.5$.
The upper magnitude limit was chosen to match the brightest GC candidate in the ACS imaging.
A colour magnitude diagram of the Suprime-Cam GC candidates is shown in Figure~\ref{fig:subcolourmag}.
The Suprime-Cam GC candidates are listed in Table~\ref{tab:subgcc}, the full version of which appears online.

\begin{table*}
	\caption{\label{tab:acsgcc}ACS globular cluster candidates}
	\begin{tabular}{c@{\ \ }c@{\ \ }c@{\ \ }c@{\ \ }c@{\ \ }c@{\ \ }c@{\ \ }c@{\ \ }c@{\ \ }c@{\ \ }c} \hline
Name      & RA         & Dec       & $(g - z)$       & $z$              & $r_{hl}$      & $v$           & [Z/H]                   & P13            & Suprime    & Note \\
          & [deg]      & [deg]     & [mag]           & [mag]            & [pc]          & [km s$^{-1}$] & [dex]                   &                & Cam        &      \\
(1)       & (2)        & (3)       & (4)             & (5)              & (6)           & (7)           & (8)                     & (9)            & (10)       & (11) \\ \hline
acs\_1369 & 185.015995 & 29.277228 & $1.45 \pm 0.02$ & $18.49 \pm 0.02$ & $6.1 \pm 0.3$ & $546 \pm 5$   & $-0.51_{-0.12}^{+0.26}$ & NGC4278\_GC261 & ---        & ---  \\ 
acs\_1102 & 185.033001 & 29.288592 & $1.30 \pm 0.01$ & $19.08 \pm 0.01$ & $3.0 \pm 0.2$ & $479 \pm 9$   & $-0.32_{-0.23}^{+0.22}$ & NGC4278\_GC22  & ---        & ---  \\
acs\_0324 & 185.023757 & 29.280609 & $1.33 \pm 0.02$ & $19.30 \pm 0.01$ & $4.0 \pm 0.1$ & $614 \pm 6$   & $-0.72_{-0.22}^{+0.22}$ & NGC4278\_GC96  & sub\_32436 & ---  \\
acs\_1138 & 185.036859 & 29.291440 & $1.15 \pm 0.01$ & $19.37 \pm 0.01$ & $3.2 \pm 0.2$ & $886 \pm 4$   & $-0.41_{-0.17}^{+0.16}$ & NGC4278\_GC78  & sub\_33742 & ---  \\
acs\_1115 & 185.027438 & 29.329398 & $1.35 \pm 0.02$ & $19.57 \pm 0.01$ & $3.9 \pm 0.1$ & $556 \pm 5$   & $-0.44_{-0.13}^{+0.15}$ & NGC4278\_GC65  & sub\_36928 & ---  \\
  ...     & ...        & ...       & ...             & ...              & ...           & ...           & ...                     & ...            & ...        & ...  \\ \hline
	\end{tabular}
	
	\medskip
	The full version of this table is provided in a machine readable form in the online Supporting Information.
	\emph{Notes} Column (1): GC IDs.
	Column (2) and (3): Right ascension and declination in the J2000.0 epoch, respectively.
	Column (4): $(g - z)$ colour.
	Column (5): $z$-band magnitude.
	Column (6): Half light radius in parsecs.
	Column (7): Radial velocity.
	Column (8): Calcium triplet metallicity.
	Column (9): Identifier from \citetalias{2013MNRAS.428..389P}.
	For objects that were observed spectroscopically for the first time in this work we have extended the \citetalias{2013MNRAS.428..389P} naming scheme to them.
	Column (10): Suprime-Cam identifier if Suprime-Cam candidate
	Column (11): Note: G = Spectroscopic background galaxy.
\end{table*}

\begin{table*}
	\caption{\label{tab:ucdc}ACS ultra compact dwarf candidates}
	\begin{tabular}{c c c c c c c c c c} \hline
Name      & RA         & Dec       & $(g - z)$       & $z$              & $r_{hl}$       & $v$            & [Z/H]                   & Suprime    & Note \\
          & [deg]      & [deg]     & [mag]           & [mag]            & [pc]           & [km s$^{-1}$]  & [dex]                   & Cam        &      \\
(1)       & (2)        & (3)       & (4)             & (5)              & (6)            & (7)            & (8)                     & (9)        & (10) \\ \hline
acs\_0259 & 185.004411 & 29.293119 & $0.91 \pm 0.05$ & $20.07 \pm 0.03$ & $21.1 \pm 0.6$ & $674 \pm 14$   & $-0.79_{-0.46}^{+0.29}$ & sub\_33897 & ---  \\
acs\_0320 & 185.046099 & 29.242682 & $0.84 \pm 0.05$ & $20.81 \pm 0.04$ & $23.2 \pm 0.5$ & $719 \pm 18$   & ---                     & sub\_28313 & ---  \\
acs\_1643 & 185.012480 & 29.299562 & $0.85 \pm 0.05$ & $21.14 \pm 0.03$ & $23.6 \pm 0.6$ & ---            & ---                     & sub\_34690 & ---  \\
acs\_1280 & 185.030887 & 29.251737 & $0.91 \pm 0.05$ & $21.66 \pm 0.03$ & $12.3 \pm 0.8$ & ---            & ---                     & sub\_29090 & ---  \\ 
acs\_0284 & 184.990593 & 29.264914 & $1.49 \pm 0.08$ & $21.93 \pm 0.04$ & $10.7 \pm 1.3$ & $z = 0.35$     & ---                     & sub\_30229 & G    \\
acs\_0079 & 185.005262 & 29.320994 & $0.86 \pm 0.06$ & $21.94 \pm 0.04$ & $20.3 \pm 1.1$ & ---            & ---                     & sub\_36325 & ---  \\ \hline
	\end{tabular}
	
	\medskip
	\emph{Notes} Column (1): UCD IDs.
	Column (2) and (3): Right ascension and declination in the J2000.0 epoch, respectively.
	Column (4): $(g - z)$ colour.
	Column (5): $z$-band magnitude.
	Column (6): Half light radius in parsecs.
	Column (7): Radial velocity.
	Column (8): Calcium triplet metallicity.
	Column (9): Suprime-Cam identifier if Suprime-Cam candidate
	Column (10): Note G = Spectroscopic background galaxy.
\end{table*}

\begin{table*}
	\caption{\label{tab:subgcc}Suprime-Cam globular cluster candidates}
	\begin{tabular}{c@{\ \ }c@{\ \ }c@{\ \ }c@{\ \ }c@{\ \ }c@{\ \ }c@{\ \ }c@{\ \ }c@{\ \ }c@{\ \ }c} \hline
Name       & RA         & Dec       & $(B - V)$       & $(V - I)$       & $I$              & $v$            & [Z/H]                   & P13              & ACS  & Note \\
           & [deg]      & [deg]     & [mag]           & [mag]           & [mag]            & [km s$^{-1}$]  & [dex]                   &                  &      &      \\
(1)        & (2)        & (3)       & (4)             & (5)             & (6)              & (7)            & (8)                     & (9)              & (10) & (11) \\ \hline
sub\_36926 & 185.264908 & 29.329283 & $0.94 \pm 0.04$ & $1.11 \pm 0.03$ & $18.51 \pm 0.02$ & ---            & ---                     & ---              & ---  & ---  \\
sub\_48443 & 184.858638 & 29.469100 & $0.86 \pm 0.01$ & $0.99 \pm 0.03$ & $18.58 \pm 0.02$ & ---            & ---                     & ---              & ---  & ---  \\ 
sub\_33949 & 184.963413 & 29.293525 & $0.83 \pm 0.01$ & $0.99 \pm 0.03$ & $18.63 \pm 0.02$ & $-5 \pm 6$     & ---                     & NGC4278\_stars23 & ---  & S    \\
sub\_41218 & 185.029808 & 29.385256 & $0.91 \pm 0.04$ & $1.16 \pm 0.03$ & $18.73 \pm 0.02$ & $480 \pm 6$    & $-0.72_{-0.12}^{+0.16}$ & NGC4278\_GC254   & ---  & ---  \\
sub\_36266 & 185.252633 & 29.320075 & $0.79 \pm 0.03$ & $0.95 \pm 0.03$ & $18.83 \pm 0.02$ & ---            & ---                     & ---              & ---  & ---  \\                                                    
  ...      & ...        & ...       & ...             & ...             & ...              & ...            & ...                     & ...              & ...  & ...  \\ \hline
	\end{tabular}
	
	\medskip
	The full version of this table is provided in a machine readable form in the online Supporting Information.
	\emph{Notes} Column (1): GC IDs.
	Column (2) and (3): Right ascension and declination in the J2000.0 epoch, respectively.
	Column (4): $(B - V)$ colour.
	Column (5): $(V - I)$ colour.
	Column (6): $I$-band magnitude.
	Column (7): Radial velocity.
	Column (8): Calcium triplet metallicity.
	Column (9): Identifier from \citetalias{2013MNRAS.428..389P}.
	For objects that were observed spectroscopically for the first time in this work we have extended the \citetalias{2013MNRAS.428..389P} naming scheme to them.
	Column (10): ACS identifier if ACS candidate.
	Column (10): Note: G = Spectroscopic background galaxy, S = Spectroscopic foreground star.
\end{table*}

\subsection{Keck DEIMOS spectroscopy}
In addition to the four Keck DEIMOS slitmasks observed for $\sim 2$ h each in February of 2010 and presented in \citetalias{2013MNRAS.428..389P}, we observed an additional, shallower slitmask on 2013 January 12 with three 600 s exposures.
Both the 2010 and 2013 observations used an identical instrument setup of 1 arcsec slits, the 1200 mm$^{-1}$ grating and a central wavelength of 7800 \AA{}.
Both years' observations were reduced using the \textsc{DEEP2 spec2d} pipeline \citep{Newman2012, 2012ascl.soft03003C}.
Radial velocities were measured in an identical manner to \citetalias{2013MNRAS.428..389P} by using the \textsc{IRAF} task \textsc{fxcor} to cross-correlate the observed spectra with 13 template stars observed with DEIMOS.
The uncertainty in the radial velocity was estimated in the same way as in \citetalias{2013MNRAS.428..389P} by adding in quadrature the uncertainty provided by \textsc{fxcor} with the standard deviation of the different templates.

We also measured the metallicities of the new GC spectra from the strength of the calcium triplet (CaT) spectral feature using the technique of \citetalias{2012MNRAS.426.1475U}.
Single stellar population models show the strength of the CaT is sensitive to metallicity while being insensitive to ages older than 3 Gyr, $\alpha$-element enhancement and horizontal branch morphology \citep{2003MNRAS.340.1317V, 2012ApJ...759L..33B}. 
To reduce the effects of the strong sky line residuals found in the CaT spectral region, we used the method of \citet{2010AJ....139.1566F} of masking the sky line regions and fitting a linear combination of the stellar templates to the observed spectra.
We then normalised the fitted spectra and measured the strength of the CaT lines on the normalised spectra.
A Monte Carlo resampling technique was used to estimate 68 \% confidence intervals.
To derive a relation between CaT strength and metallicity, \citetalias{2012MNRAS.426.1475U} measured the strength of the CaT on \citet{2003MNRAS.340.1317V} model spectra using the same technique as for the observed spectra; we use the same relation (their equation 8).
Further details of the metallicity measurement process can be found in \citetalias{2012MNRAS.426.1475U}.
 
\begin{figure}
	\begin{center}	
		\includegraphics[width=240pt]{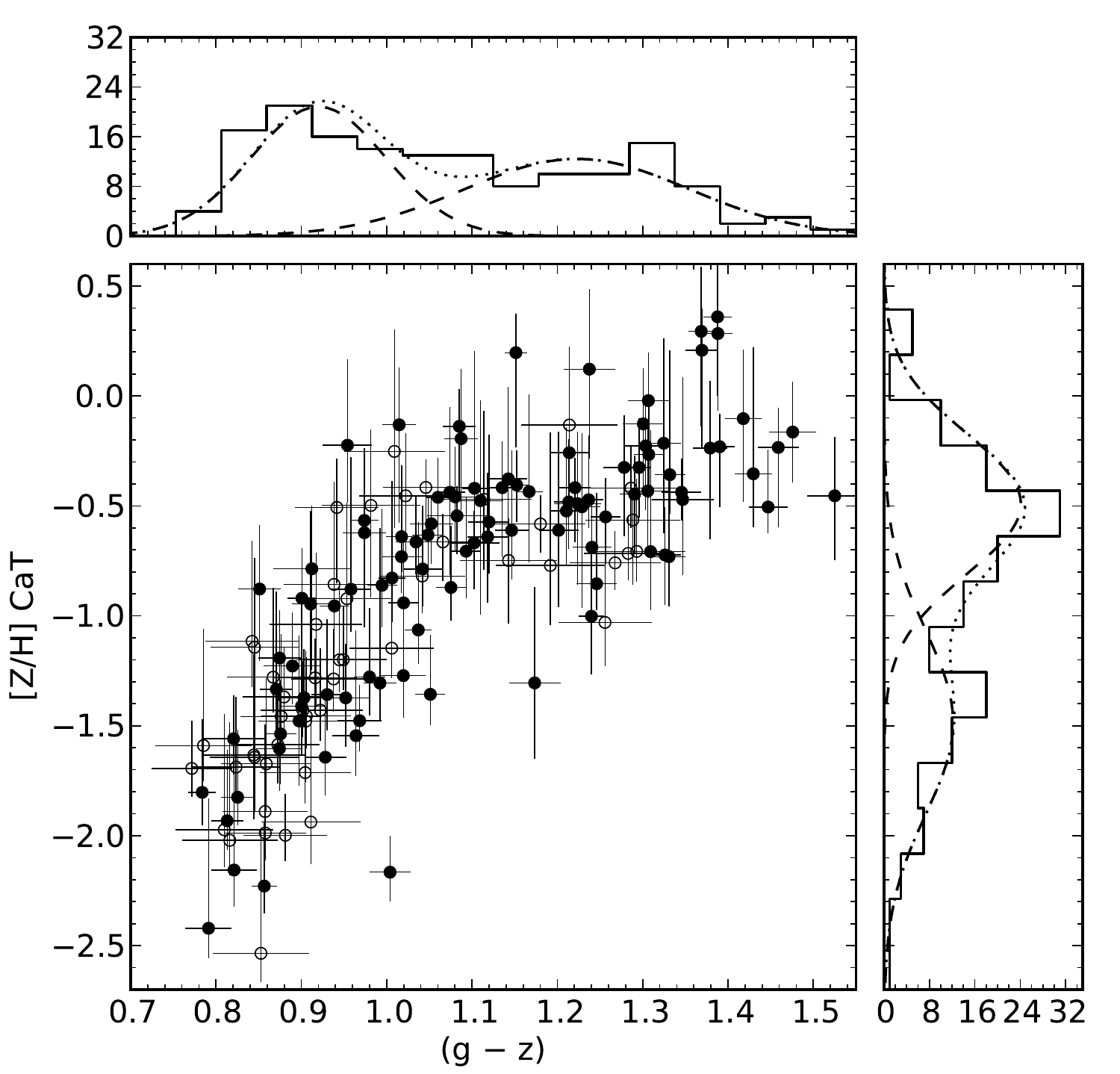}
		\caption{\label{fig:CaTplot} GC colour versus metallicity. 
On the vertical axis is the CaT-based metallicity; on the horizontal axis $(g-z)$ colour.
Solid points have ACS photometry while open points are GCs with Suprime-Cam $(V-I)$ colours transformed into $(g-z)$.
The plot to the right is the metallicity histogram; the plot on top is the colour histogram.
The bimodal \textsc{GMM} fits are plotted over the histograms.
Both the colour and metallicity distributions are bimodal.
The relationship between colour and metallicity is non-linear.}
	\end{center}
\end{figure}

In the case of GCs observed on multiple masks, the radial velocities and metallicities are consistent within observational uncertainties.
The simple means of repeated measurements are used in further analysis.
Combining the 2013 observations with those presented in \citet{2013MNRAS.428..389P} we now have 270 (previously 256) confirmed GCs with radial velocities and 155 (previously 150) with spectroscopic metallicity measurements.
We plot the CaT metallicities against colour in Figure~\ref{fig:CaTplot}.

\subsection{Comparison with previous work}
In their study of GCs using NIR photometry \citet{2011A&A...525A..19C} analysed two of the same ACS pointings as in this work.
We found an insignificant mean $(g - z)$ colour difference of $0.003 \pm 0.008$ and a mean half light radius ratio of $0.95 \pm 0.04$ between \citet{2011A&A...525A..19C} and our own results.
However we notice a significant difference in the magnitudes with those of \citet{2011A&A...525A..19C} being $0.087 \pm 0.008$ mag fainter in the $g$ band and $0.084 \pm 0.005$ in the $z$ band.
To investigate this difference we used the \textsc{IRAF} task \textsc{qphot} to measure the magnitudes of ACS GC candidates with 10 pixel apertures and used the same zero points and the same aperture corrections from 10 pixels to infinity as \citet{2011A&A...525A..19C}.
The mean difference between these \textsc{qphot} magnitudes and the magnitudes used in this paper is less than 0.02 mag.
The extinction corrections used by \citet{2011A&A...525A..19C} differ from ours by only 0.002 mag.
Thus we cannot account for the magnitude difference between \citet{2011A&A...525A..19C} and our work.
Of the 67 GCs in the \citet{2011A&A...525A..19C} catalogue we identified 66 as GC candidates.
The remaining object, GC 59 in \citet{2011A&A...525A..19C}, appears to be the nucleus of a background galaxy in the ACS imaging.

In their study of the kinematics of the GCs \citetalias{2013MNRAS.428..389P} used four of the five ACS pointings we used in this work and the same Suprime-Cam imaging.
The main difference in their analysis of the ACS data was the use of fixed aperture corrections rather than the size dependant aperture corrections used here.
For GC candidates brighter than the turnover magnitude we find the \citetalias{2013MNRAS.428..389P} colours to be on average $(g - z) = 0.010 \pm 0.001$ bluer and their magnitudes to be $z = 0.017 \pm 0.002$ brighter than ours.
They used an empirical transformation to convert the colours of objects with only Suprime-Cam photometry into $(g - z)$.

We can compare the spectroscopic classification of objects with our imaging based selection.
We plot the location of the spectroscopic objects in phase space in Figure~\ref{fig:rvradius} and list them in Appendix~\ref{spectralist}.
Two of the ACS GC candidates, acs\_2494 and acs\_2464, and one of the UCD candidates, acs\_0284, appear to be background galaxies based on the presence of significantly red-shifted emission lines while 13 of the Suprime-Cam candidates also appear to be background galaxies due to red-shifted emission lines in their spectra.
Unfortunately NGC 4278's low systemic velocity ($v_{sys} = 620$ km s$^{-1}$, \citealt{2011MNRAS.413..813C}) makes separating foreground stars from GC on the basis of their radial velocities more difficult.
Guided by ACS size measurements, \citetalias{2013MNRAS.428..389P} used a friendless algorithm to determine which objects are stars.
Two of the objects (sub\_29204 and sub\_35980) observed in January 2013 have low enough radial velocities ($27 \pm 5$ and $-181 \pm 27$ km s$^{-1}$ respectively) to unambiguously make them stars.
However a third object (sub\_27283) observed in January 2013 is more problematic as its radial velocity ($256 \pm 12$) overlaps with that of GCs at the same galactocentric radius.
It fails the Suprime-Cam GC selection criteria and is bluer ($V - I$ = 0.75) than any of the spectroscopically confirmed GCs.
In the USNO-B astrometry catalogue sub\_27283 has a non zero proper motion indicating that it is a foreground star.
Although some of the stars at larger radii have radial velocities of about 200 km s$^{-1}$ and could belong to the low velocity tail of the GC distribution, these stars are well separated from the GCs at the same radius in velocity.
None of the ACS GC candidates but 13 of the Suprime-Cam candidates have radial velocities consistent with being stars.
In our subsequent analysis, we did not remove GC candidates known spectroscopically to be stars or galaxies from our analysis nor did we remove Suprime-Cam candidates that failed the ACS candidate criteria. 
For the ACS dataset removing the two spectroscopically identified galaxies has no effect on our results.
If we removed spectroscopic and ACS contaminants from the Suprime-Cam dataset, the surface density of contaminants would be much lower in the inner regions than in the outer regions.
When we use the Suprime-Cam data we address the effects of contaminants on our results.

\begin{figure}
	\begin{center}	
		\includegraphics[width=240pt]{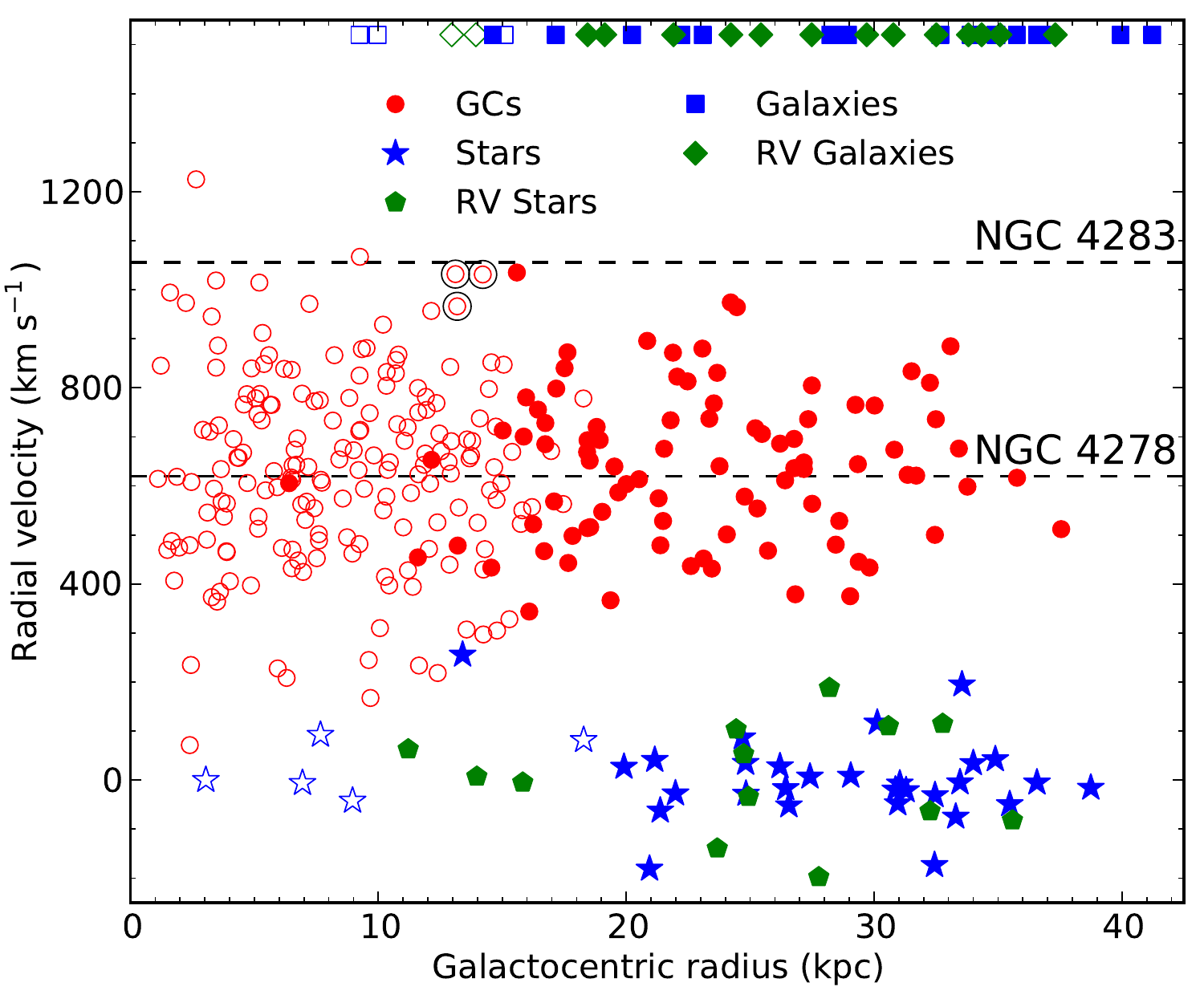}
		\caption{\label{fig:rvradius} Selection of spectroscopic GCs.
Red circles are spectroscopically confirmed GCs.
Blue stars are spectroscopically confirmed stars that do not meet the criteria to be GC candidates.
Although the star sub\_27283 ($v = 256 \pm 12$ km s$^{-1}$) has a GC like radial velocity, it has a non-zero proper motion in the USNO-B catalogue \citep{2003AJ....125..984M}.
Green pentagons are spectroscopically confirmed stars that meet the GC candidate criteria.
Blue squares are spectroscopically confirmed background galaxies that do not the meet the criteria for GC candidates.
Green diamonds are spectroscopically confirmed galaxies that meet GC candidate selection criteria.
Hollow points have ACS coverage; solid points have only Suprime-Cam coverage.
For objects in both imaging datasets we have preferred the ACS selection.
The background galaxies have been given an artificial radial velocity of 1520 km s$^{-1}$; in reality their redshifts are significantly higher.
The black dashed lines are the systemic velocities of NGC 4278 and NGC 4283.
The three circled GCs can be identified with NGC 4283.}
	\end{center}
\end{figure}

\section{Globular cluster spatial distribution and system size}
The relative number of a globular clusters in a galaxy is usually expressed by the specific frequency, which is the number of GCs normalised by the $V$-band luminosity \citep{1981AJ.....86.1627H}.
The most and least massive galaxies have higher specific frequencies on average than intermediate mass galaxies \citep{2005ApJ...635L.137F, 2008ApJ...681..197P}. 
Intriguingly both the total galaxy halo mass \citep{2009MNRAS.392L...1S, 2010MNRAS.406.1967G} and the mass of the central supermassive black hole \citep{2010ApJ...720..516B, 2011MNRAS.410.2347H, 2012AJ....144..154R} scale nearly linearly with the total number of GCs in a galaxy.
Thus the total number of GCs can be used to estimate two important galaxy properties that are difficult to directly measure.
As stellar mass, halo mass and black hole mass are all correlated, it is unclear which is the fundamental relation.

In addition to their use in estimating the size of GC populations, GC surface density profiles are an important ingredient in using GC kinematics in mass modelling.
In almost all galaxies the blue GC subpopulation is more spatially extended than the red subpopulation \citep[e.g.][]{2011MNRAS.416..155F, 2012AJ....144..164H, 2013MNRAS.428..389P}.
The radial surface density profile of the red subpopulation usually closely follows the bulge surface brightness profile while the blue surface density profile seems to be closer to the X-ray gas surface brightness profile, suggesting that it traces the galaxy's halo \citep{2012MNRAS.425...66F}.

\subsection{Spatial distribution}
\label{surfacedensity}
\begin{figure}
	\begin{center}
   
		\includegraphics[width=240pt]{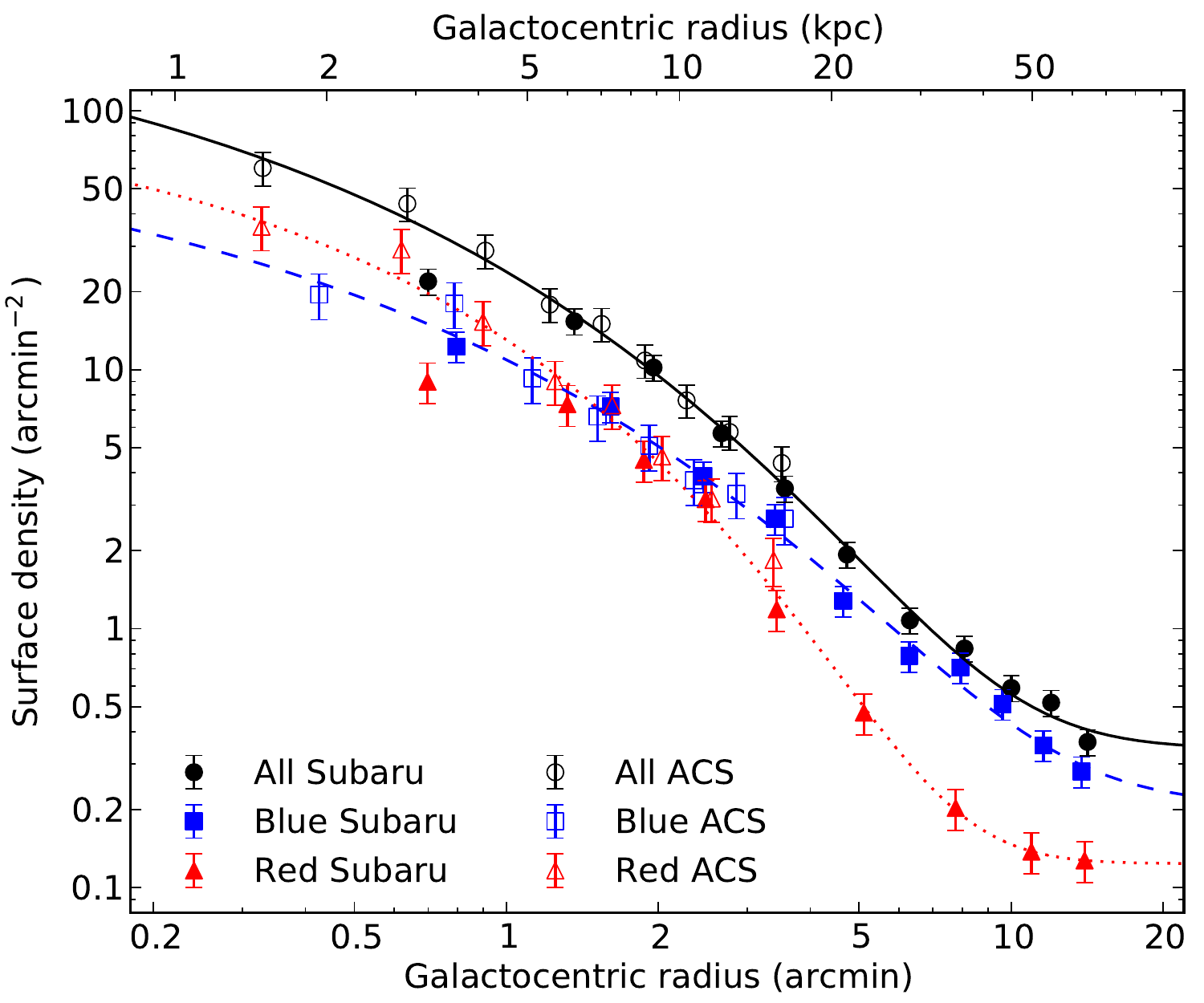}
		\caption{\label{fig:surfacedensity} Surface density profiles for the red (dotted line, triangles) and blue (dashed line, squares) GC subpopulations as well as the total population in black (solid line, circles).
ACS data are open symbols, Suprime-Cam data are solid symbols.
The innermost Suprime-Cam points are likely affected by incompleteness and are not included in the fits.
In the range of overlap between the ACS and Suprime-Cam data we see good agreement.
The solid lines are S\'ersic fits including a constant background.
The red subpopulation is more centrally concentrated than the blue subpopulation.
The red subpopulation reaches a background level of about 0.11 objects per square arcmin at about 14 arcmin.}

	\end{center}
\end{figure}

To study the radial distribution of GCs in NGC 4278 we used GC candidates brighter than the turnover magnitude to produce surface density profiles for the total GC system as well as for each of the subpopulations.
A $z$ band turnover magnitude of $z = 22.44$ was used for the ACS candidates while a $I$ band turnover of $I = 22.12$, equivalent to the $z$-band turnover for a GC with $(V - I) = 1.05$, was used for the Suprime-Cam candidates.
We used the colour splits of $(g - z) = 1.078$ and $(V - I) = 1.054$ (derived in Section~\ref{colour}) to divide the total system into red and blue subpopulations.
No attempt was made to correct for magnitude incompleteness or contamination.
Candidates were sorted by galactocentric radius and placed into bins of equal size.
The area and mean radius were calculated for each bin accounting for the incomplete azimuthal coverage of the ACS imaging at larger radii.
For the radial range with both ACS and Subprime-cam candidates we see good agreement between the two datasets, despite the different selection criteria and quality of the datasets.
The surface density profiles were fit with \citet{1963BAAA....6...41S} profiles with constant, free background terms.
The following variation of equation 1 from \citet{2005PASA...22..118G} was fit: 
\begin{equation}
N(R) = N_{e} \exp\left(-b_{n}\left[\left(\frac{R}{R_{e}}\right)^{1/n} - 1\right]\right) + bg 
\end{equation}
where $b_{n} = 1.9992n - 0.3271$.
The free parameters of the fit are the surface density $N_{e}$ at the effective radius, the effective radius $R_{e}$, the S\'ersic index $n$ and the constant background $bg$.
The innermost Suprime-Cam point for each subpopulation was not included in the fits due to the probable incompleteness.
The measured surface densities and the fitted profiles are shown in Figure~\ref{fig:surfacedensity} while the results of the fits are provided in Table~\ref{tab:surfacedensity}.
The uncertainties in the fitted parameters were estimated using bootstrapping; we note that there are significant correlations between the fitted parameters especially for the blue subpopulation.
As shown in Figure~\ref{fig:subcolourradius} the background surface density is consistent with the number of spectroscopically observed contaminants. 

As in other galaxies, the red subpopulation is more centrally concentrated than the blue subpopulation.
The red subpopulation reaches a background at a radius of about 14 arcmin while the blue subpopulation does not seem to have reached a background within the Suprime-Cam field of view.
We can check the background surface densities in two ways.
In Section~\ref{gradients} we find that the number of Suprime-Cam candidates that are shown to be contaminants spectroscopically is consistent with the background surface densities.
14 Subaru candidates fail the ACS selection criteria - 6 are likely stars and the remainder galaxies.
Using the area of the ACS observations these contaminants correspond to surface densities of $0.379 \pm 0.101$, $0.271 \pm 0.086$ and $0.108 \pm 0.054$ arcmin$^{-2}$ for all, blue and red respectively, which are consistent with the fitted values. 
We integrated the S\'ersic profiles to determine the total numbers of GCs and doubled the total to account for only having measured the surface density of GCs brighter than the turnover magnitude.
We estimate $1378^{+32}_{-194}$ GCs in total, $930^{+315}_{-176}$ blue GCs and $509^{+27}_{-40}$ red GCs.
The total, blue and red fits have $\chi^{2}$ values of 7.7 (15 degrees of freedom), 8.0 (13 degrees of freedom) and 6.9 (12 degrees of freedom) respectively indicating that they are good fits. 
Assuming a $V$-band absolute magnitude of $M_{V} = -20.9 \pm 0.2$ \citep{1991trcb.book.....D} we find a specific frequency of $6.0^{+1.2}_{-1.3}$ for NGC 4278.

In addition to the radial distribution of GCs we examined their azimuthal distribution.
In the radial region with complete ACS azimuthal coverage (galactocentric radius less than 2.5 arcmin), Kolmogorov-Smirnov (KS) tests revealed that the azimuthal distributions of all ACS candidates brighter than the turnover magnitude were consistent with being drawn from a uniform distribution ($p$-value of 0.76).
When split by colour the blue and red candidates are also each consistent with being drawn from uniform azimuthal distributions ($p$-values of 0.45 and 0.23 respectively).
Since the GC distribution is consistent with being circular, the circular galactocentric radii used through out this work are valid.
NGC 4278 is a rather round galaxy \citep[$K$-band axis ratio = 0.91,][]{2006AJ....131.1163S} and is like other galaxies in having similar ellipticities for its GC and field star populations \citep{2006ARA&A..44..193B, 2013ApJ...773L..27P}.

Remarkably, using object counts on photographic plates, \citet{1981AJ.....86.1627H} estimated a GC system size of $1075 \pm 125$, in agreement with our value of $1378^{+32}_{-194}$.
For its absolute magnitude, NGC 4278 has a relatively high specific frequency of $6.0^{+1.2}_{-1.3}$ compared to galaxies of similar luminosity in the large ACS study of \citet{2008ApJ...681..197P}.
We note that \citet{2012MNRAS.420...37B} found large differences in $S_{N}$ between HST based studies with small radial extent such as \citet{2008ApJ...681..197P} and those using wide field imaging.
Compared to the heterogeneous catalogue of specific frequencies of \citet{2013ApJ...772...82H}, NGC 4278's specific frequency is higher than most other galaxies of similar luminosity but within the scatter.
Using a mean GC mass of $4 \times 10^{5}$ M$_{\sun}$ and the ratio of halo mass to GC system mass from \citet{2009MNRAS.392L...1S}, we estimate that NGC 4278 has a total halo mass of $8 \times 10^{12}$ M$_{\sun}$.
Using equation 8 of \citet{2012AJ....144..154R} we predict a central black hole mass of $(3.8 \pm 0.5) \times 10^{8}$ M$_{\sun}$ for NGC 4278 based on the number of GCs.

The uncertainty in the specific frequency is affected by the uncertainty of the distance.
The distance has two effects on the specific frequency.
First, it changes the GC turnover magnitude.
If too faint a turnover magnitude is used, the number of GC will be overestimated while a too bright turnover magnitude would cause the number of GCs to be underestimated.
Second, the distance effects absolute magnitude.
A higher absolute magnitude reduces the specific frequency.
Since both these act in the same direction their effects partially cancel each other out.
A distance modulus 0.2 mag too low would result in an over-estimation of the GC population by 19 \% and an under-estimation of the luminosity by 17 \%, resulting in a specific frequency only 1 \% too high. 
Beyond the distance based uncertainty, \citet{2010ApJ...717..603V} showed that there is cosmic scatter in the GC turnover magnitude at a level of 0.2 mag; they found that the mean absolute turnover magnitudes of the Virgo and Fornax clusters disagree by 0.2 mag.

\subsection{The distance to NGC 4278}
In this work we adopted a distance modulus of $(m - M) = 30.97$ (15.6 Mpc) which is based on the \citet{2001ApJ...546..681T} SBF distance of $(m - M) = 31.03 \pm 0.20$ shifted by 0.06 to match the mean of the HST based SBF distances in the Virgo Cluster from \citet{2007ApJ...655..144M}.
The distance to NGC 4278 has also been estimated by using the GC luminosity function by \citet{1996AJ....112.1409F} and \citet{2001AJ....121.2950K} who found distance moduli of $30.61 \pm 0.14$ and $31.11 \pm 0.11$.
While the \citet{2001AJ....121.2950K} value is consistent with the SBF distance, the \citet{1996AJ....112.1409F} distance is significantly lower.
We note that the \citet{1996AJ....112.1409F} GC luminosity function distance to NGC 4494 is similarly lower than the SBF distance from \citet{2001ApJ...546..681T}.
As noted before, \citet{2010ApJ...717..603V} showed there seems to be significant scatter caused by using the GC luminosity function as a distance indicator.
Additionally, \citet{1996ApJ...462....1J} used the planetary nebula (PN) luminosity function to measure the distance to NGC 4278 as $30.04 \pm 0.16$.
However their luminosity function was based on only 23 PN candidates and they noted that NGC 4278 has an unusual PN luminosity function with an overabundance of bright PN candidates.
Furthermore it is known that using PN luminosity functions give systematically lower distances than SBF distances \citep{2002ApJ...577...31C, 2012Ap&SS.341..151C}.
Although they are less precise, distances based on the fundamental plane are consistent with our adopted distance \citep{2001MNRAS.327.1004B, 2013MNRAS.432.1709C}.

\begin{figure}
	\begin{center}
   
		\includegraphics[width=240pt]{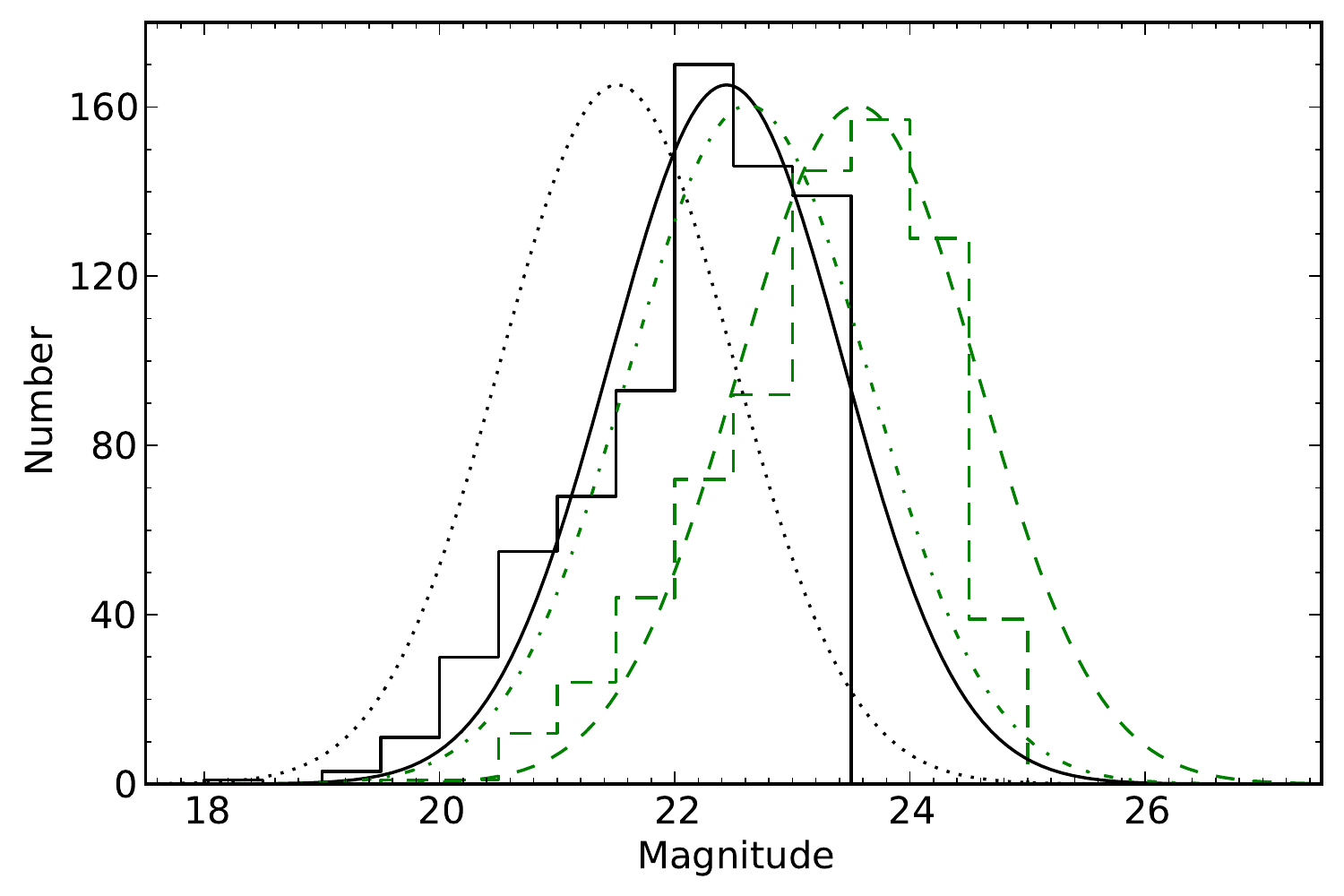}
		\caption{\label{fig:GCLF} GC luminosity functions. The green dashed-line histogram is the observed $g$-band magnitude histogram; the black solid-line histogram is the $z$-band magnitude histogram.
Note that GC candidates are limited to be brighter than $z = 23.44$.
The green dashed-line and black solid line curves are the Gaussian luminosity functions for the $g$ and $z$-bands respectively based on the turnover magnitudes ($\mu_{g} = -7.41$, $\mu_{z} = - 8.53$) and dispersions ($\sigma_{g} = 1.02$, $\sigma_{z} = 0.99$) from \citet{2010ApJ...717..603V} for a galaxy of NGC 4278's absolute magnitude and our adopted distance modulus of ($m - M = 30.97$).
Our adopted distance seems to be consistent with observed GC luminosity functions.
The green dot-dashed-line and black dotted line curves are the Gaussian luminosity functions for the $g$ and $z$-bands respectively using the same absolute turnover magnitude and dispersions but the PN luminosity function distance modulus of \citet{1996ApJ...462....1J}.
This distance is inconsistent with the observed GC luminosity function.
}

	\end{center}
\end{figure}

With our dataset we can test our adopted distance through two independent techniques.
First, by assuming that the mean size of GCs is the same in galaxies of similar mass (an assumption supported by observations e.g. \citealt{2010ApJ...715.1419M}, see Section~\ref{size}) we can use the mean angular size of the GCs ($0.0349 \pm 0.0007$ arcsec) as a geometric distance estimator.
Using the mean GC size in the Virgo and Fornax galaxies of similar luminosity to NGC 4278 from \citet{2010ApJ...715.1419M} we get a distance to NGC 4278 of $16.0 \pm 1.0$ Mpc while using the mean GC size in the Milky Way (free of any uncertainty in the extragalactic distance scale) from \citet{2005ApJ...634.1002J} gives a distance of $15.0 \pm 0.7$ Mpc.
Both of these are in perfect agreement with our adopted distance.
Second, we can use the GC luminosity function to constrain the distance.
In Figure \ref{fig:GCLF} we plot the observed $g$ and $z$-band luminosity functions along with the luminosity functions predicted from our adopted distance and the turnover magnitudes and magnitude dispersions from \citet{2010ApJ...717..603V}.  
The observed luminosity functions appear to be consistent with the predicted luminosity functions.
Although due to incompleteness the observed turnover magnitude could be brighter than its true value, the observations do not support as bright a turnover magnitude as would be required for the PN luminosity function distance or the \citet{1996AJ....112.1409F} GC luminosity function distance.
On the basis of GC sizes and the GC luminosity function we can conclude that our adopted SBF-based distance is reasonable.
We note that only two of our results are based on the distance to NGC 4278: the absolute size of the GCs and the total number of GCs.

\begin{table}
	\caption{\label{tab:surfacedensity}Globular cluster surface density fits}
	\begin{tabular}{l c c c c c}
		Sample & $N_{e}$             & $r_{e}$             & $n$                 & $bg$                   & Total               \\
		(1)    & (2)                 & (3)                 & (4)                 & (5)                    & (6)                 \\ \hline
		All    & $5.1^{+2.0}_{-0.5}$ & $2.8^{+0.1}_{-0.5}$ & $2.2^{+0.2}_{-0.5}$ & $0.35^{+0.11}_{-0.02}$ & $1378^{+29}_{-195}$ \\
		Blue   & $1.5^{+0.8}_{-0.6}$ & $4.3^{+1.6}_{-1.0}$ & $2.3^{+0.8}_{-0.6}$ & $0.21^{+0.07}_{-0.08}$ & $930^{+237}_{-188}$ \\
		Red    & $5.4^{+0.8}_{-1.0}$ & $1.8^{+0.1}_{-0.1}$ & $1.7^{+0.3}_{-0.3}$ & $0.12^{+0.02}_{-0.02}$ & $509^{+25}_{-41}$
	\end{tabular}

	\medskip
	\emph{Notes}
	Column (1): GC subpopulation.
	Column (2): GC surface density at the effective radius in arcmin$^{-2}$.
	Column (3): Effective radius in arcmin.
	Column (4): S\'ersic index.
	Column (5): Background surface density in arcmin$^{-2}$.
	Column (6): Total number of GC in each subpopulation. 

\end{table}

\subsection{The globular cluster system of NGC 4283}

We also investigated the contribution of NGC 4283's GC system to NGC 4278's GC system.
NGC 4283 is at a projected distance of 16 kpc in the sky and has a SBF distance consistent with NGC 4278's distance \citep{2001ApJ...546..681T}.
The systemic velocities of the two galaxies (NGC 4278 $v_{sys} = 620$ km s$^{-1}$, NGC 4283 $ v_{sys} = 1056$ km s$^{-1}$ \citealt{2011MNRAS.413..813C}) differ by 436 km s$^{-1}$.
Adopting an absolute magnitude of $M_{V} = -19.0$ \citep{1991trcb.book.....D} and a specific frequency of 1.3, typical of galaxies of this absolute magnitude (see table 3 of \citealt{2008ApJ...681..197P}), we would expect NGC 4283 to have only 26 GCs brighter than the turnover magnitude.

While we do not see an over-density of Suprime-Cam GC candidates near NGC 4283 we do see a slight over-density in ACS GC candidates near NGC 4283.
\citet{1981AJ.....86.1627H} also did not find an obvious over-density of sources around NGC 4283 in their photographic plate.
We used the GC surface density profile measured above to estimate how many GCs brighter than the turnover we would expect near NGC 4283 from NGC 4278's GC system.
Within 1.5 kpc of NGC 4283 we would only expect 1.1 GCs, although we see 5 GCs candidates in the same area.
In 100 000 realisations of the GC system we only saw this many GCs within 1.5 kpc of NGC 4283 in 578 realisations.
We expect 11.3 GCs within 5 kpc; we see 15 GC candidates within 5 kpc.
Making the rough assumption that the half number radius of NGC 4283's GC system is 5 kpc, we can estimate that NGC 4283 has a total of $16 \pm 12$ GCs. 

If NGC 4283 GCs made a significant contribution to NGC 4278's GC system we would expect NGC 4283 GCs to show up in phase space as interlopers.
Unfortunately the velocity difference between the galaxies is not large enough to cleanly separate the GC populations.
Using the velocity RMS of NGC 7457 (68 km s$^{-1}$) from \citetalias{2013MNRAS.428..389P}, the galaxy closest to NGC 4283 in $K$-band luminosity in \citetalias{2013MNRAS.428..389P}, as an upper limit on NGC 4283's GC velocity RMS, three GCs - acs\_0641 (1031 km s$^{-1}$), acs\_1900 (966 km s$^{-1}$) and acs\_2206 (1032 km s$^{-1}$) - lie within 5 kpc on the sky and within 136 km s$^{-1}$ of NGC 4283.
These three GCs are highlighted in Figure~\ref{fig:rvradius}.
This suggests that the three GCs may be part of the NGC 4283 GC system.
However, due to the small velocity difference between the galaxies we cannot rule out that they are part of the NGC 4278 GC system.
Assuming that all three are truly NGC 4283 GCs and that the radial velocity observations sampled both galaxy's populations with equal likelihood, we would only expect NGC 4283 to have $3 / 267 = 1.1 \%$ as many GCs as NGC 4278.
Above we determine that NGC 4278 has 1378 GCs; using this radial velocity argument we can estimate that NGC 4283 has $15 \pm 9$ GCs.

The number of NGC 4283 GCs is insignificant compared to the NGC 4278 population.
We found no effect on the results of the colour--magnitude (Section~\ref{bluetilt}), colour--radius (Section~\ref{gradients}) or size (Section~\ref{size}) relations nor on the surface density profiles (Section~\ref{surfacedensity}) due to NGC 4283.
If NGC 4283 has $16 \pm 8$ GCs, it would have a specific frequency of $0.4 \pm 0.2$.
While this is lower than the mean of 1.3 for galaxies of similar luminosities in \citet{2008ApJ...681..197P} it is within the range of values they found.
Likewise it is within the range of specific frequencies observed for galaxies of similar luminosity in the \citet{2013ApJ...772...82H} catalogue.

In the Virgo cluster the dwarf galaxies closest to giant galaxies show the lowest specific frequencies of any galaxies in the \citeauthor{2008ApJ...681..197P} study.
These galaxies are thought to have lost GCs through stripping.
If NGC 4283 has interacted with NGC 4278 it could have lost some of its GC population to NGC 4278.
To double the specific frequency of NGC 4283 by having it further away than our adopted distance would require it to be twice as distant, a distance strongly disfavoured by the SBF distance.

\section{Globular cluster colours}
\label{colour}
As can be seen in Figures \ref{fig:acscolourmag} and \ref{fig:subcolourmag}, NGC 4278 possesses the GC colour bimodality found in almost all massive galaxies \citep[e.g.][]{2001AJ....121.2950K, 2001AJ....121.2974L}.
This colour bimodality was previously noted by \citetalias{2013MNRAS.428..389P} and \citetalias{2012MNRAS.426.1475U} who used much the same data as this paper.
We used the Gaussian Mixture Modelling (\textsc{GMM}) code of \citet{2010ApJ...718.1266M} to confirm colour bimodality.
\textsc{GMM} finds both the best fitting unimodal and bimodal distributions to the data before using a parametric bootstrap to determine whether or not the bimodal distribution is an improved fit over the unimodal.
However as \citet{2010ApJ...718.1266M} note, \textsc{GMM} is more a test of non-Gaussianity than of bimodality.
They suggest using the relative separation of the fitted peaks and the kurtosis of the distribution to help determine whether a distribution is bimodal.
We adopted their requirements of relative peak separation of $D = 2$ and a negative kurtosis, in addition to a high probability from \textsc{GMM}, to call a distribution bimodal. 

Results of the \textsc{GMM} tests are given in Table~\ref{tab:gmm}.
We confirm colour bimodality in the GC candidates, spectroscopically confirmed GCs and the GCs with CaT measurements, in line with previous conclusions by \citetalias{2013MNRAS.428..389P} and \citetalias{2012MNRAS.426.1475U}.
The peak locations, peak widths and the fraction of blue clusters are all in agreement with the previous work.
For ACS candidates we found equal probability of being red or blue at $(g - z) = 1.078$ while for the Suprime-Cam candidates we found equal probability of being in either colour subpopulation at $(V - I) = 1.054$.
We used these colour splits to divide the GC candidates into red and blue subpopulations.
The fraction of blue and red Suprime-Cam candidates are consistent with numbers of blue and red GCs calculated from the surface density profiles in Section~\ref{surfacedensity}.
The $(g - z)$ colour peaks and the fraction of blue GCs of NGC 4278 agree with the relations as a function of galaxy $B$-band absolute magnitude of \citet{2006AJ....132.2333S} and \citet{2006ApJ...639...95P}.

Contamination affects the overall colour distributions, masking or mimicking bimodality.
The ACS dataset has a low level of contamination since the size cut removes foreground stars and extended background galaxies and since the ACS candidates are close to the centre of the galaxy where the surface density of GCs is much higher than that of the contaminants.
Therefore the observed ACS colour distributions likely reflect the true colour distributions.
However the Suprime-Cam dataset has a significant level of contamination as both stars and galaxies overlap with GCs in colour--colour space.
Additionally the Suprime-Cam data extend to large radii where the surface density of contaminants is comparable to that of GCs.
Using the surface density of contaminants from Section~\ref{surfacedensity} we can estimate that 39 \% and 40 \% respectively of the blue and red Suprime-Cam GC candidates that are brighter than turnover are contaminants.
We ran \textsc{GMM} on Suprime-Cam candidates brighter than the turnover magnitude within 30 kpc (more than twice the half number radius of the GC system; see Table~\ref{tab:surfacedensity}) of the centre of the galaxy.
The \textsc{GMM} results of this sample were consistent with those of all Suprime-Cam candidates brighter than turnover despite the contaminant rate being $\sim 4$ times lower.
This suggests that the general shape of the Suprime-Cam colour distribution is not affected by contamination.
The colour distributions of the spectroscopically confirmed GCs is quite similar to those of candidates brighter than turnover for both the ACS and Suprime-Cam datasets (KS probabilities of 0.29 and 0.87 respectively, within the same radial ranges, that the spectroscopically confirmed GCs are drawn from the same distributions as the candidates).

We also ran \textsc{GMM} on the CaT metallicity distribution, finding clear evidence for bimodality.
Unsurprisingly, adding only 5 new CaT measurements does not change the results from \citetalias{2012MNRAS.426.1475U}.
Likewise the colours of the GCs with CaT measurements show bimodality.
The CaT metallicity and colour distributions along with their GMM fits may be seen in Figure~\ref{fig:CaTplot}.
The relationship between GC colour and metallicity is clearly non-linear in NGC 4278 as has been seen in previous observational studies such as \citet{2006ApJ...639...95P} and \citet{2012MNRAS.426.1475U}, and theoretical models such as \citet{2009ApJ...699..486C} and \citet{2011ApJ...743..150Y}. 
Due to this non-linear relation the shape of the colour distribution is different than the shape of the metallicity distribution.

\begin{table*}
	\caption{\label{tab:gmm}Colour and metallicity distribution bimodality results from \textsc{GMM} }
	\begin{tabular}{c@{\ \ }c@{\ \ }c@{\ \ }c@{\ \ }c@{\ \ }c@{\ \ }c@{\ \ }c@{\ \ }c@{\ \ }c@{\ \ }c} \hline
	Sample      & Colour  & $N$  & $\mu_{blue}$      & $\sigma_{blue}$   & $\mu_{red}$       & $\sigma_{red}$    & $f_{blue}$        & $p$       & $D$             & $k$     \\
	            & [mag]   &      & [mag]             & [mag]             & [mag]             & [mag]             &                   &           &                 &         \\
	(1)         & (2)     & (3)  & (4)               & (5)               & (6)               & (7)               & (8)               & (9)       & (10)            & (11)    \\ \hline
	ACS         & $(g-z)$ & 716  & $0.916 \pm 0.016$ & $0.096 \pm 0.010$ & $1.294 \pm 0.024$ & $0.147 \pm 0.013$ & $0.478 \pm 0.049$ & $> 0.999$ & $3.05 \pm 0.20$ & $-1.11$ \\
	All         &         &      &                   &                   &                   &                   &                   &           &                 &         \\
	ACS         & $(g-z)$ & 410  & $0.948 \pm 0.023$ & $0.102 \pm 0.015$ & $1.330 \pm 0.031$ & $0.107 \pm 0.020$ & $0.558 \pm 0.068$ & $> 0.999$ & $3.66 \pm 0.31$ & $-1.29$ \\
	Turnover    &         &      &                   &                   &                   &                   &                   &           &                 &         \\
	ACS         & $(g-z)$ & 181  & $0.912 \pm 0.046$ & $0.074 \pm 0.028$ & $1.212 \pm 0.061$ & $0.150 \pm 0.032$ & $0.466 \pm 0.160$ & $> 0.999$ & $2.54 \pm 0.43$ & $-1.07$ \\
	RV          &         &      &                   &                   &                   &                   &                   &           &                 &         \\ 
	ACS         & $(g-z)$ & 105  & $0.997 \pm 0.038$ & $0.110 \pm 0.020$ & $1.303 \pm 0.040$ & $0.096 \pm 0.024$ & $0.605 \pm 0.132$ & $> 0.988$ & $2.96 \pm 0.38$ & $-1.04$ \\
	CaT         &         &      &                   &                   &                   &                   &                   &           &                 &         \\
	S-Cam       & $(V-I)$ & 1146 & $0.935 \pm 0.004$ & $0.067 \pm 0.003$ & $1.156 \pm 0.010$ & $0.091 \pm 0.006$ & $0.657 \pm 0.029$ & $0.990$   & $2.76 \pm 0.19$ & $-0.36$ \\
	All         &         &      &                   &                   &                   &                   &                   &           &                 &         \\
	S-Cam       & $(V-I)$ & 901  & $0.937 \pm 0.005$ & $0.063 \pm 0.003$ & $1.156 \pm 0.011$ & $0.090 \pm 0.007$ & $0.634 \pm 0.034$ & $> 0.999$ & $2.82 \pm 0.22$ & $-0.46$ \\
	Turnover    &         &      &                   &                   &                   &                   &                   &           &                 &         \\
	S-Cam       & $(V-I)$ & 508  & $0.943 \pm 0.006$ & $0.062 \pm 0.004$ & $1.149 \pm 0.014$ & $0.082 \pm 0.008$ & $0.581 \pm 0.040$ & $> 0.999$ & $2.84 \pm 0.24$ & $-0.67$ \\
	$<$ 30 kpc  &         &      &                   &                   &                   &                   &                   &           &                 &         \\
	S-Cam       & $(V-I)$ & 257  & $0.941 \pm 0.012$ & $0.048 \pm 0.009$ & $1.118 \pm 0.024$ & $0.085 \pm 0.012$ & $0.495 \pm 0.101$ & $> 0.999$ & $2.55 \pm 0.39$ & $-0.82$ \\
	RV          &         &      &                   &                   &                   &                   &                   &           &                 &         \\ 
	S-Cam       & $(V-I)$ & 153  & $0.924 \pm 0.011$ & $0.037 \pm 0.008$ & $1.089 \pm 0.017$ & $0.088 \pm 0.011$ & $0.358 \pm 0.091$ & $> 0.999$ & $2.45 \pm 0.35$ & $-0.84$ \\
	CaT         &         &      &                   &                   &                   &                   &                   &           &                 &         \\
	All         & $(g-z)$ & 155  & $0.920 \pm 0.035$ & $0.079 \pm 0.022$ & $1.223 \pm 0.059$ & $0.132 \pm 0.029$ & $0.501 \pm 0.147$ & $> 0.999$ & $2.79 \pm 0.54$ & $-0.99$ \\
	CaT         &         &      &                   &                   &                   &                   &                   &           &                 &         \\
	All         & [Z/H]   &	155  & $-1.49 \pm 0.14$  & $0.42 \pm 0.09$   & $-0.49 \pm 0.05$	 & $0.32 \pm 0.06$   & $0.397 \pm 0.101$ & $> 0.999$   & $2.68 \pm 0.52$ & $-0.45$ \\
	CaT         &         &      &                   &                   &                   &                   &                   &           &                 &         \\

	\end{tabular}

\medskip
\emph{Notes} Column (1): GC sample.
Column (2): Colour. 
Column (3): Number of GCs.
Column (4): Mean colour of the blue subpopulation.
Column (5): Dispersion of the blue subpopulation.
Column (6): Mean colour of the red subpopulation.
Column (7): Dispersion of the red subpopulation.
Column (8): Fraction of the GCs in the blue subpopulation.
Column (9): p-value that a bimodal fit is preferred over a unimodal fit.
Column (10): Separation of the \textsc{GMM} peaks normalised by their width.
Column (11): The kurtosis of the sample.
For GCs with CaT measurements but no $(g-z)$ colours we used Equation~\ref{eq:vigz} to convert $(V-I)$ into $(g-z)$.

\end{table*}

\begin{figure}
	\begin{center}
		\includegraphics[width=240pt]{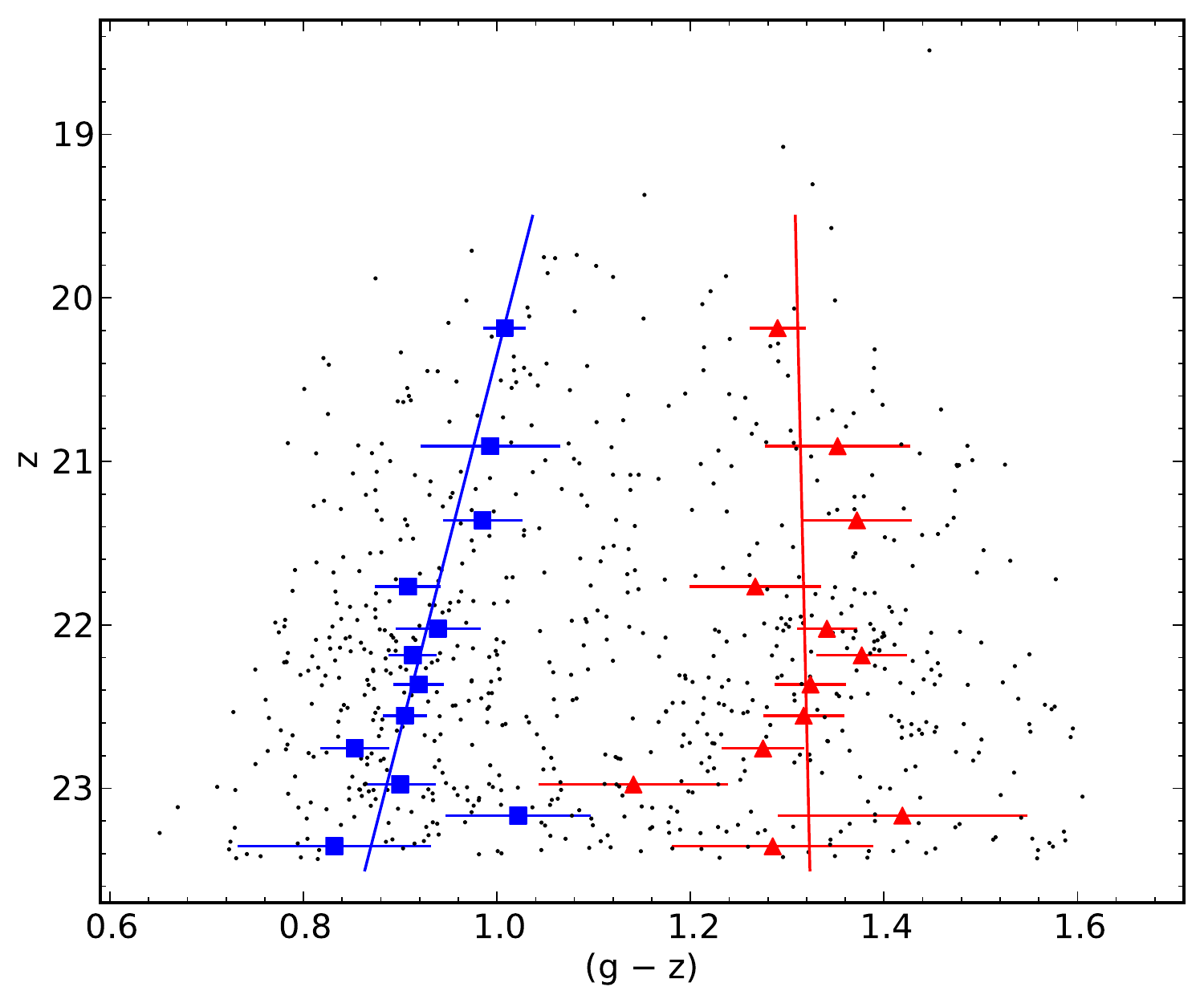}
		\caption{\label{fig:acsbluetilt} Colour--magnitude diagram of ACS GC candidates.
The bimodal peaks found by \textsc{GMM} for each magnitude bin are over-plotted with error bars in blue and red.
The blue and red lines are the weighted least squares fits to the blue and red peaks as a function of magnitude.
NGC 4278 shows a significant blue tilt but does not show any evidence for a red tilt.
Fainter than $z = 21.58$ $(M_{z} = -9.39)$ there is no evidence of a colour magnitude relation.}
	\end{center}
\end{figure}

\subsection{The blue tilt}
\label{bluetilt}
First described by \citet{2006AJ....132.1593S}, \citet{2006AJ....132.2333S} and \citet{2006ApJ...636...90H}, the blue tilt is a colour--magnitude relation where the brighter GCs in the blue subpopulation are redder than the fainter blue subpopulation GCs.
This relation is found in most but not all \citep[i.e. NGC 4472, ][]{2006AJ....132.1593S} galaxies with sufficient numbers of GCs.
No clear colour--magnitude relation has been seen for the red subpopulations.
The blue tilt is seen in both high mass and low mass galaxies but is weaker in low mass galaxies \citep{2010ApJ...710.1672M}.
The strength of the blue tilt decreases at fainter magnitudes \citep{2009ApJ...699..254H}.
Since colour traces metallicity and magnitude traces mass, the blue tilt is a mass metallicity relationship.
\citet{2008AJ....136.1828S} and \citet{2009ApJ...695.1082B} explain the observed mass--metallicity relations as self enrichment where the more massive GCs are able to hold onto metals from the first generation of GC stars and incorporate the metals into later stellar generation(s).
Less massive GCs are not able to hold onto the metals, explaining the lack of a blue tilt at fainter magnitudes.
Since the red GCs form with a much higher metallicity they experience a much smaller fractional change in metallicity due to self enrichment and would be expected to show a much weaker colour magnitude relation compared to the blue subpopulation.

To quantify any colour--magnitude relation we split the ACS GC candidates by $z$-band magnitude into twelve bins of equal size.
In each bin we used \textsc{GMM} to find the peaks of the red and blue subpopulations. 
We then fit colour as a function of $z$ band magnitude using the mean magnitude of the GC candidates in each bin.
The peaks found by \textsc{GMM} along with the fitted colour--magnitude relations are plotted on the colour--magnitude diagram in Figure~\ref{fig:acsbluetilt}.
The blue subpopulation shows a clear blue tilt:
\begin{equation}
(g - z)_{blue} = (-0.043 \pm 0.008) M_{z} + (1.015 \pm 0.018)
\end{equation}
while the red subpopulation does not show evidence for a colour--magnitude relation:
\begin{equation}
(g - z)_{red} = (-0.004 \pm 0.014) M_{z} + (1.310 \pm 0.028) .
\end{equation}
These results are insensitive to the bin size used.
The $\chi^{2}$ value of the blue fit (7.0, 10 degrees of freedom) is significantly less than the $\chi^{2}$ value for a constant colour (25.0, 11 degrees of freedom).
The red fit and a constant colour for the red subpopulation both have $\chi^{2}$ values of 9.6 and 9.7 respectively (10 and 11 degrees of freedom respectively).
We repeated the fits using only the nine faintest bins which corresponds to fitting all candidates fainter than $z = 21.58$ (or less massive than $5.8 \times 10^{5}$ M$_{\sun}$ using a mean mass-to-light ratio $\Upsilon_{z} = 1.9$ calculated from the stellar population models of \citealt{2009ApJ...699..486C} for an age of 12.6 Gyrs, a Kroupa initial mass function and the observed GC colour distribution) and found no evidence for a colour--magnitude relation in either subpopulation (blue slope $-0.018 \pm 0.027$ mag mag$^{-1}$, red slope $-0.051 \pm 0.043$ mag mag$^{-1}$).
We also repeated the fits using only the brightest six bins which corresponds to fitting all candidates brighter than $z = 22.27$ (or more massive than $3.1 \times 10^{5}$ M$_{\sun}$) finding a blue slope ($-0.048 \pm 0.009$ mag mag$^{-1}$) and a red slope ($0.029 \pm 0.015$ mag mag$^{-1}$) slightly steeper but consistent with the slopes from the fits to all candidates. 
Additionally we spilt the candidates into two equally sized bins by galactocentric radius.
In the inner bin, with a mean galactocentric radius of 4.0 kpc, we found a blue slope of $-0.068 \pm 0.021$ mag mag$^{-1}$ and a red slope of $0.004 \pm 0.025$ mag mag$^{-1}$.
In the outer bin, with a mean galactocentric radius of 11.2 kpc, we found a blue slope of $-0.006 \pm 0.026$ mag mag$^{-1}$ and a red slope of $0.010 \pm 0.032$ mag mag$^{-1}$.

The slopes of the colour--magnitude relations for both GC subpopulations are consistent with those observed by \citet{2010ApJ...710.1672M}.
Using the colour--metallicity relation of \citetalias{2012MNRAS.426.1475U} and by assuming a constant mass to light ratio within each subpopulation, the slopes correspond to metallicity--mass slopes of $Z \propto M^{0.28 \pm 0.06}$ and $Z \propto M^{0.01 \pm 0.12}$ for the blue and red subpopulations respectively.
Using the colour--metallicity relation of \citet{2006ApJ...639...95P} the mass--metallicity relations are $Z \propto M^{0.55 \pm 0.10}$ and $Z \propto M^{0.02 \pm 0.06}$.  
The lack of a colour--magnitude relation at magnitudes fainter than $(M_{z} = -9.39)$ is in line with the observations of \citet{2009ApJ...699..254H} and \citet{2010ApJ...710.1672M} who both saw significantly weaker blue tilts for GCs fainter than $M_{z} \sim -9.5$.
This suggests a mass threshold for the self enrichment that causes the blue tilt.
Like in other galaxies, the blue tilt in NGC 4278 is much stronger closer to the centre of the galaxy than further away \citep{2006ApJ...653..193M, 2010ApJ...710.1672M, 2012MNRAS.420...37B}.
That the blue tilt is weaker both in lower mass galaxies and the outer parts of higher mass galaxies suggests the environment in which the GC forms effects the amount of self enrichment that occurs.

\subsection{Colour gradients}
\label{gradients}
\begin{figure}
	\begin{center}
		\includegraphics[width=240pt]{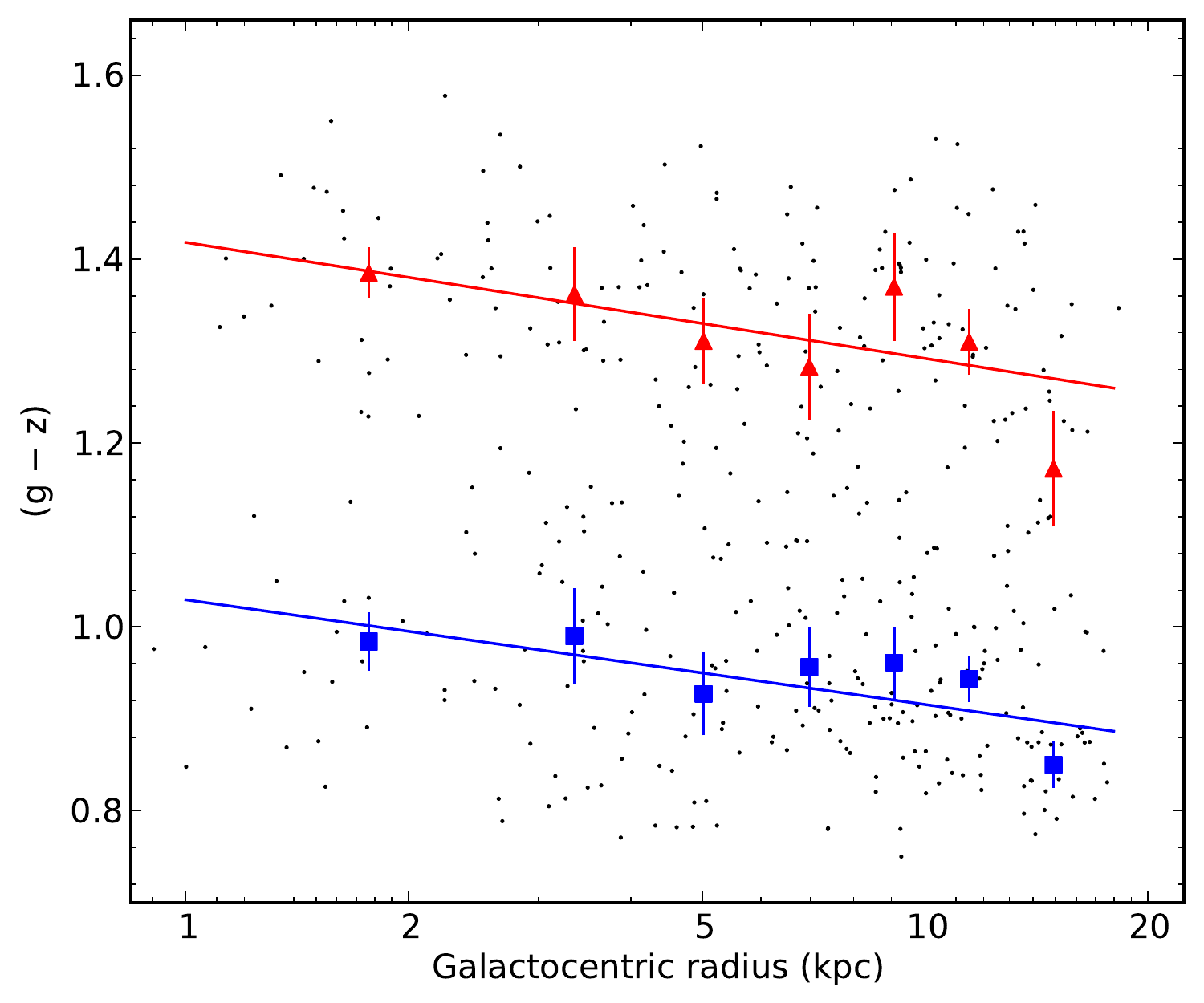}
		\caption{\label{fig:acscolourradius} Colour versus galactocentric radius for ACS GC candidates brighter than the turnover magnitude.
The bimodal peaks found by \textsc{GMM} for each radius bin are over-plotted with error bars in blue and red.
The curved blue and red lines are the linear weighted least square fits to the blue and red peaks as a function of the logarithm of radius.
The GC subpopulations in NGC 4278 show significant radial colour gradients.}
	\end{center}
\end{figure}

Although GC systems usually show global radial colour gradients (since  the blue subpopulations are usually more spatially extended than the red subpopulation) radial colour gradients within both or one of the subpopulations have been seen in several galaxies including NGC 4472 \citep{1996AJ....111.1529G},  NGC 1399 \citep{2001AJ....121.1992F, 2006A&A...451..789B}, NGC 4486 \citep{2009ApJ...703..939H, 2011ApJS..197...33S, 2012MNRAS.421..635F}, NGC 1407 \citep{2011MNRAS.413.2943F}, NGC 3923 \citep{2011MNRAS.416..155F}, NGC 3115 \citep{2011MNRAS.416..155F, 2011ApJ...736L..26A} and the six massive ellipticals in the study of \citet{2009ApJ...699..254H}, as well in a large sample of early-type galaxies in the Virgo and Fornax clusters \citep{2011ApJ...728..116L}.
Some galaxies such as NGC 4486 \citep{2011ApJS..197...33S} and NGC 4649 \citep{2012ApJ...760...87S} show a clear negative gradient in the blue subpopulation while showing evidence of substructure in the red subpopulation.
Since GC colour traces metallicity \citep[e.g.][]{2012ApJ...759L..33B} these colour gradients represent metallicity gradients.
Metallicity gradients have been seen within the GC subpopulations in the Milky Way \citep{1985ApJ...293..424Z, 2001stcl.conf..223H, 2009ApJ...699..254H} and in the metal poor subpopulation of M31 \citep{2008MNRAS.385.1973F}.
In a handful of galaxies, including NGC 1407 \citep{2011MNRAS.413.2943F}, NGC 4486 \citep{2009ApJ...703..939H, 2011ApJS..197...33S}, and the Milky Way \citep{2001stcl.conf..223H}, the colour or metallicity gradients flatten at large radii.
This was interpreted by \citet{2011MNRAS.413.2943F} as evidence for two phase galaxy assembly \cite[e.g.][]{2010ApJ...725.2312O} where the centre of the galaxy was built up early by dissipative collapse and the galaxy halo later by minor mergers.

To look for radial colour gradients within each subpopulation we split the ACS GC candidates brighter than the turnover magnitude into seven equally sized bins in radius.
Within each bin we used \textsc{GMM} to find the peaks of the red and blue subpopulations.
In each bin \textsc{GMM} found strong evidence for bimodality.
We then fit colour as a function of the logarithm of the galactocentric radius using the mean galactocentric radius of the GC candidates in each bin.
The peaks found by \textsc{GMM} along with the fitted colour--radius relations are plotted in Figure~\ref{fig:acscolourradius}.
We found colour gradients within both the blue subpopulation:
\begin{equation}
(g - z)_{blue} = (-0.114 \pm 0.047) \log R + (1.029 \pm 0.044)
\end{equation}
and the red subpopulation:
\begin{equation}
(g - z)_{red} = (-0.126 \pm 0.047) \log R + (1.418 \pm 0.035) .
\end{equation}
These results are independent of the number of bins used.
The $\chi^{2}$ values of the fits are lower than the $\chi^{2}$ values for constant colour with radius for both the blue (7.2 versus 15.7, 5 and 6 degrees of freedom respectively) and red (4.9 versus 12.2, 5 and 6 degrees of freedom respectively) subpopulations.
The $(g - z)$ gradients of the two subpopulations are consistent with one another. 

\begin{figure}
	\begin{center}
		\includegraphics[width=240pt]{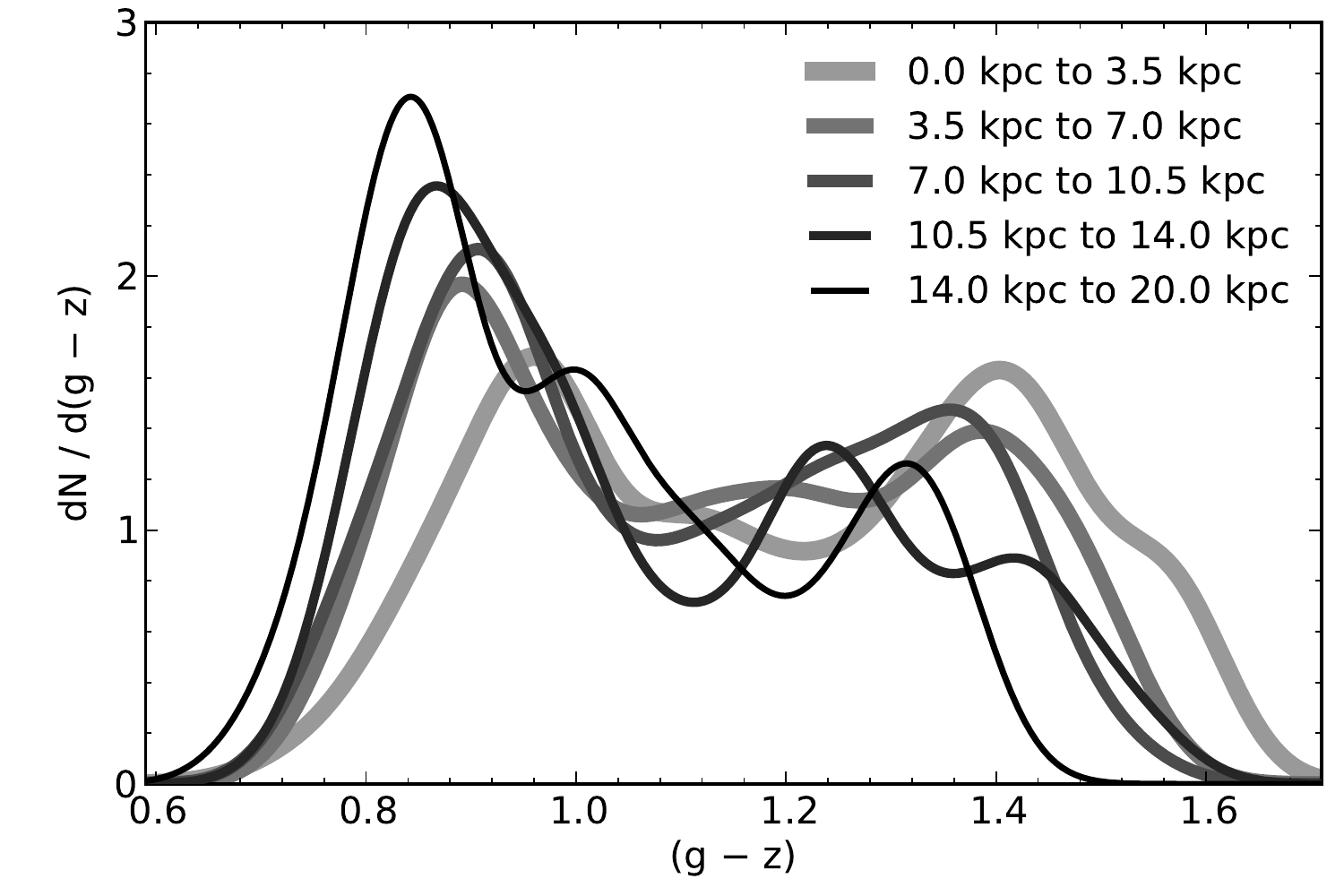}
		\caption{\label{fig:acsradialcolourkde} Gaussian kernel density estimation of the ACS colour distribution in radial bins.
We used 3.5 kpc wide bins and a kernel size of $(g - z) = 0.05$.
Lighter grey and wider lines corresponds to further from the galaxy.
As in Figure~\ref{fig:acscolourradius} the GC subpopulations shift to the blue with increasing radius.}
	\end{center}
\end{figure}

We also used Gaussian kernel density estimation to examine the relationship between colour and radius as plotted in Figure~\ref{fig:acsradialcolourkde}.
We see similar results to those from \textsc{GMM} as the blue and red peaks move blueward with increasing radius.
Using the peaks of the Gaussian kernel density estimates we found colour gradients of $-0.115 \pm 0.025$ mag dex$^{-1}$ and $-0.126 \pm 0.059$ mag dex$^{-1}$ for the blue and red subpopulations respectively, consistent with the values obtained using \textsc{GMM}.
\citet{2012ApJ...760...87S} noted that in NGC 4649, GCs brighter than the turnover showed different colour gradients than those in the magnitude range  $21 < z < 23$ due to variations in the blue tilt with radius.
Since the strength of the blue tilt varies with radius within NGC 4278 (Section~\ref{bluetilt}), we measured the colour gradients of ACS candidates with magnitudes in the range $21 < z < 23$ and found gradients consistent with the colour gradients of GC candidates brighter than the turnover magnitude.

\begin{figure}
	\begin{center}
		\includegraphics[width=240pt]{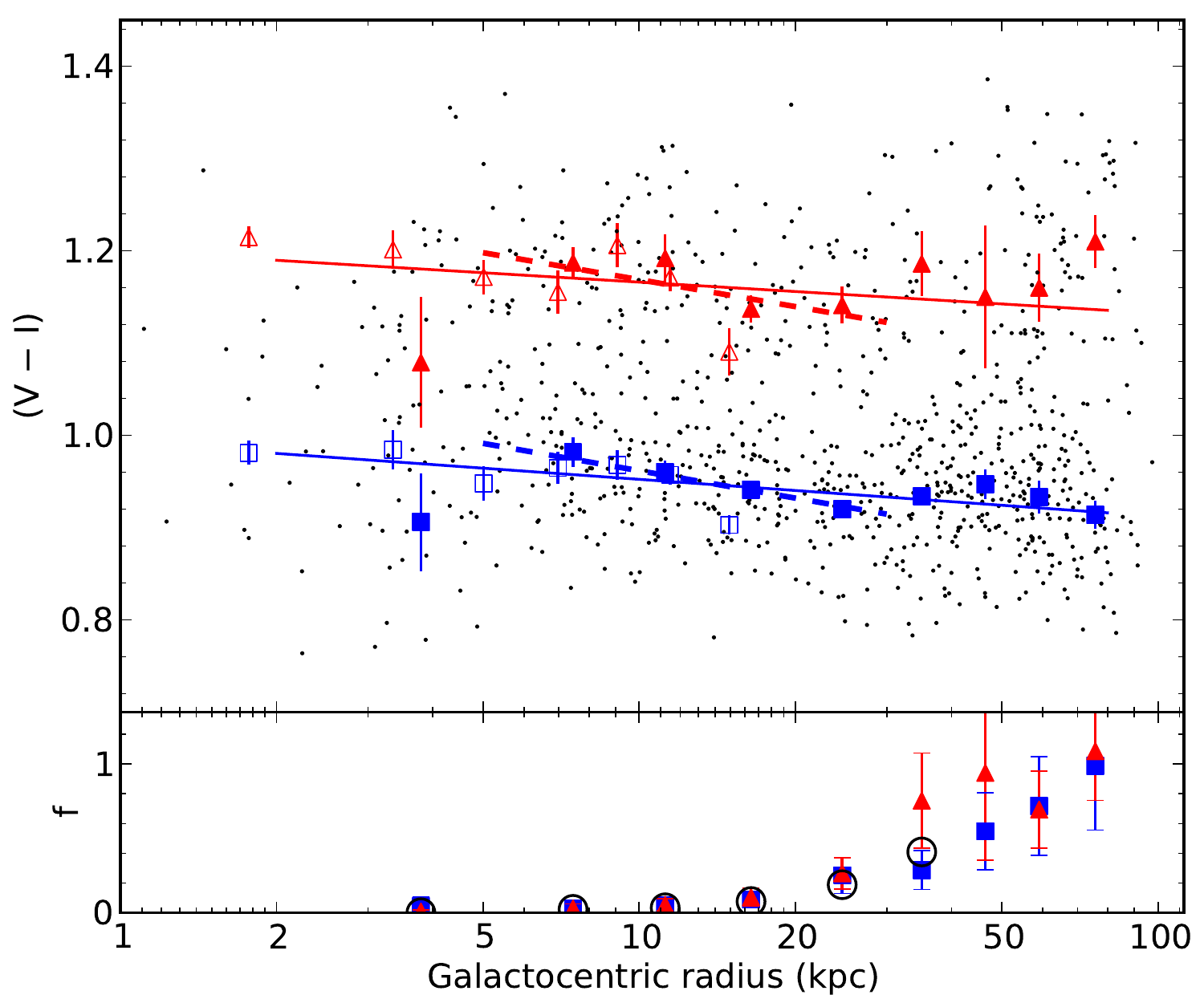}
		\caption{\label{fig:subcolourradius} Suprime-Cam GC colour gradients. \emph{Top} Colour versus galactocentric radius for Suprime-Cam GC candidates brighter than the turnover magnitude.
The bimodal peaks found by \textsc{GMM} for each magnitude bin are plotted with error bars as solid blue squares and red triangles.
The curved blue and red lines are the linear weighted least square fits to the blue and red peaks as a function of the logarithm of radius.
The dashed lines shows the fits to candidates between 5 and 30 pc.
The open points show the \textsc{GMM} results for the ACS candidates. 
The GCs in NGC 4278 show significant radial colour gradients.
The over-density of red candidates at 80 kpc is not spatially collocated.
\emph{Bottom} Estimated contamination fraction for the two subpopulations in Suprime-Cam imaging.
In each radial bin the number background object predicted from the fitted background surface density and the bin's area are divided by the number of GCs \textsc{GMM} determines in each subpopulation.
At 40 kpc more than 50 \% of the candidates are likely background objects.
The black open circles are the ratio of spectroscopic contaminants to all spectroscopically observed candidates in each bin.
The number of spectroscopically observed candidates agrees with the number predicted from the surface density fitting.
}
	\end{center}
\end{figure}

We also looked for colour gradients within the Suprime-Cam dataset.
As with the ACS data, we split Suprime-Cam candidates brighter than the turnover magnitude into nine equally sized bins by radius.
Within each bin we used \textsc{GMM} to find the peaks of the red and blue subpopulations. 
The peaks found by \textsc{GMM} along with the fitted colour--radius relations are plotted on the colour--radius diagram in the top part of Figure~\ref{fig:subcolourradius}.
\textsc{GMM} did not favour a two Gaussian fit over a one Gaussian fit in the innermost bin; in the sixth and seventh bins the kurtosis was positive.
We fit colour as a function of the logarithm of the galactocentric radius using the mean galactocentric radius of the GC candidates in each bin.
We found colour gradients within both the blue subpopulation:
\begin{equation}
(V - I)_{blue} = (-0.040 \pm 0.018) \log R + (0.992 \pm 0.026)
\end{equation}
and the red subpopulation:
\begin{equation}
(V - I)_{red} = (-0.034 \pm 0.038) \log R + (1.225 \pm 0.040) .
\end{equation}
The outermost point was not included in the red fit.
Within the radial range covered by the ACS imaging we see agreement between the Suprime-Cam and ACS gradients.
At large radii, where the Suprime-Cam data are most useful, contamination by foreground stars and background galaxies becomes an important consideration.
Although we do not expect the colour of the contaminants to vary with radius, the contaminants could mask any colour gradient. 
We used the background surface densities from Section~\ref{surfacedensity} to estimate what fraction of GCs are contaminants in each subpopulation  at different radii.
We plot the fraction in the lower part of Figure~\ref{fig:subcolourradius}.
At 40 kpc more than half of the candidates are likely contaminants.
We used the ratio of spectroscopically confirmed contaminants to all spectroscopically observed candidates in each bin to check the contaminant densities.
The number of spectroscopic contaminants in each bin is consistent with the predicted level of contamination.

We repeated the fits for only candidates between 5 and 30 kpc of the galaxy split into five radial bins.
We found slopes of $-0.099 \pm 0.015$ mag dex$^{-1}$ and $-0.097 \pm 0.025$ mag dex$^{-1}$ for the blue and the red subpopulations respectively.
Converted into $(g-z)$ these slopes are steeper but consistent than those measured from the ACS dataset and predict much redder colours within 5 pc than is observed from the ACS dataset.

We see clear evidence for radial colour gradients in both GC subpopulations in the inner regions.
Using the colour--metallicity relation of \citetalias{2012MNRAS.426.1475U} we found that the ACS colour gradients correspond to metallicity gradients of $-0.29 \pm 0.12$ mag dex$^{-1}$ and $-0.32 \pm 0.12$ mag dex$^{-1}$ for the blue and red subpopulation respectively.
Using the colour--metallicity relation of \citet{2006ApJ...639...95P} these correspond to $-0.59 \pm 0.24$ mag dex$^{-1}$ and $-0.23 \pm 0.09$  mag dex$^{-1}$.
Unfortunately the increasing level of contamination at larger radii makes it unclear whether we are observing a flat colour gradient in the outer regions or just the effects of a constant background in the Suprime-Cam data. 
We also note that a constant gradient in log radius appears to flatten at large radii in linear radius.
While the red colour gradient is less clear than that of the blue, unlike in NGC 4649 \citep{2012ApJ...760...87S} and in NGC 4486 \citep{2011ApJS..197...33S}, we do not see clear evidence for substructure in the red subpopulation.
Low number statistics likely limit our ability to detect substructure. 

The strengths of NGC 4278's GC colour gradients are steeper than the mean gradients observed in Virgo and Fornax early-type galaxies by \citet{2011ApJ...728..116L} but consistent with the observed scatter.
Likewise NGC 4278's gradients are steeper than those observed in six early-type galaxies by \citet{2009ApJ...699..254H}.
However, NGC 4278's gradients are consistent with those of studies such as \citet[NGC 1407]{2011MNRAS.413.2943F}, \citet[NGC 3115]{2011ApJ...736L..26A}, \citet[NGC 4365]{2012MNRAS.420...37B} and \citet[NGC 4649]{2012ApJ...760...87S}.
For NGC 4278 and these studies, the strengths of the blue and red subpopulation gradients are consistent with one another.
The connection between GC colour gradients and their host galaxies' properties will be discussed in depth in future SLUGGS survey papers.

\section{Globular cluster sizes}
\label{size}
Previous observations \citep[e.g.][]{2001AJ....121.2950K, 2001AJ....121.2974L,2005ApJ...634.1002J, 2010ApJ...715.1419M} have shown that the sizes of GCs are consistent between galaxies of similar luminosities so much so that GC sizes have been suggested as a distance indicator \citep{2001AJ....121.2950K}.
The mean size of the GCs of a galaxy varies weakly with galaxy luminosity; luminous galaxies have slightly smaller GCs than faint galaxies \citep{2005ApJ...634.1002J, 2009MNRAS.392..879G, 2010ApJ...715.1419M}.
Remarkably, GC sizes are constant across more than two orders of magnitude of GC luminosity \citep[e.g.][]{2000ApJ...539..618M, 2005ApJ...634.1002J}.
Only the brightest GCs show any size dependence on luminosity \citep[e.g.][]{2009ApJ...699..254H}.  
 
Although GC half light radii are independent of magnitude \citep{2005ApJ...634.1002J} for all but the brightest GCs \citep{2009ApJ...699..254H}, the sizes of GCs in the Milky Way have long been known to increase with distance from the Galactic centre \citep{1956ZA.....41...61V}.
Similar radial trends are seen in other galaxies \citep[e.g.][]{1991ApJ...375..594V, 2007ApJ...670L.105G, 2006AJ....132.1593S, 2009ApJ...699..254H, 2012MNRAS.420...37B}.
Since tidal fields influence GC sizes and tidal fields are stronger closer to the centre of a galaxy, they provide a natural explanation for why GCs are smaller closer to the centres of galaxies \citep[e.g.][]{2012ApJ...756..167M}.
This could also explain why dwarf galaxies have larger GCs due to the weaker tidal forces in less massive galaxies.
Alternatively the larger sizes at larger radii could be an effect of the initial sizes being larger, as GC formation is still poorly understood.

\citet{1998AJ....116.2841K} first noticed that red GCs in NGC 3115 were $\sim 20 \%$ smaller than blue ones. 
Similar differences in the half light radii between the colour subpopulations have been seen in several studies including \citet{2001AJ....121.2974L}, \citet{2005ApJ...634.1002J}, \citet{2009ApJ...699..254H} and \citet{2010ApJ...715.1419M}.
Since the red subpopulation is in general more centrally concentrated (see Section~\ref{surfacedensity}) than the blue, and GC sizes increase with galactocentric radius, \citet{2003ApJ...593..340L} argued that projection effects cause the apparent size difference between the subpopulations.
However, \citet{2004ApJ...613L.117J} proposed that mass segregation and the effects of metallicity on stellar evolution can produce the difference in half light radius.
Additionally, the initial sizes of GCs could be metallicity dependent.

\begin{figure*}
	\begin{center}
		\includegraphics[width=504pt]{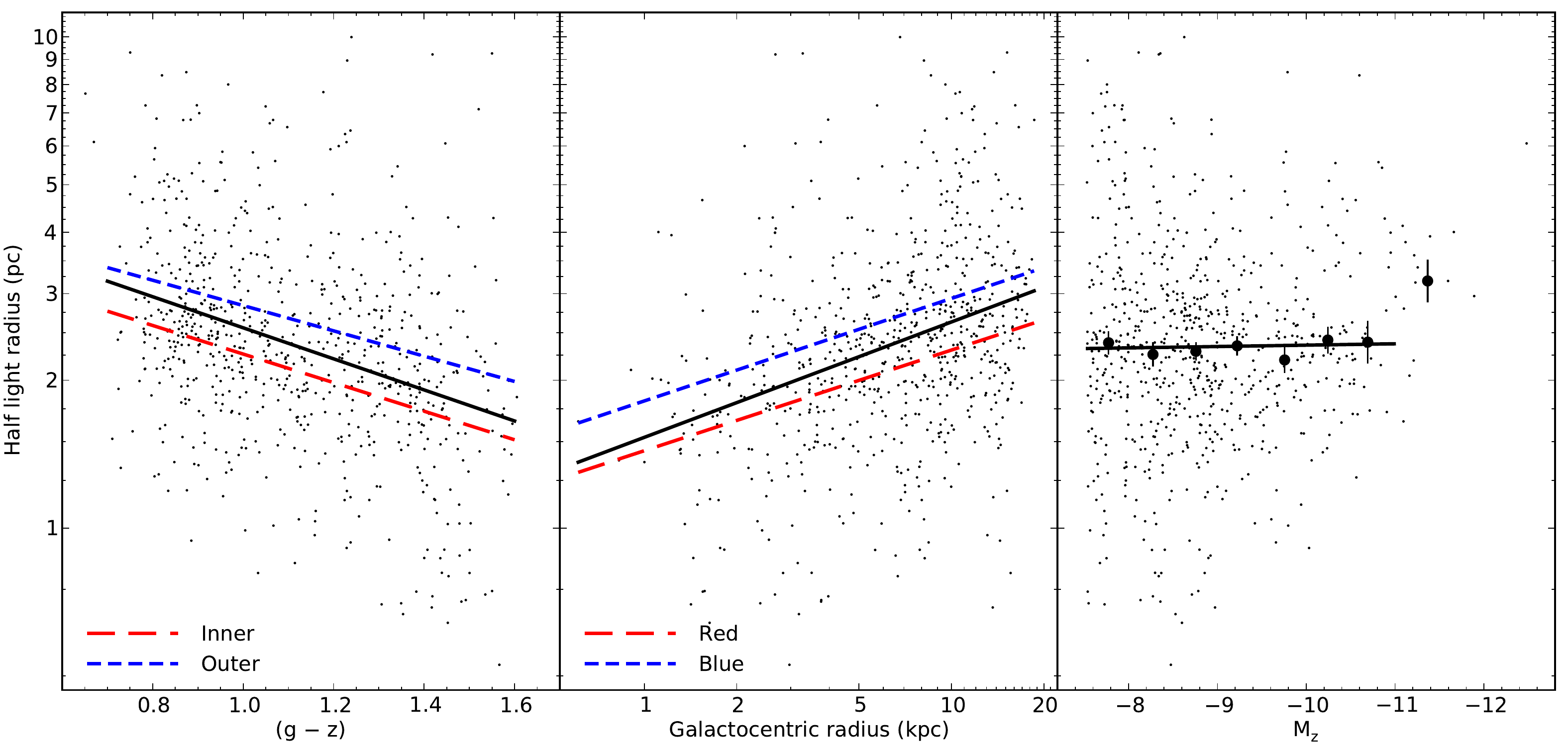}
		\caption{\label{fig:acssizes} GC candidate sizes from ACS.
\emph{Left} GC candidate half light radius versus colour.
The black line is the linear fit of the logarithm of the radius as a function of the colour.
The short dashed blue line is the outer half of the candidates while the long dashed red line is the inner half of the candidates.
Blue (metal poor) GCs are larger than red (metal rich) GCs, independent of galactocentric radius.
\emph{Middle} Half light radius versus galactocentric radius.
The black line is the power law fit of half light radius as a function of galactocentric radius.
The short dashed blue line and the long dashed red line are the same fit for the blue and red subpopulations respectively.
Both blue and red GCs are larger further from the centre of the galaxy.
\emph{Right} Half light radius versus absolute magnitude.
The black circles are the median half light radius in 0.5 magnitude bins; the brightest bin is all GCs brighter than $M_{z} = -11$.
The black line is the linear fit of the logarithm of the radius as a function of magnitude restricted to GC candidates fainter than $M_{z} = -11$.
The brightest GCs show a higher mean size than the bulk of the GC population.}
	\end{center}
\end{figure*}

\begin{figure}
	\begin{center}
		\includegraphics[width=240pt]{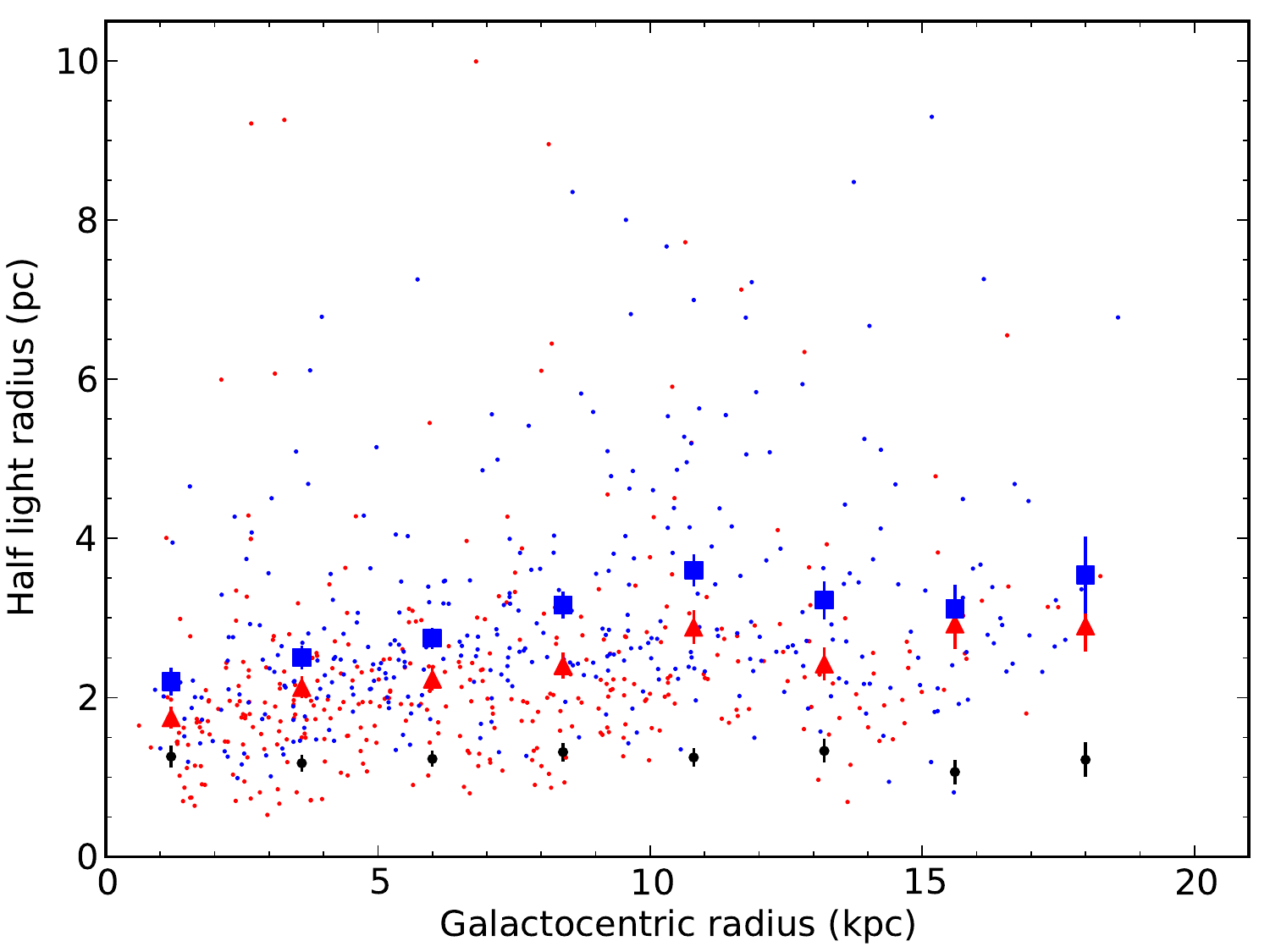}
		\caption{\label{fig:acsradiussize} Half light radius versus galactocentric radius for ACS GC candidates.
Points are colour coded based on whether they are members of the blue or red subpopulations.
The blue squares and red triangles are the mean half light radii of the blue and red GCs respectively in 2.4 kpc wide bins.
The black points are the ratio of the mean blue half light radii to the mean red half light radii.
The ratio is constant with galactocentric radius suggesting that the difference in GC size between the two subpopulations is independent of galactocentric radius.
Both subpopulations show a clear trend of increasing size with radius within 12 kpc ($5 R_{e}$); beyond this radius the trend flattens.
}
	\end{center}
\end{figure}

Using the sizes calculated in Section~\ref{sizemeasurements} we find that the ACS GC candidates have a mean half light radius of $2.64 \pm 0.05$ pc.
This value agrees with previous work at our adopted distance and host galaxy luminosity \citep{2005ApJ...634.1002J, 2010ApJ...715.1419M}. 
Splitting the sample in two by colour we found that the red GC candidates have a smaller mean half light radius of $2.31 \pm 0.07$ pc and the blue GC candidates have a larger mean half light radius of $2.99 \pm 0.07$ pc.

Using the ACS GC candidates, we explored the dependence of GC size on colour, galactocentric radius and luminosity.
The relationships of colour, radius and 
luminosity with size are plotted in Figure~\ref{fig:acssizes}.
We fitted a linear relation between the logarithm of the half light radius and the colour:
\begin{equation}
\log r_{hl} = (-0.316^{+0.035}_{-0.033}) \times (g - z) + (0.724^{+0.037}_{-0.039})
\end{equation}
The uncertainties in the fit were calculated using bootstrapping and the RMS is 0.191 dex.
We also split the candidates into two equal sized bins by galactocentric radius and fitted the relationship between the logarithm of the half light radius and the colour for each bin, finding the slopes in each bin to be smaller but consistent with that of the entire population.

We fitted a linear relation between the logarithm of the half light radius and the logarithm of the galactocentric radius:
\begin{equation}
\log r_{hl} = (0.235^{+0.023}_{-0.023}) \log R + (0.185_{-0.021}^{+0.020})
\end{equation}
The uncertainties in the fit were calculated using bootstrapping.
We also fitted the relationship for the red and blue subpopulations separately, finding the slopes of the relations to be smaller but consistent with that of the entire population.
Since colour changes with radius we fitted size as a function of both colour and radius simultaneously:
\begin{multline}
\log r_{hl} = (0.193_{-0.022}^{+0.023}) \times \log R \\ + (-0.263_{-0.033}^{+0.031}) \times (g - z) + (0.511_{-0.041}^{+0.040}).
\end{multline}
The RMS of this relation is 0.183 dex.
Since we believe the size measurement errors are reasonable (from the differences in the sizes measured for GCs appearing in multiple pointings) and much lower than this scatter (the mean error in log radius is 0.036 dex) this indicates that there is a large scatter in GC size at fixed colour and projected galactocentric radius.

In the right panel of Figure~\ref{fig:acssizes} we plot the median half light radius in magnitude bins.
The size of GCs appears to be constant with magnitude, apart from the brightest GCs. 
We fitted a linear relationship between the logarithm of the half light radius and the $z$ band absolute magnitude for candidates fainter than $M_{z} = -11$, finding no significant relation:
\begin{equation}
\log r_{hl} = (0.007^{+0.022}_{-0.025}) M_{z} + (0.472_{-0.217}^{+0.195})
\end{equation}
The uncertainties in the fit were calculated using bootstrapping and the relation has a RMS of 0.204 dex.
A Spearman rank correlation test gives a correlation value of $\rho = -0.0059$ which corresponds to a probability of $p = 0.88$ that an uncorrelated population of the same size would produce the observed correlation value.
A KS test shows a probability of $5.4 \times 10^{-4}$ that GCs brighter than $M_{z} = -11$ are drawn from same distribution as the fainter GC candidates.

In Figure~\ref{fig:acsradiussize} we plot the mean half light radii of the blue and red GCs in radial bins. 
The ratio of the blue sizes to red sizes is independent of radius and has a mean value $1.21 \pm 0.03$.
Using the median size rather than the mean size gave identical results as did using the mean of the log size.
That the ratio is independent of galactocentric radius indicates that the apparent size difference between metal poor and metal rich GCs is caused by the effects of metallicity, not projection effects.
Recent multi-pointing HST studies, such as those of NGC 4365 \citep{2012MNRAS.420...37B} and NGC 4649 \citep{2012ApJ...760...87S},  have similarly seen size differences at all radii.
In the case of NGC 4649 \citep{2012ApJ...760...87S} the size differences are observed at all radii despite the size being constant or decreasing beyond 15 kpc.
\citet{2012ApJ...759L..39W} simulated the effects of projection on their observations of GCs in NGC 4486.
They concluded that projection effects cannot explain the observed size difference.
Theoretical studies of GC dynamical evolution which include the effects of stellar evolution have produced a $\sim 20$ \% difference in half light radius between metal poor and metal rich GC using both Monte Carlo techniques \citep{2012MNRAS.425.2234D} and direct N-body models \citep{2012MNRAS.427..167S}.
The difference in half light radius thus could be explained as a combination of mass segregation and the differences in stellar evolution of stars of different metallicities.

We observe a GC size--galactocentric radius relationship of $r_{hl} \propto R^{0.19 \pm 0.02}$ after taking the effects of colour into account.
Previous studies have revealed a range of power law indices ($\alpha$) for this relation with the Milky Way showing $\alpha = 0.46 \pm 0.05$ \citep{2012ApJ...746...93W}, the six giant galaxies studied by \citep{2009ApJ...699..254H} showing $\alpha = 0.11$, NGC 4594 \citep{2010MNRAS.401.1965H} showing $\alpha = 0.17 \pm 0.02$ and NGC 4365 \citep{2012MNRAS.420...37B} showing $\alpha = 0.47 \pm 0.02$.
In both NGC 4649 \citep{2012ApJ...760...87S} and in NGC 1399 \citep{2011ApJ...736...90P} only the inner ($< 15$ kpc) GCs show a size--radius relation.
In NGC 4278 the relationship between size and radius appears to be stronger within 12 kpc ($5 R_{e}$) than beyond.
This is consistent with the models of \citet{2012ApJ...756..167M} where the relationship between size and radius flattens at large galactocentric radii.
Since the tidal history of a GC affects its size, the sizes of GCs in a galaxy are dependent on its assembly history.
Detailed models of GC sizes, spatial distributions and kinematics could be compared with observations to place constraints on the assembly history of a galaxy. 
We note that the tidal history, not just the tidal forces at the GC's current location, affect a GC's size.
This at least partially explains the scatter in GC sizes observed even when accounting for galactocentric distance and colour.
Additionally projection effects likely cause some scatter as tidal forces are dependent on the true 3 dimensional galactocentric radius, not on the observed projected radius. 

In line with other galaxies \citep[e.g.][]{2009ApJ...699..254H}, we found that GC size is independent of luminosity below $M_{z} \sim -11$ while brighter GCs have larger mean sizes.
In common with massive early-type galaxies like NGC 4486 \citep{2011AJ....142..199B} and NGC 4649 \citep{2012ApJ...760...87S}, NGC 4278 has a handful of UCD and UCD candidates (see Table \ref{tab:ucdc}).
One UCD (acs\_0320) is similar in half-light radius (23 pc), colour ($V - I = 0.92$) and absolute magnitude ($M_V = -9.6$) to the peculiar Milky Way GC NGC 2419 \citep{2011ApJ...738..186I, 2012ApJ...760...86C}.
The confirmed UCDs in NGC 4278 are discussed in more detail in \citet{Forbes2013}.

\section{Summary}
\label{summary}
We used five HST ACS pointings and Subaru Suprime-Cam wide field imaging to study the GC system of NGC 4278.
We used colours and sizes to select GC candidates from the ACS imaging and a colour--colour selection to identify GC candidates in the Suprime-Cam imaging.
We also identified a handful of UCD candidates in the ACS imaging.
Using new spectroscopy we confirm two of the UCD candidates, and when combined with \citetalias{2013MNRAS.428..389P}, bring the total number of spectroscopically confirmed GCs and UCDs in NGC 4278 to 270.

We measured the surface density profile of NGC 4278 GC candidates, finding that as in other galaxies, NGC 4278's blue GCs are more spatially extended than the red GCs.
By integrating the surface density profile we estimated the total number of GCs to be $1378^{+32}_{-194}$.
This corresponds to a specific frequency of $6.0^{+1.2}_{-1.3}$, making the GC system of NGC 4278 rich for its luminosity.
Using the GC surface density profile we were able to show that the nearby galaxy NGC 4283 makes an insignificant contribution to the NGC 4278 GC system and that NGC 4283 likely has a low number of GCs for its luminosity.
Although NGC 4278's and NGC 4283's GC systems are rich and poor respectively for their luminosities, they are both within the observed scatter of the large sample of \citet{2013ApJ...772...82H}.

Like almost all large galaxies, NGC 4278's GC system shows colour bimodality.
A clear blue tilt is observed along with evidence that the blue tilt is stronger in the inner regions of the galaxy than in the outer regions.
We observe clear radial colour gradients in both GC subpopulations.
The gradients are present in both the ACS and Suprime-Cam data.
Unfortunately, due to the level of contamination in the Suprime-Cam data we could not reliably use it to probe the metallicity gradient much further than the coverage of the ACS data.
This means that we cannot see if the gradients flatten at large radii as is the case in NGC 1407.
When compared to colour gradients in other galaxies, NGC 4278's gradients are typical.

We also studied the relationships between GCs' half light radii and their colours and positions within NGC 4278.
We found that like other galaxies, GC sizes decrease with redder colours and increase with distance from NGC 4278.
For GCs fainter than $M_{z} = -11$ we found no relationship between size and GC luminosity; GCs brighter than $M_{z} = -11$ have larger sizes than the fainter GCs.
We found that the ratio of the sizes of blue GCs to red GCs is independent of galactocentric distance indicating that the apparent size difference between red and blue GCs is due to dynamical and stellar evolution and not to projection effects.
Although NGC 4278 is a less luminous galaxy than previous multi-pointing studies of GC systems (e.g. NGC 4594, NGC 4365 and NGC 4649), the properties of NGC 4278's GC system are similar.

\section*{Acknowledgments}
We wish to thank the referee for his useful comments which greatly improved the manuscript.
We thank Vincenzo Pota and Christina Blom for their careful reading of the manuscript.
We thank Caroline Foster, Samantha Penny, Jacob Arnold, Christina Blom, Vincenzo Pota, Sreeja Kartha, Nicola Pastorello, Anna Sippel, Juan Madrid and Gonzalo D\'iaz for useful discussions.
We acknowledge the usage of the HyperLeda database (http://leda.univ-lyon1.fr).
This publication makes use of data products from the Two Micron All Sky Survey, which is a joint project of the University of Massachusetts and the Infrared Processing and Analysis Center/California Institute of Technology, funded by the National Aeronautics and Space Administration and the National Science Foundation.
This research has made use of the NASA/IPAC Extragalactic Database (NED) which is operated by the Jet Propulsion Laboratory, California Institute of Technology, under contract with the National Aeronautics and Space Administration. 
The analysis pipeline used to reduce the DEIMOS data was developed at UC Berkeley with support from NSF grant AST-0071048.
This publication made use of \textsc{PyRAF} and \textsc{PyFITS} which are products of the Space Telescope Science Institute, which is operated by AURA for NASA.
This research made use of \textsc{TOPCAT} \citep{2005ASPC..347...29T}.
Some of the data presented herein were obtained at the W. M. Keck Observatory, operated as a scientific partnership among the California Institute of Technology, the University of California and the National Aeronautics and Space Administration, and made possible by the generous financial support of the W. M. Keck Foundation.
The authors wish to recognize and acknowledge the very significant cultural role and reverence that the summit of Mauna Kea has always had within the indigenous Hawaiian community.  We are most fortunate to have the opportunity to conduct observations from this mountain. 
This research is based in part on data collected at Subaru Telescope, which is operated by the National Astronomical Observatory of Japan. 
This research is based on observations made with the NASA/ESA Hubble Space Telescope, obtained from the data archive at the Space Telescope Science Institute. STScI is operated by the Association of Universities for Research in Astronomy, Inc. under NASA contract NAS 5-26555.
This material is based upon work supported by the Australian Research Council under grant DP130100388 and by the National Science Foundation under grants AST-0909237, AST-1109878 and AST-1211995.

\bibliographystyle{mn2e}
\bibliography{n4278gc}

\begin{thebibliography}{122}
\expandafter\ifx\csname natexlab\endcsname\relax\def\natexlab#1{#1}\fi

\bibitem[{{Abazajian} {et~al.}(2009){Abazajian}, {Adelman-McCarthy},
  {Ag{\"u}eros}, {Allam}, {Allende Prieto}, {An}, {Anderson}, {Anderson},
  {Annis}, {Bahcall}, \& et~al.}]{2009ApJS..182..543A}
{Abazajian} K.~N., {Adelman-McCarthy} J.~K., {Ag{\"u}eros} M.~A., {Allam}
  S.~S., {Allende Prieto} C., {An} D., {Anderson} K.~S.~J., {Anderson} S.~F.,
  {Annis} J., {Bahcall} N.~A., et~al., 2009, \apjs, 182, 543

\bibitem[{{Alves-Brito} {et~al.}(2011){Alves-Brito}, {Hau}, {Forbes},
  {Spitler}, {Strader}, {Brodie}, \& {Rhode}}]{2011MNRAS.417.1823A}
{Alves-Brito} A., {Hau} G.~K.~T., {Forbes} D.~A., {Spitler} L.~R., {Strader}
  J., {Brodie} J.~P., {Rhode} K.~L., 2011, \mnras, 417, 1823

\bibitem[{{Arnold} {et~al.}(2011){Arnold}, {Romanowsky}, {Brodie}, {Chomiuk},
  {Spitler}, {Strader}, {Benson}, \& {Forbes}}]{2011ApJ...736L..26A}
{Arnold} J.~A., {Romanowsky} A.~J., {Brodie} J.~P., {Chomiuk} L., {Spitler}
  L.~R., {Strader} J., {Benson} A.~J., {Forbes} D.~A., 2011, \apjl, 736, L26

\bibitem[{{Baba} {et~al.}(2002){Baba}, {Yasuda}, {Ichikawa}, {Yagi}, {Iwamoto},
  {Takata}, {Horaguchi}, {Taga}, {Watanabe}, {Ozawa}, \&
  {Hamabe}}]{2002ASPC..281..298B}
{Baba} H., {Yasuda} N., {Ichikawa} S., {Yagi} M., {Iwamoto} N., {Takata} T.,
  {Horaguchi} T., {Taga} M., {Watanabe} M., {Ozawa} T., {Hamabe} M., 2002, in
  Astronomical Society of the Pacific Conference Series, Vol. 281, Astronomical
  Data Analysis Software and Systems XI, {D.~A.~Bohlender, D.~Durand, \&
  T.~H.~Handley}, ed., pp. 298--302

\bibitem[{{Bailin} \& {Harris}(2009)}]{2009ApJ...695.1082B}
{Bailin} J., {Harris} W.~E., 2009, \apj, 695, 1082

\bibitem[{{Bassino} {et~al.}(2006){Bassino}, {Faifer}, {Forte}, {Dirsch},
  {Richtler}, {Geisler}, \& {Schuberth}}]{2006A&A...451..789B}
{Bassino} L.~P., {Faifer} F.~R., {Forte} J.~C., {Dirsch} B., {Richtler} T.,
  {Geisler} D., {Schuberth} Y., 2006, \aap, 451, 789

\bibitem[{{Beasley} {et~al.}(2008){Beasley}, {Bridges}, {Peng}, {Harris},
  {Harris}, {Forbes}, \& {Mackie}}]{2008MNRAS.386.1443B}
{Beasley} M.~A., {Bridges} T., {Peng} E., {Harris} W.~E., {Harris} G.~L.~H.,
  {Forbes} D.~A., {Mackie} G., 2008, \mnras, 386, 1443

\bibitem[{{Blakeslee} {et~al.}(2001){Blakeslee}, {Lucey}, {Barris}, {Hudson},
  \& {Tonry}}]{2001MNRAS.327.1004B}
{Blakeslee} J.~P., {Lucey} J.~R., {Barris} B.~J., {Hudson} M.~J., {Tonry}
  J.~L., 2001, \mnras, 327, 1004

\bibitem[{{Blom} {et~al.}(2012){Blom}, {Spitler}, \&
  {Forbes}}]{2012MNRAS.420...37B}
{Blom} C., {Spitler} L.~R., {Forbes} D.~A., 2012, \mnras, 420, 37

\bibitem[{{Brassington} {et~al.}(2009){Brassington}, {Fabbiano}, {Kim},
  {Zezas}, {Zepf}, {Kundu}, {Angelini}, {Davies}, {Gallagher}, {Kalogera},
  {Fragos}, {King}, {Pellegrini}, \& {Trinchieri}}]{2009ApJS..181..605B}
{Brassington} N.~J., {Fabbiano} G., {Kim} D.-W., {Zezas} A., {Zepf} S., {Kundu}
  A., {Angelini} L., {Davies} R.~L., {Gallagher} J., {Kalogera} V., {Fragos}
  T., {King} A.~R., {Pellegrini} S., {Trinchieri} G., 2009, \apjs, 181, 605

\bibitem[{{Brodie} {et~al.}(2011){Brodie}, {Romanowsky}, {Strader}, \&
  {Forbes}}]{2011AJ....142..199B}
{Brodie} J.~P., {Romanowsky} A.~J., {Strader} J., {Forbes} D.~A., 2011, \aj,
  142, 199

\bibitem[{{Brodie} \& {Strader}(2006)}]{2006ARA&A..44..193B}
{Brodie} J.~P., {Strader} J., 2006, \araa, 44, 193

\bibitem[{{Brodie} {et~al.}(2012){Brodie}, {Usher}, {Conroy}, {Strader},
  {Arnold}, {Forbes}, \& {Romanowsky}}]{2012ApJ...759L..33B}
{Brodie} J.~P., {Usher} C., {Conroy} C., {Strader} J., {Arnold} J.~A., {Forbes}
  D.~A., {Romanowsky} A.~J., 2012, \apjl, 759, L33

\bibitem[{{Burkert} \& {Tremaine}(2010)}]{2010ApJ...720..516B}
{Burkert} A., {Tremaine} S., 2010, \apj, 720, 516

\bibitem[{{Cappellari} {et~al.}(2011){Cappellari}, {Emsellem}, {Krajnovi{\'c}},
  {McDermid}, {Scott}, {Verdoes Kleijn}, {Young}, {Alatalo}, {Bacon}, {Blitz},
  {Bois}, {Bournaud}, {Bureau}, {Davies}, {Davis}, {de Zeeuw}, {Duc},
  {Khochfar}, {Kuntschner}, {Lablanche}, {Morganti}, {Naab}, {Oosterloo},
  {Sarzi}, {Serra}, \& {Weijmans}}]{2011MNRAS.413..813C}
{Cappellari} M., {Emsellem} E., {Krajnovi{\'c}} D., {McDermid} R.~M., {Scott}
  N., {Verdoes Kleijn} G.~A., {Young} L.~M., {Alatalo} K., {Bacon} R., {Blitz}
  L., {Bois} M., {Bournaud} F., {Bureau} M., {Davies} R.~L., {Davis} T.~A., {de
  Zeeuw} P.~T., {Duc} P.-A., {Khochfar} S., {Kuntschner} H., {Lablanche} P.-Y.,
  {Morganti} R., {Naab} T., {Oosterloo} T., {Sarzi} M., {Serra} P., {Weijmans}
  A.-M., 2011, \mnras, 413, 813

\bibitem[{{Cappellari} {et~al.}(2013){Cappellari}, {Scott}, {Alatalo}, {Blitz},
  {Bois}, {Bournaud}, {Bureau}, {Crocker}, {Davies}, {Davis}, {de Zeeuw},
  {Duc}, {Emsellem}, {Khochfar}, {Krajnovi{\'c}}, {Kuntschner}, {McDermid},
  {Morganti}, {Naab}, {Oosterloo}, {Sarzi}, {Serra}, {Weijmans}, \&
  {Young}}]{2013MNRAS.432.1709C}
{Cappellari} M., {Scott} N., {Alatalo} K., {Blitz} L., {Bois} M., {Bournaud}
  F., {Bureau} M., {Crocker} A.~F., {Davies} R.~L., {Davis} T.~A., {de Zeeuw}
  P.~T., {Duc} P.-A., {Emsellem} E., {Khochfar} S., {Krajnovi{\'c}} D.,
  {Kuntschner} H., {McDermid} R.~M., {Morganti} R., {Naab} T., {Oosterloo} T.,
  {Sarzi} M., {Serra} P., {Weijmans} A.-M., {Young} L.~M., 2013, \mnras, 432,
  1709

\bibitem[{{Chies-Santos} {et~al.}(2012){Chies-Santos}, {Larsen}, {Cantiello},
  {Strader}, {Kuntschner}, {Wehner}, \& {Brodie}}]{2012A&A...539A..54C}
{Chies-Santos} A.~L., {Larsen} S.~S., {Cantiello} M., {Strader} J.,
  {Kuntschner} H., {Wehner} E.~M., {Brodie} J.~P., 2012, \aap, 539, A54

\bibitem[{{Chies-Santos} {et~al.}(2011{\natexlab{a}}){Chies-Santos}, {Larsen},
  {Kuntschner}, {Anders}, {Wehner}, {Strader}, {Brodie}, \&
  {Santos}}]{2011A&A...525A..20C}
{Chies-Santos} A.~L., {Larsen} S.~S., {Kuntschner} H., {Anders} P., {Wehner}
  E.~M., {Strader} J., {Brodie} J.~P., {Santos} J.~F.~C., 2011{\natexlab{a}},
  \aap, 525, A20+

\bibitem[{{Chies-Santos} {et~al.}(2011{\natexlab{b}}){Chies-Santos}, {Larsen},
  {Wehner}, {Kuntschner}, {Strader}, \& {Brodie}}]{2011A&A...525A..19C}
{Chies-Santos} A.~L., {Larsen} S.~S., {Wehner} E.~M., {Kuntschner} H.,
  {Strader} J., {Brodie} J.~P., 2011{\natexlab{b}}, \aap, 525, A19

\bibitem[{{Ciardullo}(2012)}]{2012Ap&SS.341..151C}
{Ciardullo} R., 2012, \apss, 341, 151

\bibitem[{{Ciardullo} {et~al.}(2002){Ciardullo}, {Feldmeier}, {Jacoby}, {Kuzio
  de Naray}, {Laychak}, \& {Durrell}}]{2002ApJ...577...31C}
{Ciardullo} R., {Feldmeier} J.~J., {Jacoby} G.~H., {Kuzio de Naray} R.,
  {Laychak} M.~B., {Durrell} P.~R., 2002, \apj, 577, 31

\bibitem[{{Cohen} \& {Kirby}(2012)}]{2012ApJ...760...86C}
{Cohen} J.~G., {Kirby} E.~N., 2012, \apj, 760, 86

\bibitem[{{Conroy} {et~al.}(2009){Conroy}, {Gunn}, \&
  {White}}]{2009ApJ...699..486C}
{Conroy} C., {Gunn} J.~E., {White} M., 2009, \apj, 699, 486

\bibitem[{{Cooper} {et~al.}(2012){Cooper}, {Newman}, {Davis}, {Finkbeiner}, \&
  {Gerke}}]{2012ascl.soft03003C}
{Cooper} M.~C., {Newman} J.~A., {Davis} M., {Finkbeiner} D.~P., {Gerke} B.~F.,
  2012, in Astrophysics Source Code Library, record ascl:1203.003, p. 3003

\bibitem[{{C{\^o}t{\'e}} {et~al.}(2004){C{\^o}t{\'e}}, {Blakeslee},
  {Ferrarese}, {Jord{\'a}n}, {Mei}, {Merritt}, {Milosavljevi{\'c}}, {Peng},
  {Tonry}, \& {West}}]{2004ApJS..153..223C}
{C{\^o}t{\'e}} P., {Blakeslee} J.~P., {Ferrarese} L., {Jord{\'a}n} A., {Mei}
  S., {Merritt} D., {Milosavljevi{\'c}} M., {Peng} E.~W., {Tonry} J.~L., {West}
  M.~J., 2004, \apjs, 153, 223

\bibitem[{{de Vaucouleurs} {et~al.}(1991){de Vaucouleurs}, {de Vaucouleurs},
  {Corwin}, {Buta}, {Paturel}, \& {Fouque}}]{1991trcb.book.....D}
{de Vaucouleurs} G., {de Vaucouleurs} A., {Corwin} Jr. H.~G., {Buta} R.~J.,
  {Paturel} G., {Fouque} P., 1991, {Third Reference Catalogue of Bright
  Galaxies}, {de Vaucouleurs, G., de Vaucouleurs, A., Corwin, H.~G., Jr., Buta,
  R.~J., Paturel, G., \& Fouque, P.}, ed. Springer-Verlag, Berlin

\bibitem[{{Downing}(2012)}]{2012MNRAS.425.2234D}
{Downing} J.~M.~B., 2012, \mnras, 425, 2234

\bibitem[{{Fabbiano} {et~al.}(2010){Fabbiano}, {Brassington}, {Lentati},
  {Angelini}, {Davies}, {Gallagher}, {Kalogera}, {Kim}, {King}, {Kundu},
  {Pellegrini}, {Richings}, {Trinchieri}, {Zezas}, \&
  {Zepf}}]{2010ApJ...725.1824F}
{Fabbiano} G., {Brassington} N.~J., {Lentati} L., {Angelini} L., {Davies}
  R.~L., {Gallagher} J., {Kalogera} V., {Kim} D.-W., {King} A.~R., {Kundu} A.,
  {Pellegrini} S., {Richings} A.~J., {Trinchieri} G., {Zezas} A., {Zepf} S.,
  2010, \apj, 725, 1824

\bibitem[{{Faifer} {et~al.}(2011){Faifer}, {Forte}, {Norris}, {Bridges},
  {Forbes}, {Zepf}, {Beasley}, {Gebhardt}, {Hanes}, \&
  {Sharples}}]{2011MNRAS.416..155F}
{Faifer} F.~R., {Forte} J.~C., {Norris} M.~A., {Bridges} T., {Forbes} D.~A.,
  {Zepf} S.~E., {Beasley} M., {Gebhardt} K., {Hanes} D.~A., {Sharples} R.~M.,
  2011, \mnras, 416, 155

\bibitem[{{Fan} {et~al.}(2008){Fan}, {Ma}, {de Grijs}, \&
  {Zhou}}]{2008MNRAS.385.1973F}
{Fan} Z., {Ma} J., {de Grijs} R., {Zhou} X., 2008, \mnras, 385, 1973

\bibitem[{{Forbes} {et~al.}(2013){Forbes}, {Pota}, {Usher}, {Strader},
  {Romanowsky}, {Brodie}, {Arnold}, \& {Spitler}}]{Forbes2013}
{Forbes} D., {Pota} V., {Usher} C., {Strader} J., {Romanowsky} A., {Brodie} J.,
  {Arnold} J., {Spitler} L., 2013, {MNRAS}, in press (arXiv:1306.5245)

\bibitem[{{Forbes}(1996)}]{1996AJ....112.1409F}
{Forbes} D.~A., 1996, \aj, 112, 1409

\bibitem[{{Forbes}(2005)}]{2005ApJ...635L.137F}
---, 2005, \apjl, 635, L137

\bibitem[{{Forbes} {et~al.}(2001){Forbes}, {Beasley}, {Brodie}, \&
  {Kissler-Patig}}]{2001ApJ...563L.143F}
{Forbes} D.~A., {Beasley} M.~A., {Brodie} J.~P., {Kissler-Patig} M., 2001,
  \apjl, 563, L143

\bibitem[{{Forbes} {et~al.}(1996){Forbes}, {Franx}, {Illingworth}, \&
  {Carollo}}]{1996ApJ...467..126F}
{Forbes} D.~A., {Franx} M., {Illingworth} G.~D., {Carollo} C.~M., 1996, \apj,
  467, 126

\bibitem[{{Forbes} {et~al.}(2012){Forbes}, {Ponman}, \&
  {O'Sullivan}}]{2012MNRAS.425...66F}
{Forbes} D.~A., {Ponman} T., {O'Sullivan} E., 2012, \mnras, 425, 66

\bibitem[{{Forbes} {et~al.}(2011){Forbes}, {Spitler}, {Strader}, {Romanowsky},
  {Brodie}, \& {Foster}}]{2011MNRAS.413.2943F}
{Forbes} D.~A., {Spitler} L.~R., {Strader} J., {Romanowsky} A.~J., {Brodie}
  J.~P., {Foster} C., 2011, \mnras, 413, 2943

\bibitem[{{Forte} {et~al.}(2001){Forte}, {Geisler}, {Ostrov}, {Piatti}, \&
  {Gieren}}]{2001AJ....121.1992F}
{Forte} J.~C., {Geisler} D., {Ostrov} P.~G., {Piatti} A.~E., {Gieren} W., 2001,
  \aj, 121, 1992

\bibitem[{{Forte} {et~al.}(2012){Forte}, {Vega}, \&
  {Faifer}}]{2012MNRAS.421..635F}
{Forte} J.~C., {Vega} E.~I., {Faifer} F., 2012, \mnras, 421, 635

\bibitem[{{Foster} {et~al.}(2010){Foster}, {Forbes}, {Proctor}, {Strader},
  {Brodie}, \& {Spitler}}]{2010AJ....139.1566F}
{Foster} C., {Forbes} D.~A., {Proctor} R.~N., {Strader} J., {Brodie} J.~P.,
  {Spitler} L.~R., 2010, \aj, 139, 1566

\bibitem[{{Geisler} {et~al.}(1996){Geisler}, {Lee}, \&
  {Kim}}]{1996AJ....111.1529G}
{Geisler} D., {Lee} M.~G., {Kim} E., 1996, \aj, 111, 1529

\bibitem[{{Georgiev} {et~al.}(2010){Georgiev}, {Puzia}, {Goudfrooij}, \&
  {Hilker}}]{2010MNRAS.406.1967G}
{Georgiev} I.~Y., {Puzia} T.~H., {Goudfrooij} P., {Hilker} M., 2010, \mnras,
  406, 1967

\bibitem[{{Georgiev} {et~al.}(2009){Georgiev}, {Puzia}, {Hilker}, \&
  {Goudfrooij}}]{2009MNRAS.392..879G}
{Georgiev} I.~Y., {Puzia} T.~H., {Hilker} M., {Goudfrooij} P., 2009, \mnras,
  392, 879

\bibitem[{{Giroletti} {et~al.}(2005){Giroletti}, {Taylor}, \&
  {Giovannini}}]{2005ApJ...622..178G}
{Giroletti} M., {Taylor} G.~B., {Giovannini} G., 2005, \apj, 622, 178

\bibitem[{{G{\'o}mez} \& {Woodley}(2007)}]{2007ApJ...670L.105G}
{G{\'o}mez} M., {Woodley} K.~A., 2007, \apjl, 670, L105

\bibitem[{{Graham} \& {Driver}(2005)}]{2005PASA...22..118G}
{Graham} A.~W., {Driver} S.~P., 2005, \pasa, 22, 118

\bibitem[{{Gregory} \& {Thompson}(1977)}]{1977ApJ...213..345G}
{Gregory} S.~A., {Thompson} L.~A., 1977, \apj, 213, 345

\bibitem[{{Hargis} \& {Rhode}(2012)}]{2012AJ....144..164H}
{Hargis} J.~R., {Rhode} K.~L., 2012, \aj, 144, 164

\bibitem[{{Harris} \& {Harris}(2011)}]{2011MNRAS.410.2347H}
{Harris} G.~L.~H., {Harris} W.~E., 2011, \mnras, 410, 2347

\bibitem[{{Harris}(2001)}]{2001stcl.conf..223H}
{Harris} W.~E., 2001, in Saas-Fee Advanced Course 28: Star Clusters,
  {L.~Labhardt \& B.~Binggeli}, ed., p. 223

\bibitem[{{Harris}(2009{\natexlab{a}})}]{2009ApJ...699..254H}
---, 2009{\natexlab{a}}, \apj, 699, 254

\bibitem[{{Harris}(2009{\natexlab{b}})}]{2009ApJ...703..939H}
---, 2009{\natexlab{b}}, \apj, 703, 939

\bibitem[{{Harris} {et~al.}(2013){Harris}, {Harris}, \&
  {Alessi}}]{2013ApJ...772...82H}
{Harris} W.~E., {Harris} G.~L.~H., {Alessi} M., 2013, \apj, 772, 82

\bibitem[{{Harris} {et~al.}(2010){Harris}, {Spitler}, {Forbes}, \&
  {Bailin}}]{2010MNRAS.401.1965H}
{Harris} W.~E., {Spitler} L.~R., {Forbes} D.~A., {Bailin} J., 2010, \mnras,
  401, 1965

\bibitem[{{Harris} \& {van den Bergh}(1981)}]{1981AJ.....86.1627H}
{Harris} W.~E., {van den Bergh} S., 1981, \aj, 86, 1627

\bibitem[{{Harris} {et~al.}(2006){Harris}, {Whitmore}, {Karakla}, {Oko{\'n}},
  {Baum}, {Hanes}, \& {Kavelaars}}]{2006ApJ...636...90H}
{Harris} W.~E., {Whitmore} B.~C., {Karakla} D., {Oko{\'n}} W., {Baum} W.~A.,
  {Hanes} D.~A., {Kavelaars} J.~J., 2006, \apj, 636, 90

\bibitem[{{Ibata} {et~al.}(2011){Ibata}, {Sollima}, {Nipoti}, {Bellazzini},
  {Chapman}, \& {Dalessandro}}]{2011ApJ...738..186I}
{Ibata} R., {Sollima} A., {Nipoti} C., {Bellazzini} M., {Chapman} S.~C.,
  {Dalessandro} E., 2011, \apj, 738, 186

\bibitem[{{Ivezi{\'c}} {et~al.}(2007){Ivezi{\'c}}, {Smith}, {Miknaitis}, {Lin},
  {Tucker}, {Lupton}, {Gunn}, {Knapp}, {Strauss}, {Sesar}, {Doi}, {Tanaka},
  {Fukugita}, {Holtzman}, {Kent}, {Yanny}, {Schlegel}, {Finkbeiner},
  {Padmanabhan}, {Rockosi}, {Juri{\'c}}, {Bond}, {Lee}, {Stoughton}, {Jester},
  {Harris}, {Harding}, {Morrison}, {Brinkmann}, {Schneider}, \&
  {York}}]{2007AJ....134..973I}
{Ivezi{\'c}} {\v Z}., {Smith} J.~A., {Miknaitis} G., {Lin} H., {Tucker} D.,
  {Lupton} R.~H., {Gunn} J.~E., {Knapp} G.~R., {Strauss} M.~A., {Sesar} B.,
  {Doi} M., {Tanaka} M., {Fukugita} M., {Holtzman} J., {Kent} S., {Yanny} B.,
  {Schlegel} D., {Finkbeiner} D., {Padmanabhan} N., {Rockosi} C.~M.,
  {Juri{\'c}} M., {Bond} N., {Lee} B., {Stoughton} C., {Jester} S., {Harris}
  H., {Harding} P., {Morrison} H., {Brinkmann} J., {Schneider} D.~P., {York}
  D., 2007, \aj, 134, 973

\bibitem[{{Jacoby} {et~al.}(1996){Jacoby}, {Ciardullo}, \&
  {Harris}}]{1996ApJ...462....1J}
{Jacoby} G.~H., {Ciardullo} R., {Harris} W.~E., 1996, \apj, 462, 1

\bibitem[{{Jord{\'a}n}(2004)}]{2004ApJ...613L.117J}
{Jord{\'a}n} A., 2004, \apjl, 613, L117

\bibitem[{{Jord{\'a}n} {et~al.}(2007){Jord{\'a}n}, {Blakeslee}, {C{\^o}t{\'e}},
  {Ferrarese}, {Infante}, {Mei}, {Merritt}, {Peng}, {Tonry}, \&
  {West}}]{2007ApJS..169..213J}
{Jord{\'a}n} A., {Blakeslee} J.~P., {C{\^o}t{\'e}} P., {Ferrarese} L.,
  {Infante} L., {Mei} S., {Merritt} D., {Peng} E.~W., {Tonry} J.~L., {West}
  M.~J., 2007, \apjs, 169, 213

\bibitem[{{Jord{\'a}n} {et~al.}(2005){Jord{\'a}n}, {C{\^o}t{\'e}}, {Blakeslee},
  {Ferrarese}, {McLaughlin}, {Mei}, {Peng}, {Tonry}, {Merritt},
  {Milosavljevi{\'c}}, {Sarazin}, {Sivakoff}, \& {West}}]{2005ApJ...634.1002J}
{Jord{\'a}n} A., {C{\^o}t{\'e}} P., {Blakeslee} J.~P., {Ferrarese} L.,
  {McLaughlin} D.~E., {Mei} S., {Peng} E.~W., {Tonry} J.~L., {Merritt} D.,
  {Milosavljevi{\'c}} M., {Sarazin} C.~L., {Sivakoff} G.~R., {West} M.~J.,
  2005, \apj, 634, 1002

\bibitem[{{King}(1962)}]{1962AJ.....67..471K}
{King} I., 1962, \aj, 67, 471

\bibitem[{{Kundu} \& {Whitmore}(1998)}]{1998AJ....116.2841K}
{Kundu} A., {Whitmore} B.~C., 1998, \aj, 116, 2841

\bibitem[{{Kundu} \& {Whitmore}(2001)}]{2001AJ....121.2950K}
---, 2001, \aj, 121, 2950

\bibitem[{{Kuntschner} {et~al.}(2010){Kuntschner}, {Emsellem}, {Bacon},
  {Cappellari}, {Davies}, {de Zeeuw}, {Falc{\'o}n-Barroso}, {Krajnovi{\'c}},
  {McDermid}, {Peletier}, {Sarzi}, {Shapiro}, {van den Bosch}, \& {van de
  Ven}}]{2010MNRAS.408...97K}
{Kuntschner} H., {Emsellem} E., {Bacon} R., {Cappellari} M., {Davies} R.~L.,
  {de Zeeuw} P.~T., {Falc{\'o}n-Barroso} J., {Krajnovi{\'c}} D., {McDermid}
  R.~M., {Peletier} R.~F., {Sarzi} M., {Shapiro} K.~L., {van den Bosch}
  R.~C.~E., {van de Ven} G., 2010, \mnras, 408, 97

\bibitem[{{Landolt}(1992)}]{1992AJ....104..340L}
{Landolt} A.~U., 1992, \aj, 104, 340

\bibitem[{{Larsen}(1999)}]{1999A&AS..139..393L}
{Larsen} S.~S., 1999, \aaps, 139, 393

\bibitem[{{Larsen} \& {Brodie}(2003)}]{2003ApJ...593..340L}
{Larsen} S.~S., {Brodie} J.~P., 2003, \apj, 593, 340

\bibitem[{{Larsen} {et~al.}(2001){Larsen}, {Brodie}, {Huchra}, {Forbes}, \&
  {Grillmair}}]{2001AJ....121.2974L}
{Larsen} S.~S., {Brodie} J.~P., {Huchra} J.~P., {Forbes} D.~A., {Grillmair}
  C.~J., 2001, \aj, 121, 2974

\bibitem[{{Liu} {et~al.}(2011){Liu}, {Peng}, {Jord{\'a}n}, {Ferrarese},
  {Blakeslee}, {C{\^o}t{\'e}}, \& {Mei}}]{2011ApJ...728..116L}
{Liu} C., {Peng} E.~W., {Jord{\'a}n} A., {Ferrarese} L., {Blakeslee} J.~P.,
  {C{\^o}t{\'e}} P., {Mei} S., 2011, \apj, 728, 116

\bibitem[{{Madrid} {et~al.}(2012){Madrid}, {Hurley}, \&
  {Sippel}}]{2012ApJ...756..167M}
{Madrid} J.~P., {Hurley} J.~R., {Sippel} A.~C., 2012, \apj, 756, 167

\bibitem[{{Masters} {et~al.}(2010){Masters}, {Jord{\'a}n}, {C{\^o}t{\'e}},
  {Ferrarese}, {Blakeslee}, {Infante}, {Peng}, {Mei}, \&
  {West}}]{2010ApJ...715.1419M}
{Masters} K.~L., {Jord{\'a}n} A., {C{\^o}t{\'e}} P., {Ferrarese} L.,
  {Blakeslee} J.~P., {Infante} L., {Peng} E.~W., {Mei} S., {West} M.~J., 2010,
  \apj, 715, 1419

\bibitem[{{McLaughlin}(2000)}]{2000ApJ...539..618M}
{McLaughlin} D.~E., 2000, \apj, 539, 618

\bibitem[{{Mei} {et~al.}(2007){Mei}, {Blakeslee}, {C{\^o}t{\'e}}, {Tonry},
  {West}, {Ferrarese}, {Jord{\'a}n}, {Peng}, {Anthony}, \&
  {Merritt}}]{2007ApJ...655..144M}
{Mei} S., {Blakeslee} J.~P., {C{\^o}t{\'e}} P., {Tonry} J.~L., {West} M.~J.,
  {Ferrarese} L., {Jord{\'a}n} A., {Peng} E.~W., {Anthony} A., {Merritt} D.,
  2007, \apj, 655, 144

\bibitem[{{Mieske} {et~al.}(2006){Mieske}, {Jord{\'a}n}, {C{\^o}t{\'e}},
  {Kissler-Patig}, {Peng}, {Ferrarese}, {Blakeslee}, {Mei}, {Merritt}, {Tonry},
  \& {West}}]{2006ApJ...653..193M}
{Mieske} S., {Jord{\'a}n} A., {C{\^o}t{\'e}} P., {Kissler-Patig} M., {Peng}
  E.~W., {Ferrarese} L., {Blakeslee} J.~P., {Mei} S., {Merritt} D., {Tonry}
  J.~L., {West} M.~J., 2006, \apj, 653, 193

\bibitem[{{Mieske} {et~al.}(2010){Mieske}, {Jord{\'a}n}, {C{\^o}t{\'e}},
  {Peng}, {Ferrarese}, {Blakeslee}, {Mei}, {Baumgardt}, {Tonry}, {Infante}, \&
  {West}}]{2010ApJ...710.1672M}
{Mieske} S., {Jord{\'a}n} A., {C{\^o}t{\'e}} P., {Peng} E.~W., {Ferrarese} L.,
  {Blakeslee} J.~P., {Mei} S., {Baumgardt} H., {Tonry} J.~L., {Infante} L.,
  {West} M.~J., 2010, \apj, 710, 1672

\bibitem[{{Miyazaki} {et~al.}(2002){Miyazaki}, {Komiyama}, {Sekiguchi},
  {Okamura}, {Doi}, {Furusawa}, {Hamabe}, {Imi}, {Kimura}, {Nakata}, {Okada},
  {Ouchi}, {Shimasaku}, {Yagi}, \& {Yasuda}}]{2002PASJ...54..833M}
{Miyazaki} S., {Komiyama} Y., {Sekiguchi} M., {Okamura} S., {Doi} M.,
  {Furusawa} H., {Hamabe} M., {Imi} K., {Kimura} M., {Nakata} F., {Okada} N.,
  {Ouchi} M., {Shimasaku} K., {Yagi} M., {Yasuda} N., 2002, \pasj, 54, 833

\bibitem[{{Monet} {et~al.}(2003){Monet}, {Levine}, {Canzian}, {Ables}, {Bird},
  {Dahn}, {Guetter}, {Harris}, {Henden}, {Leggett}, {Levison}, {Luginbuhl},
  {Martini}, {Monet}, {Munn}, {Pier}, {Rhodes}, {Riepe}, {Sell}, {Stone},
  {Vrba}, {Walker}, {Westerhout}, {Brucato}, {Reid}, {Schoening}, {Hartley},
  {Read}, \& {Tritton}}]{2003AJ....125..984M}
{Monet} D.~G., {Levine} S.~E., {Canzian} B., {Ables} H.~D., {Bird} A.~R.,
  {Dahn} C.~C., {Guetter} H.~H., {Harris} H.~C., {Henden} A.~A., {Leggett}
  S.~K., {Levison} H.~F., {Luginbuhl} C.~B., {Martini} J., {Monet} A.~K.~B.,
  {Munn} J.~A., {Pier} J.~R., {Rhodes} A.~R., {Riepe} B., {Sell} S., {Stone}
  R.~C., {Vrba} F.~J., {Walker} R.~L., {Westerhout} G., {Brucato} R.~J., {Reid}
  I.~N., {Schoening} W., {Hartley} M., {Read} M.~A., {Tritton} S.~B., 2003,
  \aj, 125, 984

\bibitem[{{Morganti} {et~al.}(2006){Morganti}, {de Zeeuw}, {Oosterloo},
  {McDermid}, {Krajnovi{\'c}}, {Cappellari}, {Kenn}, {Weijmans}, \&
  {Sarzi}}]{2006MNRAS.371..157M}
{Morganti} R., {de Zeeuw} P.~T., {Oosterloo} T.~A., {McDermid} R.~M.,
  {Krajnovi{\'c}} D., {Cappellari} M., {Kenn} F., {Weijmans} A., {Sarzi} M.,
  2006, \mnras, 371, 157

\bibitem[{{Muratov} \& {Gnedin}(2010)}]{2010ApJ...718.1266M}
{Muratov} A.~L., {Gnedin} O.~Y., 2010, \apj, 718, 1266

\bibitem[{{Newman} {et~al.}(2012){Newman}, {Cooper}, {Davis}, {Faber}, {Coil},
  {Guhathakurta}, {Koo}, {Phillips}, {Conroy}, {Dutton}, {Finkbeiner}, {Gerke},
  {Rosario}, {Weiner}, {Willmer}, {Yan}, {Harker}, {Kassin}, {Konidaris},
  {Lai}, {Madgwick}, {Noeske}, {Wirth}, {Connolly}, {Kaiser}, {Kirby},
  {Lemaux}, {Lin}, {Lotz}, {Luppino}, {Marinoni}, {Matthews}, {Metevier}, \&
  {Schiavon}}]{Newman2012}
{Newman} J.~A., {Cooper} M.~C., {Davis} M., {Faber} S.~M., {Coil} A.~L.,
  {Guhathakurta} P., {Koo} D.~C., {Phillips} A.~C., {Conroy} C., {Dutton}
  A.~A., {Finkbeiner} D.~P., {Gerke} B.~F., {Rosario} D.~J., {Weiner} B.~J.,
  {Willmer} C.~N.~A., {Yan} R., {Harker} J.~J., {Kassin} S.~A., {Konidaris}
  N.~P., {Lai} K., {Madgwick} D.~S., {Noeske} K.~G., {Wirth} G.~D., {Connolly}
  A.~J., {Kaiser} N., {Kirby} E.~N., {Lemaux} B.~C., {Lin} L., {Lotz} J.~M.,
  {Luppino} G.~A., {Marinoni} C., {Matthews} D.~J., {Metevier} A., {Schiavon}
  R.~P., 2012, {ApJS}, submitted (arXiv:1203.3192)

\bibitem[{{Oser} {et~al.}(2010){Oser}, {Ostriker}, {Naab}, {Johansson}, \&
  {Burkert}}]{2010ApJ...725.2312O}
{Oser} L., {Ostriker} J.~P., {Naab} T., {Johansson} P.~H., {Burkert} A., 2010,
  \apj, 725, 2312

\bibitem[{{Ouchi} {et~al.}(2004){Ouchi}, {Shimasaku}, {Okamura}, {Furusawa},
  {Kashikawa}, {Ota}, {Doi}, {Hamabe}, {Kimura}, {Komiyama}, {Miyazaki},
  {Miyazaki}, {Nakata}, {Sekiguchi}, {Yagi}, \& {Yasuda}}]{2004ApJ...611..660O}
{Ouchi} M., {Shimasaku} K., {Okamura} S., {Furusawa} H., {Kashikawa} N., {Ota}
  K., {Doi} M., {Hamabe} M., {Kimura} M., {Komiyama} Y., {Miyazaki} M.,
  {Miyazaki} S., {Nakata} F., {Sekiguchi} M., {Yagi} M., {Yasuda} N., 2004,
  \apj, 611, 660

\bibitem[{{Paolillo} {et~al.}(2011){Paolillo}, {Puzia}, {Goudfrooij}, {Zepf},
  {Maccarone}, {Kundu}, {Fabbiano}, \& {Angelini}}]{2011ApJ...736...90P}
{Paolillo} M., {Puzia} T.~H., {Goudfrooij} P., {Zepf} S.~E., {Maccarone} T.~J.,
  {Kundu} A., {Fabbiano} G., {Angelini} L., 2011, \apj, 736, 90

\bibitem[{{Park} \& {Lee}(2013)}]{2013ApJ...773L..27P}
{Park} H.~S., {Lee} M.~G., 2013, \apjl, 773, L27

\bibitem[{{Peng} {et~al.}(2006){Peng}, {Jord{\'a}n}, {C{\^o}t{\'e}},
  {Blakeslee}, {Ferrarese}, {Mei}, {West}, {Merritt}, {Milosavljevi{\'c}}, \&
  {Tonry}}]{2006ApJ...639...95P}
{Peng} E.~W., {Jord{\'a}n} A., {C{\^o}t{\'e}} P., {Blakeslee} J.~P.,
  {Ferrarese} L., {Mei} S., {West} M.~J., {Merritt} D., {Milosavljevi{\'c}} M.,
  {Tonry} J.~L., 2006, \apj, 639, 95

\bibitem[{{Peng} {et~al.}(2008){Peng}, {Jord{\'a}n}, {C{\^o}t{\'e}},
  {Takamiya}, {West}, {Blakeslee}, {Chen}, {Ferrarese}, {Mei}, {Tonry}, \&
  {West}}]{2008ApJ...681..197P}
{Peng} E.~W., {Jord{\'a}n} A., {C{\^o}t{\'e}} P., {Takamiya} M., {West} M.~J.,
  {Blakeslee} J.~P., {Chen} C., {Ferrarese} L., {Mei} S., {Tonry} J.~L., {West}
  A.~A., 2008, \apj, 681, 197

\bibitem[{{Pota} {et~al.}(2013){Pota}, {Forbes}, {Romanowsky}, {Brodie},
  {Spitler}, {Strader}, {Foster}, {Arnold}, {Benson}, {Blom}, {Hargis},
  {Rhode}, \& {Usher}}]{2013MNRAS.428..389P}
{Pota} V., {Forbes} D.~A., {Romanowsky} A.~J., {Brodie} J.~P., {Spitler} L.~R.,
  {Strader} J., {Foster} C., {Arnold} J.~A., {Benson} A., {Blom} C., {Hargis}
  J.~R., {Rhode} K.~L., {Usher} C., 2013, \mnras, 428, 389

\bibitem[{{Puzia} {et~al.}(2005){Puzia}, {Kissler-Patig}, {Thomas}, {Maraston},
  {Saglia}, {Bender}, {Goudfrooij}, \& {Hempel}}]{2005A&A...439..997P}
{Puzia} T.~H., {Kissler-Patig} M., {Thomas} D., {Maraston} C., {Saglia} R.~P.,
  {Bender} R., {Goudfrooij} P., {Hempel} M., 2005, \aap, 439, 997

\bibitem[{{Rhode}(2012)}]{2012AJ....144..154R}
{Rhode} K.~L., 2012, \aj, 144, 154

\bibitem[{{Schlegel} {et~al.}(1998){Schlegel}, {Finkbeiner}, \&
  {Davis}}]{1998ApJ...500..525S}
{Schlegel} D.~J., {Finkbeiner} D.~P., {Davis} M., 1998, \apj, 500, 525

\bibitem[{{S{\'e}rsic}(1963)}]{1963BAAA....6...41S}
{S{\'e}rsic} J.~L., 1963, Boletin de la Asociacion Argentina de Astronomia La
  Plata Argentina, 6, 41

\bibitem[{{Shapiro} {et~al.}(2010){Shapiro}, {Falc{\'o}n-Barroso}, {van de
  Ven}, {de Zeeuw}, {Sarzi}, {Bacon}, {Bolatto}, {Cappellari}, {Croton},
  {Davies}, {Emsellem}, {Fakhouri}, {Krajnovi{\'c}}, {Kuntschner}, {McDermid},
  {Peletier}, {van den Bosch}, \& {van der Wolk}}]{2010MNRAS.402.2140S}
{Shapiro} K.~L., {Falc{\'o}n-Barroso} J., {van de Ven} G., {de Zeeuw} P.~T.,
  {Sarzi} M., {Bacon} R., {Bolatto} A., {Cappellari} M., {Croton} D., {Davies}
  R.~L., {Emsellem} E., {Fakhouri} O., {Krajnovi{\'c}} D., {Kuntschner} H.,
  {McDermid} R.~M., {Peletier} R.~F., {van den Bosch} R.~C.~E., {van der Wolk}
  G., 2010, \mnras, 402, 2140

\bibitem[{{Sippel} {et~al.}(2012){Sippel}, {Hurley}, {Madrid}, \&
  {Harris}}]{2012MNRAS.427..167S}
{Sippel} A.~C., {Hurley} J.~R., {Madrid} J.~P., {Harris} W.~E., 2012, \mnras,
  427, 167

\bibitem[{{Sirianni} {et~al.}(2005){Sirianni}, {Jee}, {Ben{\'{\i}}tez},
  {Blakeslee}, {Martel}, {Meurer}, {Clampin}, {De Marchi}, {Ford}, {Gilliland},
  {Hartig}, {Illingworth}, {Mack}, \& {McCann}}]{2005PASP..117.1049S}
{Sirianni} M., {Jee} M.~J., {Ben{\'{\i}}tez} N., {Blakeslee} J.~P., {Martel}
  A.~R., {Meurer} G., {Clampin} M., {De Marchi} G., {Ford} H.~C., {Gilliland}
  R., {Hartig} G.~F., {Illingworth} G.~D., {Mack} J., {McCann} W.~J., 2005,
  \pasp, 117, 1049

\bibitem[{{Skrutskie} {et~al.}(2006){Skrutskie}, {Cutri}, {Stiening},
  {Weinberg}, {Schneider}, {Carpenter}, {Beichman}, {Capps}, {Chester},
  {Elias}, {Huchra}, {Liebert}, {Lonsdale}, {Monet}, {Price}, {Seitzer},
  {Jarrett}, {Kirkpatrick}, {Gizis}, {Howard}, {Evans}, {Fowler}, {Fullmer},
  {Hurt}, {Light}, {Kopan}, {Marsh}, {McCallon}, {Tam}, {Van Dyk}, \&
  {Wheelock}}]{2006AJ....131.1163S}
{Skrutskie} M.~F., {Cutri} R.~M., {Stiening} R., {Weinberg} M.~D., {Schneider}
  S., {Carpenter} J.~M., {Beichman} C., {Capps} R., {Chester} T., {Elias} J.,
  {Huchra} J., {Liebert} J., {Lonsdale} C., {Monet} D.~G., {Price} S.,
  {Seitzer} P., {Jarrett} T., {Kirkpatrick} J.~D., {Gizis} J.~E., {Howard} E.,
  {Evans} T., {Fowler} J., {Fullmer} L., {Hurt} R., {Light} R., {Kopan} E.~L.,
  {Marsh} K.~A., {McCallon} H.~L., {Tam} R., {Van Dyk} S., {Wheelock} S., 2006,
  \aj, 131, 1163

\bibitem[{{Spitler} \& {Forbes}(2009)}]{2009MNRAS.392L...1S}
{Spitler} L.~R., {Forbes} D.~A., 2009, \mnras, 392, L1

\bibitem[{{Spitler} {et~al.}(2006){Spitler}, {Larsen}, {Strader}, {Brodie},
  {Forbes}, \& {Beasley}}]{2006AJ....132.1593S}
{Spitler} L.~R., {Larsen} S.~S., {Strader} J., {Brodie} J.~P., {Forbes} D.~A.,
  {Beasley} M.~A., 2006, \aj, 132, 1593

\bibitem[{{Stetson}(1992)}]{1992ASPC...25..297S}
{Stetson} P.~B., 1992, in Astronomical Society of the Pacific Conference
  Series, Vol.~25, Astronomical Data Analysis Software and Systems I,
  {D.~M.~Worrall, C.~Biemesderfer, \& J.~Barnes}, ed., pp. 297--+

\bibitem[{{Stetson}(2000)}]{2000PASP..112..925S}
---, 2000, \pasp, 112, 925

\bibitem[{{Strader} {et~al.}(2007){Strader}, {Beasley}, \&
  {Brodie}}]{2007AJ....133.2015S}
{Strader} J., {Beasley} M.~A., {Brodie} J.~P., 2007, \aj, 133, 2015

\bibitem[{{Strader} {et~al.}(2005){Strader}, {Brodie}, {Cenarro}, {Beasley}, \&
  {Forbes}}]{2005AJ....130.1315S}
{Strader} J., {Brodie} J.~P., {Cenarro} A.~J., {Beasley} M.~A., {Forbes} D.~A.,
  2005, \aj, 130, 1315

\bibitem[{{Strader} {et~al.}(2006){Strader}, {Brodie}, {Spitler}, \&
  {Beasley}}]{2006AJ....132.2333S}
{Strader} J., {Brodie} J.~P., {Spitler} L., {Beasley} M.~A., 2006, \aj, 132,
  2333

\bibitem[{{Strader} {et~al.}(2012){Strader}, {Fabbiano}, {Luo}, {Kim},
  {Brodie}, {Fragos}, {Gallagher}, {Kalogera}, {King}, \&
  {Zezas}}]{2012ApJ...760...87S}
{Strader} J., {Fabbiano} G., {Luo} B., {Kim} D.-W., {Brodie} J.~P., {Fragos}
  T., {Gallagher} J.~S., {Kalogera} V., {King} A., {Zezas} A., 2012, \apj, 760,
  87

\bibitem[{{Strader} {et~al.}(2011){Strader}, {Romanowsky}, {Brodie}, {Spitler},
  {Beasley}, {Arnold}, {Tamura}, {Sharples}, \&
  {Arimoto}}]{2011ApJS..197...33S}
{Strader} J., {Romanowsky} A.~J., {Brodie} J.~P., {Spitler} L.~R., {Beasley}
  M.~A., {Arnold} J.~A., {Tamura} N., {Sharples} R.~M., {Arimoto} N., 2011,
  \apjs, 197, 33

\bibitem[{{Strader} \& {Smith}(2008)}]{2008AJ....136.1828S}
{Strader} J., {Smith} G.~H., 2008, \aj, 136, 1828

\bibitem[{{Taylor}(2005)}]{2005ASPC..347...29T}
{Taylor} M.~B., 2005, in Astronomical Society of the Pacific Conference Series,
  Vol. 347, Astronomical Data Analysis Software and Systems XIV, {Shopbell} P.,
  {Britton} M., {Ebert} R., eds., p.~29

\bibitem[{{Tonry} {et~al.}(2001){Tonry}, {Dressler}, {Blakeslee}, {Ajhar},
  {Fletcher}, {Luppino}, {Metzger}, \& {Moore}}]{2001ApJ...546..681T}
{Tonry} J.~L., {Dressler} A., {Blakeslee} J.~P., {Ajhar} E.~A., {Fletcher}
  A.~B., {Luppino} G.~A., {Metzger} M.~R., {Moore} C.~B., 2001, \apj, 546, 681

\bibitem[{{Tully} \& {Fisher}(1988)}]{1988cng..book.....T}
{Tully} R.~B., {Fisher} J.~R., 1988, {Catalog of Nearby Galaxies}

\bibitem[{{Usher} {et~al.}(2012){Usher}, {Forbes}, {Brodie}, {Foster},
  {Spitler}, {Arnold}, {Romanowsky}, {Strader}, \&
  {Pota}}]{2012MNRAS.426.1475U}
{Usher} C., {Forbes} D.~A., {Brodie} J.~P., {Foster} C., {Spitler} L.~R.,
  {Arnold} J.~A., {Romanowsky} A.~J., {Strader} J., {Pota} V., 2012, \mnras,
  426, 1475

\bibitem[{{van den Bergh}(1956)}]{1956ZA.....41...61V}
{van den Bergh} S., 1956, \zap, 41, 61

\bibitem[{{van den Bergh} {et~al.}(1991){van den Bergh}, {Morbey}, \&
  {Pazder}}]{1991ApJ...375..594V}
{van den Bergh} S., {Morbey} C., {Pazder} J., 1991, \apj, 375, 594

\bibitem[{{Vazdekis} {et~al.}(2003){Vazdekis}, {Cenarro}, {Gorgas}, {Cardiel},
  \& {Peletier}}]{2003MNRAS.340.1317V}
{Vazdekis} A., {Cenarro} A.~J., {Gorgas} J., {Cardiel} N., {Peletier} R.~F.,
  2003, \mnras, 340, 1317

\bibitem[{{Villegas} {et~al.}(2010){Villegas}, {Jord{\'a}n}, {Peng},
  {Blakeslee}, {C{\^o}t{\'e}}, {Ferrarese}, {Kissler-Patig}, {Mei}, {Infante},
  {Tonry}, \& {West}}]{2010ApJ...717..603V}
{Villegas} D., {Jord{\'a}n} A., {Peng} E.~W., {Blakeslee} J.~P., {C{\^o}t{\'e}}
  P., {Ferrarese} L., {Kissler-Patig} M., {Mei} S., {Infante} L., {Tonry}
  J.~L., {West} M.~J., 2010, \apj, 717, 603

\bibitem[{{Webb} {et~al.}(2012{\natexlab{a}}){Webb}, {Harris}, \&
  {Sills}}]{2012ApJ...759L..39W}
{Webb} J.~J., {Harris} W.~E., {Sills} A., 2012{\natexlab{a}}, \apjl, 759, L39

\bibitem[{{Webb} {et~al.}(2012{\natexlab{b}}){Webb}, {Sills}, \&
  {Harris}}]{2012ApJ...746...93W}
{Webb} J.~J., {Sills} A., {Harris} W.~E., 2012{\natexlab{b}}, \apj, 746, 93

\bibitem[{{Yagi} {et~al.}(2002){Yagi}, {Yoshihiko}, {Ogasawara}, {Kosugi},
  {Takata}, {Ishihara}, {Yokono}, {Morita}, {Nakamoto}, {Watanabe}, \&
  {Ukawa}}]{2002SPIE.4847..322Y}
{Yagi} M., {Yoshihiko} M., {Ogasawara} R., {Kosugi} G., {Takata} T., {Ishihara}
  Y., {Yokono} Y., {Morita} Y., {Nakamoto} H., {Watanabe} N., {Ukawa} K., 2002,
  in Society of Photo-Optical Instrumentation Engineers (SPIE) Conference
  Series, Vol. 4847, Society of Photo-Optical Instrumentation Engineers (SPIE)
  Conference Series, {J.-L.~Starck \& F.~D.~Murtagh}, ed., pp. 322--331

\bibitem[{{Yoon} {et~al.}(2006){Yoon}, {Yi}, \& {Lee}}]{2006Sci...311.1129Y}
{Yoon} S., {Yi} S.~K., {Lee} Y., 2006, Science, 311, 1129

\bibitem[{{Yoon} {et~al.}(2011){Yoon}, {Lee}, {Blakeslee}, {Peng}, {Sohn},
  {Cho}, {Kim}, {Chung}, {Kim}, \& {Lee}}]{2011ApJ...743..150Y}
{Yoon} S.-J., {Lee} S.-Y., {Blakeslee} J.~P., {Peng} E.~W., {Sohn} S.~T., {Cho}
  J., {Kim} H.-S., {Chung} C., {Kim} S., {Lee} Y.-W., 2011, \apj, 743, 150

\bibitem[{{Younes} {et~al.}(2010){Younes}, {Porquet}, {Sabra}, {Grosso},
  {Reeves}, \& {Allen}}]{2010A&A...517A..33Y}
{Younes} G., {Porquet} D., {Sabra} B., {Grosso} N., {Reeves} J.~N., {Allen}
  M.~G., 2010, \aap, 517, A33+

\bibitem[{{Zinn}(1985)}]{1985ApJ...293..424Z}
{Zinn} R., 1985, \apj, 293, 424

\end{thebibliography}

\appendix
\section{Spectroscopically confirmed objects}
\label{spectralist}

\begin{landscape}

\begin{table*}

\caption{\label{tab:spectra} Spectroscopically confirmed globular clusters, galaxies and stars}
\begin{tabular}{l@{  }c@{  }c@{  }c@{  }c@{  }c@{  }c@{  }c@{  }c@{  }c@{  }c} \hline
Name            & RA         & Dec       & $(g-z)$         & $z$              & $r_{hl}$      & $(B-V)$         & $(V-I)$         & $I$              & $v$          & [Z/H]                   \\
                & [deg]      & [deg]     & [mag]           & [mag]            & [pc]          & [mag]           & [mag]           & [mag]            & [km s$^{-1}$ & [dex]                   \\
(1)             & (2)        & (3)       & (4)             & (5)              & (6)           & (7)             & (8)             & (9)              & (10)         & (11)                    \\ \hline
NGC4278\_GC1    & 185.079334 & 29.258192 & $0.91 \pm 0.03$ & $22.19 \pm 0.02$ & $3.4 \pm 0.3$ & $0.72 \pm 0.05$ & $1.06 \pm 0.04$ & $21.66 \pm 0.03$ & $307 \pm 10$ & ---                     \\              
NGC4278\_GC2    & 185.018457 & 29.295004 & $0.84 \pm 0.02$ & $22.08 \pm 0.01$ & $2.0 \pm 0.1$ & $0.83 \pm 0.06$ & $0.91 \pm 0.05$ & $21.69 \pm 0.04$ & $668 \pm 13$ & ---                     \\
NGC4278\_GC3    & 185.038466 & 29.276498 & $0.81 \pm 0.02$ & $20.95 \pm 0.01$ & $2.9 \pm 0.1$ & $0.73 \pm 0.04$ & $0.90 \pm 0.04$ & $20.33 \pm 0.04$ & $1225 \pm 8$ & $-1.93_{-0.13}^{+0.32}$ \\
NGC4278\_GC4    & 185.001351 & 29.315286 & $0.95 \pm 0.03$ & $21.22 \pm 0.01$ & $5.6 \pm 0.2$ & $0.81 \pm 0.04$ & $0.99 \pm 0.03$ & $20.73 \pm 0.02$ & $394 \pm 11$ & $-1.37_{-0.22}^{+0.25}$ \\
NGC4278\_GC5    & 185.032302 & 29.289058 & $1.15 \pm 0.01$ & $20.13 \pm 0.01$ & $2.1 \pm 0.0$ & $0.95 \pm 0.04$ & $1.08 \pm 0.03$ & $19.86 \pm 0.02$ & $235 \pm 8$  & $ 0.20_{-0.43}^{+0.18}$ \\
...             & ...        & ...       & ...             & ...              & ...           & ...             & ...             & ...              & ...          & ...                     \\
NGC4278\_gal1   & 185.002305 & 29.331285 & $0.99 \pm 0.16$ & $24.66 \pm 0.10$ & $2.9 \pm 1.7$ & ---             & ---             & ---              & $z = 0.74$   & ---                     \\
NGC4278\_gal2   & 185.061485 & 29.242873 & $1.16 \pm 0.05$ & $23.12 \pm 0.03$ & $1.9 \pm 0.4$ & ---             & ---             & ---              & $z = 0.43$   & ---                     \\
NGC4278\_gal3   & 184.903458 & 29.276536 & ---             & ---              & ---           & $0.66 \pm 0.04$ & $0.86 \pm 0.03$ & $21.09 \pm 0.02$ & $z = 0.20$   & ---                     \\
NGC4278\_gal4   & 185.084196 & 29.296686 & $0.83 \pm 0.03$ & $22.22 \pm 0.02$ & $5.3 \pm 0.2$ & $0.83 \pm 0.05$ & $0.78 \pm 0.04$ & $21.85 \pm 0.03$ & $z = 0.44$   & ---                     \\
NGC4278\_gal5   & 185.139871 & 29.253525 & ---             & ---              & ---           & $0.73 \pm 0.04$ & $0.79 \pm 0.03$ & $21.75 \pm 0.03$ & $z = 0.15$   & ---                     \\
...             & ...        & ...       & ...             & ...              & ...           & ...             & ...             & ...              & ...          & ...                     \\
NGC4278\_stars1 & 185.065617 & 29.286417 & $3.84 \pm 0.05$ & $21.27 \pm 0.01$ & $0.1 \pm 0.1$ & $0.33 \pm 0.10$ & $2.93 \pm 0.08$ & $21.28 \pm 0.05$ & $-42 \pm 18$ & ---                     \\
NGC4278\_stars2 & 185.027350 & 29.252617 & $0.98 \pm 0.02$ & $20.56 \pm 0.01$ & $0.0 \pm 0.1$ & $0.89 \pm 0.04$ & $0.99 \pm 0.03$ & $20.15 \pm 0.02$ & $ 92 \pm 11$ & ---                     \\
NGC4278\_stars3 & 185.006250 & 29.264183 & $2.77 \pm 0.02$ & $17.69 \pm 0.01$ & $0.0 \pm 0.0$ & $1.72 \pm 0.06$ & $2.15 \pm 0.05$ & $17.51 \pm 0.04$ & $ -6 \pm 5$  & ---                     \\
NGC4278\_stars4 & 185.090008 & 29.184019 & ---             & ---              & ---           & $0.74 \pm 0.03$ & $0.92 \pm 0.03$ & $18.01 \pm 0.02$ & $117 \pm 5$  & ---                     \\
NGC4278\_stars5 & 185.074267 & 29.188356 & ---             & ---              & ---           & $1.01 \pm 0.03$ & $1.08 \pm 0.03$ & $17.01 \pm 0.02$ & $  7 \pm 6$  & ---                     \\
...             & ...        & ...       & ...             & ...              & ...           & ...             & ...             & ...              & ...          & ...                     \\ \hline

\end{tabular}

\medskip
The full version of this table is provided in a machine readable form in the online Supporting Information.
\emph{Notes} Column (1): Identifier from \citetalias{2013MNRAS.428..389P}.
Globular cluster designations start with NGC4278\_GC, galaxy designations with NGC4278\_gal and star designations with NGC4278\_stars.
For objects that were observed spectroscopically for the first time in this work we have extended the \citetalias{2013MNRAS.428..389P} naming scheme to them.
Column (2) and (3): Right ascension and declination in the J2000.0 epoch, respectively.
Column (4): ACS $(g - z)$ colour.
Column (5): ACS $z$-band magnitude.
Column (6): Half light radius in parsecs if the object was at our adopted distance of 15.6 Mpc.
Column (4): Suprime-Cam $(B - V)$ colour.
Column (5): Suprime-Cam $(V - I)$ colour.
Column (6): Suprime-Cam $I$-band magnitude.
Column (7): Radial velocity. For galaxies this is the estimated redshift.
Column (8): Calcium triplet metallicity.
Column (9): Identifier from \citetalias{2013MNRAS.428..389P}.

\end{table*}

\end{landscape}

\end{document}